\documentstyle[12pt,epsf,epsfig,amsmath]{article}
\newfont{\thiplo}{msbm10 scaled\magstep 2}
\newfont{\gothic}{eufb10 scaled\magstep 2}
\newfont{\unc}{eurb10} 
\newskip\humongous \humongous=0pt plus 1000pt minus 1000pt
\def\caja{\mathsurround=0pt}
\def\eqalign#1{\,\vcenter{\openup1\jot \caja
        \ialign{\strut \hfil$\displaystyle{##}$&$
        \displaystyle{{}##}$\hfil\crcr#1\crcr}}\,}
\newif\ifdtup

\def\eqright #1\cr{\noalign{\hfill$\displaystyle{{}#1}$}}
\def\eqleft #1\cr{\noalign{\noindent$\displaystyle{{}#1}$\hfill}}

\def\oldreffmt#1{\rlap{[#1]} \hbox to 2\parindent{}}

\def\figfmt#1{\rlap{Figure {#1}} \hbox to 1in{}}

\def\VEV#1{\left\langle #1\right\rangle}

\def\sectioneq{\def\theequation{\thesection.\arabic{equation}}{\let
\holdsection=\section\def\section{\setcounter{equation}{0}\holdsection}}}%

\newcounter{holdequation}

\def\auto{\eqno(\refstepcounter{equation}\theequation)}
\def\begineq #1\endeq{$$ \refstepcounter{equation}\eqalign{#1}\eqno
	(\theequation) $$}
\def\contlimit{\,{\hbox{$\longrightarrow$}\kern-1.8em\lower1ex
\hbox{${\scriptstyle (a\rightarrow0)}$}}\,}
\def\centeron#1#2{{\setbox0=\hbox{#1}\setbox1=\hbox{#2}\ifdim
\wd1>\wd0\kern.5\wd1\kern-.5\wd0\fi
\copy0\kern-.5\wd0\kern-.5\wd1\copy1\ifdim\wd0>\wd1
\kern.5\wd0\kern-.5\wd1\fi}}
\def\centerover#1#2{\centeron{#1}{\setbox0=\hbox{#1}\setbox
1=\hbox{#2}\raise\ht0\hbox{\raise\dp1\hbox{\copy1}}}}
\def\centerunder#1#2{\centeron{#1}{\setbox0=\hbox{#1}\setbox
1=\hbox{#2}\lower\dp0\hbox{\lower\ht1\hbox{\copy1}}}}
\def\lsim{\;\centeron{\raise.35ex\hbox{$<$}}{\lower.65ex\hbox
{$\sim$}}\;}
\def\gsim{\;\centeron{\raise.35ex\hbox{$>$}}{\lower.65ex\hbox
{$\sim$}}\;}
\def\st#1{\centeron{$#1$}{$/$}}

\def\super#1{\ifmmode \hbox{\textsuper{#1}}\else\textsuper{#1}\fi}
\def\textsuper#1{\newcount\holdspacefactor\holdspacefactor=\spacefactor
$^{#1}$\spacefactor=\holdspacefactor}

\def\getcite#1,{\advance\citenumber by1
\def\getcitearg{#1}\def\lastarg{@}
\ifnum\citenumber=1
\ref{#1}\let\next=\getcite\else\ifx\getcitearg\lastarg\let\next=\relax
\else ,\ref{#1}\let\next=\getcite\fi\fi\next}

\def\pom{{\rm P\kern -0.53em\llap I\,}}
\def\spom{{\rm P\kern -0.36em\llap \small I\,}}
\def\sspom{{\rm P\kern -0.33em\llap \footnotesize I\,}}

\relax
\def\contlimit{\,{\hbox{$\longrightarrow$}\kern-1.8em\lower1ex
\hbox{${\scriptstyle (a\rightarrow0)}$}}\,}
\def\upon #1/#2 {{\textstyle{#1\over #2}}}
\relax
\renewcommand{\thefootnote}{\fnsymbol{footnote}} 

\def\mainhead#1{\setcounter{equation}{0}\addtocounter{section}{1}
  \vbox{\begin{center}\large\bf #1\end{center}}\nobreak\par}
\sectioneq
\def\subhead#1{\bigskip\vbox{\noindent\bf #1}\nobreak\par}

\def\til#1{\centeron{\hbox{$#1$}}{\lower 2ex\hbox{$\char'176$}}}
\def\tild#1{\centeron{\hbox{$\,#1$}}{\lower 2.5ex\hbox{$\char'176$}}}
\def\sumtil{\centeron{\hbox{$\displaystyle\sum$}}{\lower
-1.5ex\hbox{$\widetilde{\phantom{xx}}$}}}

\def\gam{\hat{\gamma}}
\def\kt{\hat{k}}
\def\pt{\hat{p}}
\def\qt{\hat{q}}
\def\Pit{\hat{\Pi}}

\newcommand{\bit}{\begin{itemize}}
\newcommand{\eit}{\end{itemize}}

\newcommand{\beq}{\begin{equation}}
\newcommand{\eeq}{\end{equation}}
\newcommand{\beqa}{\begin{eqnarray}}
\newcommand{\eeqa}{\end{eqnarray}}

\begin{document} 

\begin{titlepage} 

\rightline{\vbox{\halign{&#\hfil\cr
&ANL-HEP-PR-03-95\cr
&\today\cr}}} 
\vspace{0.25in} 

\begin{center} 
  
{\large\bf Electroweak High-Energy Scattering}

{\large\bf and the Chiral Anomaly }\footnote{Work 
supported by the U.S.
Department of Energy under Contract
W-31-109-ENG-38} 

\medskip

Alan. R. White\footnote{arw@hep.anl.gov }

\vskip 0.6cm

\centerline{Argonne National Laboratory}
\centerline{9700 South Cass, Il 60439, USA.}
\vspace{0.5cm}

\end{center}

\begin{abstract} 

The effect of perturbative QCD interactions
on the high energy scattering of electroweak vector bosons,
when the exchanged channel has pion quantum numbers, is considered. 
The chiral anomaly is shown to appear in the couplings of particular 
transverse momentum diagrams, 
producing an enhancement of the scattering amplitude
by a power of the energy. At $O(\alpha_s)$
a single large transverse momentum gluon is involved and, within the transverse
momentum diagram framework, there is no cancelation.
In higher orders, soft gluons, carrying both normal and 
anomalous color parity, are also present.

The manipulation of a transverse momentum cut-off 
to replace the ultra-violet anomaly divergence by
infra-red divergences that can lead to confinement and chiral symmetry breaking
is briefly discussed. Possible implications for electroweak symmetry breaking
are noted.

\end{abstract}

\renewcommand{\thefootnote}{\arabic{footnote}} \end{titlepage}

\mainhead{1. INTRODUCTION}

The cancelation of the chiral anomaly in the electroweak
sector of the Standard Model is crucial for the existence of the model as a
well-defined short-distance field theory. In perturbation theory, the anomaly
is a large momentum contribution in axial vector triangle diagrams
that, if uncanceled, destroys the renormalizability of a left-handed gauge
theory such as the electroweak sector of the Standard Model. 
In the regge limit the feynman diagrams of standard perturbation theory
contract to form transverse momentum diagrams and produce 
a new perturbation expansion that can be organized
into reggeon diagrams\cite{fkl}-\cite{arw93}. Beyond leading-log order,
the external couplings of the 
transverse momentum diagrams (which are also the couplings of the reggeon
diagrams) contain contracted loop diagrams involving
``effective vertices'' that result from the contraction. The 
effective vertices can then produce new ``anomalies'', not present in the normal
perturbation expansion. For QCD, this is demonstrated in the next-to-leading log
calculation\cite{fl1} of gluon scattering, in which  
an infra-red (gluon) triangle anomaly is responsible for the helicity
non-conservation that occurs, and in 
our work on the contribution of infra-red quark loop anomalies to
pion/pomeron\cite{arw02} and triple pomeron\cite{arw021} vertices.

In the electroweak scattering problem considered in this paper, 
the underlying left-handed theory contains
elementary axial vector vertices. It is natural, therefore, that (components of)
these elementary vertices will appear also in the
regge limit effective vertices. A priori, therefore, large momentum contributions, 
directly analagous to the familiar triangle anomaly, can be expected within the
loop diagrams that contribute to (beyond-leading-order) transverse momentum diagram
couplings. Indeed, as we shall see, internal effective vertices,
resulting from longitudinal vector meson exchange, also
appear which are quark current components that have point  
interactions only at infinite momentum. Such current components  
do not appear in the original lagrangian and do not couple to 
leptons. Consequently, in the electroweak regge limit,
we can anticipate a significantly expanded ``anomaly problem'' which the well-known
short distance cancelations, between quarks and leptons, 
will not be sufficient to remove.

In this paper we will show that the triangle anomaly does indeed
appear in the couplings of 
transverse momentum diagrams that describe the high-energy 
scattering of $W^{\pm,0}$ vector mesons. All the diagrams we consider 
describe the exchange of a quark-antiquark pair, with the flavor quantum numbers
of the pion, together with some number of gluons.
We choose pion quantum numbers because our ultimate goal 
is to understand the relationship between the anomaly and chiral symmetry breaking
in the context of high-energy scattering.
Since two fermion exchange is involved, we would expect 
the energy dependence to be, at most, logarithmic. (Pion exchange
would have no energy dependence.) The signal of the anomaly 
will be a power divergence of transverse momentum integrals that
produces an additional power of the energy in the full amplitude. 

Since the anomaly phenomenon we discuss 
involves longitudinal vector meson states it is natural
to expect that 
the underlying gauge invariance of the electroweak theory will
be responsible for some form of cancelation. We discuss this possibility
at some length in an Appendix
at the end of the paper. While it appears that
the anomaly is not completely eliminated, 
these identities do produce cancelations that 
can not be straightforwardly expressed in terms of
transverse momentum diagram divergences. Moreover, it is clear that
the large transverse momentum region producing the anomaly  
could also contribute in an important way within superficially non-leading
feynman diagrams. If there is finally a cancelation,
then it most likely means 
the failure of the transverse momentum diagram formalism for
the electroweak theory unless (as, in any case, we strongly advocate) a
transverse momentum cut-off is imposed from the outset.
In the main body of the paper our purpose will be
to study contributions to transverse momentum diagrams and, apart from
the discussion in Appendix D, and the related discussion in sub-section
{\bf 5.3}, we 
will make only brief references to the possibility
that there could be important contributions outside of the transverse 
momentum diagram formalism.
 
The anomaly occurs only in the even signature amplitude, which is a sum 
of scattering amplitudes for vector mesons with opposite and same sign
helicities, i.e.
$$
\eqalign{A^+(S)~&=~A^+(P_+P_-)~=~A_{-+}(P_+,P_-)~+~ A_{++}(-P_+,P_-)\cr
&=~A_{-+}(S)~+~ A_{++}(-S)}
\auto\label{esa0}
$$
with 
$$
~A_{-+}(S)~\centerunder{$\to$}{\raisebox{-4mm}{${\scriptstyle S\to \infty}$}}
~~c~ S~,~~~~~~ 
A_{++}(S)~\centerunder{$\to$}{\raisebox{-4mm}{${\scriptstyle S\to \infty}$}}~~-c~S
\auto\label{esa01}
$$
In a vector theory the amplitudes $A_{-+}(S)$ and $A_{++}(S)$ would simply add
in a single helicity amplitude and the anomaly would cancel.

Our calculations are carried out in a theory that is very close to, but is 
not quite, the Standard Model. For simplicity, 
as we discuss further in Section 2, we ignore both
leptons and the photon and consider only one doublet of quarks. We
discuss the general framework for our analysis in Section 3
and isolate the simplest diagram,
which is $O(\alpha_s)$, 
that potentially gives an enhancement. Section 4 is devoted to a detailed 
demonstration that the enhancement does indeed occur in this diagram.
It is generated 
by the combination of an effective vertex due to the left-handed coupling of
a scattering vector meson, a quark-antiquark effective vertex
due to a longitudinal
massive vector intermediate state, and a single (large transverse momentum) gluon
vertex. 
At $O(\alpha_s)$, there is only a small 
number of possibilities for the anomaly to occur within the
transverse momentum diagram formalism and, as we discuss in Section 5, 
it is clear that it does not cancel. 

As we show in Section 6, at $O(\alpha_s^2)$ there are contributions
in which an additional soft gluon 
plays no kinematical role in the occurrence of 
the anomaly and simply accompanies the $O(\alpha_s)$ process.
Not surprisingly, the soft gluon  
produces a transverse momentum infra-red 
divergence in individual diagram contributions. In tracing the corresponding 
cancelation, we find new (but closely related)
processes which occur first only at $O(\alpha_s)^2$ and for which
infra-red properties of the anomaly are needed to fully determine their
contribution. 
Also at $O(\alpha_s^2)$, ``anomalous'' (odd) color charge 
parity two gluon exchange appears, involving one finite and one large
transverse momentum gluon. 

In this paper we will frequently refer to a multigluon transverse
momentum state which carries color zero and anomalous color charge 
parity (not equal to the gluon number) as ``anomalous gluons''.
In the present context, such a state first appears at $O(\alpha_s^3)$. Three 
gluons with even color parity and (separately) large, finite, and soft 
transverse momentum are involved. In higher orders 
various combinations of soft and finite transverse momentum gluons can
accompany the large transverse momentum gluon. Additional gluons could also share
the large transverse momentum, but we do not discuss this possibility. 

Reggeized gluon exchanges that are the outcome of perturbative
calculations\cite{fkl}-\cite{arw93} in a vector theory
carry normal color parity (even/odd for
an even/odd number of gluons).
However, we have argued (for a very long time\cite{arw84}) that 
anomalous gluons play a crucial role\cite{arw02} 
in the emergence of QCD confinement and chiral symmetry breaking,
in the context of high-energy scattering and reggeon diagrams.
In particular, we have argued that the 
pomeron is formed from anomalous gluons and that 
configurations of this kind
are an essential component of regge limit pions and nucleons. However,
while we have been able to show how\cite{arw021}
anomalies provide triple pomeron and pion/pomeron interactions 
involving the anomalous gluons, 
it has proven very difficult to find a simple 
starting point in which the anomalous gluons couple directly and 
from which a detailed description of hadron 
amplitudes can be developed. It is very encouraging,
therefore, to see that anomalous gluons appear straightforwardly in 
the anomaly contributions that dominate the electroweak scattering
amplitudes we consider.

We have not explored the full consequences of the 
power enhancement (\ref{esa01}) nor, as we discussed above, is it clear that
the possibilities for cancelation have been exhausted. 
While no unitarity bound is violated,
we, nevertheless, believe that the enhancement severely threatens the 
unitarity of the theory (at least in the $t$ - channel)
and should not be present
in physical amplitudes. Rather than looking for further cancelations,
we will argue (only very briefly in this paper) that  
although the enhancement is actually unphysical, it selects the
physically relevant diagrams and, in doing so, anticipates chiral symmetry
breaking and confinement. In fact there is no sign, in the anomaly
amplitudes that give (\ref{esa01}), of either the
$s$ - channel or the $t$ - channel intermediate states 
that are present in the diagrams from which they are calculated.
In Section 7, we suggest that the enhancement is obtained by using
a ``wrong procedure'' to evaluate the regge limit
contribution of diagrams. If a transverse momentum cut-off is initially imposed,
the energy enhancement will be eliminated. Instead, because the cut-off
produces a Ward identity violation, the anomaly diagrams 
dominate  because of infra-red transverse momentum divergences that appear
and infra-red properties of the anomaly come into play. We will argue that 
these divergences should be analysed, and ``physical amplitudes'' extracted,
before the cut-off is removed. This is emphasized as a major conclusion
of the paper in Section 8, which also contains other conclusions.

Initially,   
a study of the infra-red anomaly contributions of diagrams, that
matches the present study of ultra-violet contributions, will be required. 
After this, we anticipate, the analysis of infra-red divergences will
parallel our discussion\cite{arw02} of hadron scattering.
All-orders properties of the
divergences have to be combined with Reggeon Field Theory,
to obtain the  ``physical amplitudes'' in which the cut-off
can be removed. In this paper
we will describe only the general arguments
that we believe should be employed. We expect that
the resulting amplitudes will have both confinement 
and chiral symmetry breaking, in the sense that the scattering 
will be describable as the exchange of a color zero, Goldstone boson, pion.
Although our hadron work provides the framework for our 
general understanding, a major part of 
the logic and justification for the procedure we outline can be appreciated
directly within the present context, without reference to the pomeron problem.
That the starting point is much more straightforward 
than in the hadron case holds out the promise that 
it will be correspondingly easier to carry the procedure through in detail.

\newpage

\mainhead{2. THE ALMOST STANDARD MODEL}

For simplicity, we will consider a theory which, for our purposes, 
is sufficiently close to the Standard Model, but which is actually 
less complex. We will consider 
a ``flavor SU(2)'' triplet of vector mesons $\{W^+,W^-,W^0\}$ with mass $M$ 
and left-handed couplings to a flavor doublet of quarks $\{u,d\}$
with the usual QCD interaction. We will effectively assume that the vector 
mesons originate from a spontaneously-broken gauge theory, as in the Standard 
Model, but apart from the
discussion of reggeization in this Section, their self-interaction will play
almost no role in our analysis. (There will be no gauge dependence in our 
discussion because the vector mesons will always be on-shell and gluons
will only contribute in gauge-independent transverse momentum diagrams - although,
in effect, we evaluate gluon contributions in feynman gauge.)

We ignore the extra complications of the photon and all mixing angles,
which could only lessen the possibilities for cancelation of the anomaly
phenomenon that we find. Since there is no photon, the usual electroweak
ultraviolet anomaly is absent and so we do not need to include leptons.
In fact, as we noted in the Introduction, the anomaly we discuss involves
(components of) QCD currents to which leptons do not couple and, therefore, could
not provide any possibility for cancelation.
If we give the quarks a small mass
$m~(<< ~M)$ any potentially singular infra-red contributions will be eliminated. 
However, for much of our discussion we will be interested only in large internal 
(transverse) momenta, where ``large'' is defined relative to $M$, and so $m$ will 
be omitted. The absence of a quark mass has the technical
advantage that we will be able to exploit the considerable, regge limit,
simplifications of the feynman diagrams that describe a massless, chiral, theory.

We will study the high-energy 
scattering of the massive vector mesons via a quark-antiquark
exchange channel in which, potentially, a ``pion'' could appear as a bound state.
Perturbatively, the leading behavior of the amplitudes we study would be given
(if there were no anomaly enhancement)
by the exchange of vector mesons. However, 
since the flavor symmetry is
non-abelian, the vector mesons will be reggeized by self-interactions.
Since $CP$ is conserved, signature is well-defined and reggeized 
vector meson exchange will give high-energy behavior in the odd-signature
channel of the form
$$
A(S,0)~~\centerunder{$\sim$}{\raisebox{-4mm}{${\scriptstyle S\to \infty}$}}
~~S^{\alpha(0)}
\auto\label{rbe1}
$$
where
$$
\alpha(0) ~=~1 - \frac{g^2}{16\pi^2} ~+~O(g^4) ~~<~1
\auto\label{rbe11}
$$
The even signature channel will be dominated by the exchange of two
reggeized vector mesons for which (apart from a logarithmic factor)
$$
A(S,0)~~\centerunder{$\sim$}{\raisebox{-4mm}{${\scriptstyle S\to \infty}$}}
~~S^{2\alpha(0)-1}~~<~~S^{\alpha(0)}
\auto\label{rbe2}
$$
Therefore, if we sum
(in principle at least) all diagrams producing all self-interaction
reggeization effects then
the contribution of (any number of) exchanged vector mesons to flavor
exchange amplitudes will be smaller than the anomaly enhanced
quark-antiquark exchange amplitudes we discuss which give
$$
A(S,0)~~\
\centerunder{$\sim$}{\raisebox{-4mm}{${\scriptstyle S\to \infty}$}}~~ 
\frac{S}{M^2} 
\auto\label{fpb}
$$

Since the running of $\alpha_s$ does not enter our calculations, the relative value
of $M$, compared to the QCD scale, does not appear in our discussion. Therefore,
the instability of the vector mesons is not an issue. More generally 
it also should not be a very significant issue. We can, of course, define
vector meson scattering amplitudes by going to complex poles and any 
undesirable features of these amplitudes will feed back into the scattering
amplitudes of physical particles. Alternatively, we could exploit the 
reggeization property of vector mesons and, although 
it would be very obscure
for most readers, we could carry out our discussion in terms of reggeon amplitudes.
In this case, it would be rather straightforward to argue that the non-regge
nature of the anomaly enhancement energy behavior that we find will violate
$t$ - channel unitarity. However, since the simplest
diagrams we consider are already very high-order
in the electroweak coupling (~$O(\alpha_w^4)$~) their contribution is very small
at current energies and so any 
unitarity problems could only be of physical relevance
at extremely high energies. 

We could also regard our calculations as academic and say that we are simply using
left-handed vector mesons to uncover properties of QCD. From this point of
view we could make the mass $M$ as small as we like. 
Also, throughout the main body of the paper the origin of $M$
will be irrelevant and we will implicitly assume that it
originates from some mechanism which
is unrelated to the quarks we consider. It is important that the high-energy 
behavior (\ref{fpb}) is obtained with internal vector mesons on mass-shell.
Because longitudinal states are involved this behavior 
is scaled by $M^2$ and so can not be canceled by a physical
Higgs contribution of any kind.

If we ignore reggeization, or if the vector mesons
are massless, then the anomaly enhanced amplitudes
will give high-energy behavior 
comparable with that of (multiple) vector meson exchange. 
This could be an additional reason, 
to be added to those briefly mentioned in Section 5, 
why bound-states of (higher-colored) quarks and antiquarks
should actually be responsible for the vector meson mass generation. If the
electroweak scale $M$ is actually a second QCD scale, as would then be the case,
we would surely expect unitarity to be just as important for the higher scale
as for the lower scale.

\newpage 

\mainhead{3. $O(\alpha_s)$ - ONE GLUON DIAGRAMS}

In this Section we describe the general framework within which we discuss
$0(\alpha_s)$ diagrams and focus on the simplest feynman diagram that, potentially,
produces an anomaly enhancement.

\subhead{3.1 Transverse Momentum Diagrams}

As is very well known, the leading (regge limit)
high-energy behavior of a feynman diagram is typically obtained
by routing the large light-cone momenta through the diagram in such 
a way that the number of particles that are close to mass-shell and have 
large, relative, longitudinal momentum separations 
(i.e. large rapidity differences) is maximal. After
longitudinal integrations are carried out, close to the on-shell
configuration, the result is a transverse momentum 
integral multiplied by logarithms of the energy.
The transverse momentum integral corresponds to a ``transverse 
momentum diagram'' obtained by contracting all of the (close to) on-shell lines.
In general, there is one logarithm and one transverse
momentum loop for each large rapidity difference. Consequently,
the leading-log amplitude contains a transverse momentum 
diagram with the maximal number of loops. 
In Appendix B we provide a brief, non-technical, 
review\cite{rk} of known results that apply to 
the fermion exchange scattering amplitudes we will discuss.

The relationship between transverse momentum diagrams 
and the process of putting lines on-shell
in full feynman diagrams will dominate our discussion.
When two, or more, particles have finite relative rapidity,
fewer lines are placed on-shell in the reduction to a transverse momentum diagram,
and a non-leading log amplitude, with a smaller number of
transverse momentum loops, is obtained. In this case, 
the couplings and interactions in the transverse momentum diagrams have
more structure. It is in (superficially)
non-leading amplitudes of this kind that the high-energy
behavior can be enhanced by the occurence of the triangle anomaly within 
the couplings of the transverse momentum diagram.

\subhead{3.2 Double Logs}

A well-known
extra complication, in the application of the transverse momentum diagram
formalism to fermion exchange amplitudes, is
that transverse loops involving only fermion
propagators are generally logarithmically divergent\cite{rk}
at large transverse momentum. 
These divergences, effectively, produce additional logarithms of the energy
and give rise to ``double logs'' that are 
associated with single rapidity differences.
In the diagrams we discuss there is, potentially, a logarithmic divergence 
of this kind but it is overwhelmed by the anomaly power 
divergence that we find. Therefore, we will not be directly interested in 
logarithmic transverse momentum 
divergences and, in the main body of the paper, will refer to them only 
for reasons of completeness. In Appendix B 
we briefly discuss the possible physical
relevance of the anomaly with respect to the double logs. 

From the general viewpoint of this paper, however, 
it is important that, because we can regard 
the double logs as described by transverse momentum diagrams,
they do not represent high-energy behavior that is not anticipated
by this formalism. (Even though it might not be the most efficient method
for studying properties of the double logs.)
It is, perhaps, worth noting that, 
since the divergences do not occur in reggeization
diagrams, they do not affect the reorganization of transverse momentum diagrams
into reggeon diagrams. In fact, this reorganization reduces the degree of 
divergence. The divergences occur only in reggeon diagram loops containing
just reggeized quarks and antiquarks and, if the leading log form of the 
trajectory
function is used, the presence of the reggeon propagator reduces the divergence
from log to log[log] form. 

\subhead{3.3 The Enhanced Transverse Momentum Diagram }

As we will elaborate below, the lowest-order appearance of the
anomaly enhancement is associated with the 
transverse momentum diagram shown in Fig.~1.
{\begin{center}
$~$\hspace{1in}
\epsfxsize=2.2in
\epsffile{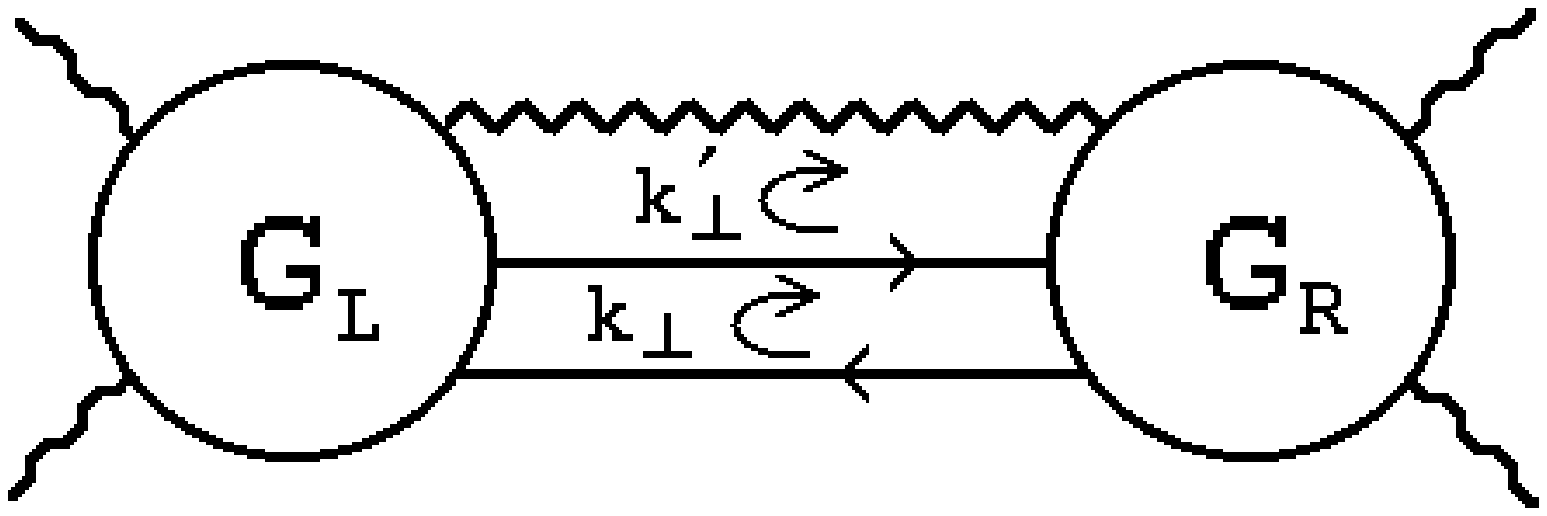}
\hspace{0.6in}
\parbox{1.4in}{\epsfxsize=1.3in
\epsffile{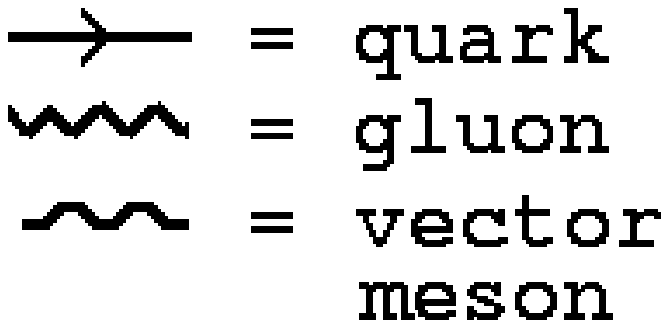}
}

Fig.~1 A Transverse Momentum Diagram
\end{center}
The remainder of this Section, and the following two Sections, 
will be devoted to the
study of $O(\alpha_s)$ feynman diagrams which give a contribution 
to the high-energy scattering of vector mesons
that contains this transverse momentum integral.

We will use the diagrammatic notation of Fig.~1 - 
for quarks, gluons, and vector mesons -
throughout the paper, in both feynman diagrams and transverse momentum diagrams. 
(Almost all of our discussion will be concerned with feynman diagrams 
and so there should be no confusion as to which kind of diagram is under
consideration.) For simplicity, in this and the following two Sections, 
we will omit flavor and color quantum numbers and consider just the momentum 
and spin structure of diagrams. In this case, a ``gluon''is effectively
a ``photon'', i.e. a massless vector particle with a vector coupling to a massless
(for most of the discussion) ``quark-antiquark'' pair. A  
``vector meson'' is a massive vector particle with a left-handed (right-handed)
coupling to the quark (antiquark). Because of the left-handed coupling, 
the high-energy scattering of vector bosons with definite helicity
has a particularly simple diagrammatic structure.

To avoid the introduction of an extra momentum scale, 
we will consider forward scattering, i.e. zero momentum transfer. We should
emphasize, however, that this does not imply that our calculations are
invalidated for the simple reason that we consider an infra-red region in 
which perturbation theory does not apply. It should become clear that, since
the phenomenon we discuss involves large internal transverse momenta, a momentum
transfer $t$ with $M^2 <<t<<S$ would not significantly affect our analysis.
In the forward direction, the integrand of Fig.~1 is a product of the 
couplings $G_L(k_{\perp},k_{\perp}')$ and $G_R(k_{\perp},k_{\perp}')$
and transverse momentum propagators for the gluon, quark, and antiquark and
the integral has the simple form 
$$
\int ~d^2k_{\perp}'~\int d^2k_{\perp} ~\frac{
Tr\bigl\{~\st{k}_{\perp}~G_L(k_{\perp},k_{\perp}')~\st{k}_{\perp}~
G_R(k_{\perp},k_{\perp}')~\bigr\}}
{{k_{\perp}'}^2~(k_{\perp}^2)^2}
\auto\label{tqqg}
$$

The $G_L$ and $G_R$
should satisfy (reggeon) Ward identities so that, as discussed further in Section
5, there is no infra-red divergence at ${k_{\perp}'}^2=0$ and a divergence 
at  ${k_{\perp}}^2=0$ would be eliminated by 
adding either a momentum transfer or a quark mass, as discussed in the previous
Section. Conventionally, since two fermion exchange is involved, we would expect 
the accompanying energy dependence to be only logarithmic. We would 
also expect the large momentum behavior of $G_L$ and $G_R$ to be such that
the full integral is, at worst, logarithmically divergent (producing an additional
energy logarithm, as discussed above). 
The signal of the anomaly will be that $G_L$ and $G_R$
actually grow at large transverse momentum, 
in a manner that produces an additional power of the energy.

\subhead{3.4 A Feynman Graph Producing the Anomaly Enhancement}

In the next Section we will see
that the feynman diagram shown in Fig.~2 has a regge limit 
contribution, involving the transverse momentum diagram of Fig.~1, 
in which each of the hatched lines is placed on-shell. 
{\begin{center}
\epsfxsize=2.2in
\epsffile{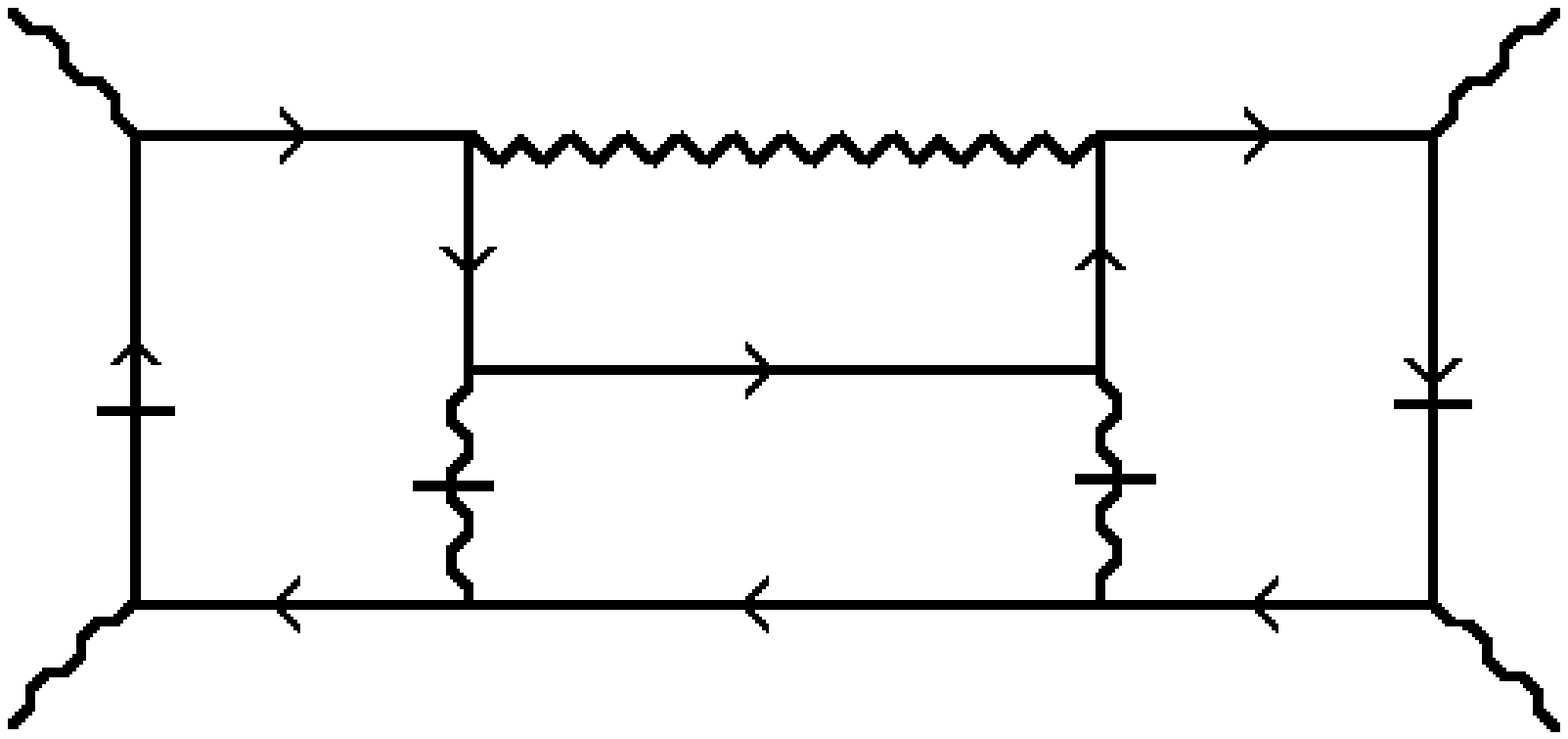}

Fig.~2 A Feynman Diagram - the Hatched Lines are On-Shell
\end{center}
In effect,
the box graphs at either end of the full graph contract to give 
triangle diagram contributions to the couplings, $G_L$ and $G_R$, 
as shown in Fig.~3. 
{\begin{center}
\epsfxsize=3.8in
\epsffile{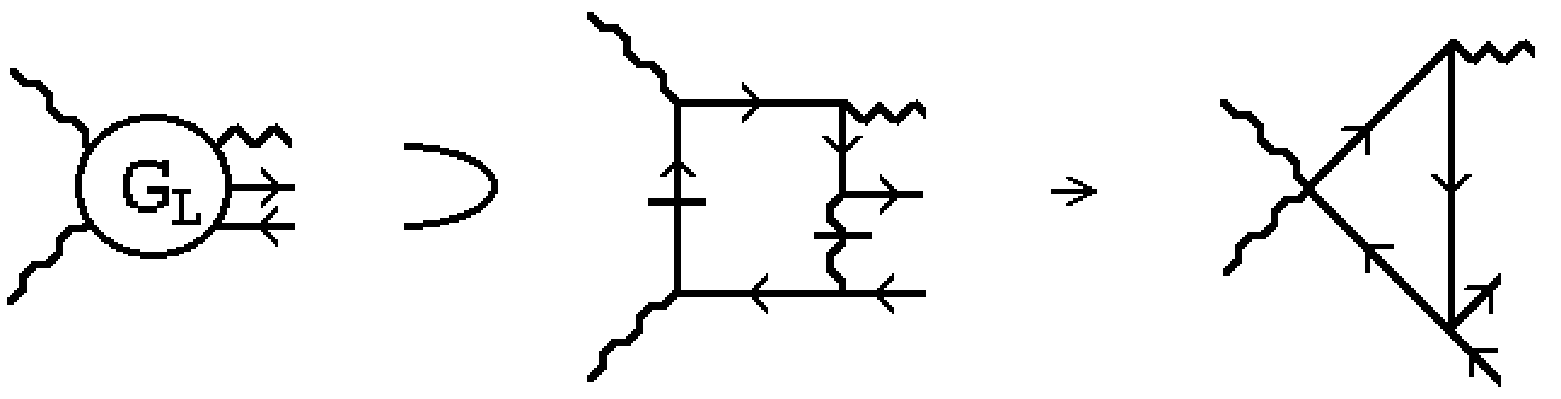}

$~$ \newline 
Fig.~3 A Triangle Diagram Coupling with ``Effective Vertices''
\end{center}
If the left-handed nature of the interactions of the
scattering vector particles leads to a left-handed (vector) ``effective
vertex'' for the triangle diagram then, naively, it would
appear that the triangle anomaly is obviously present. 
If this has the standard form of the 
ultra-violet triangle anomaly, we would expect
a linear growth with $k'_{\perp}$
that would then produce a divergence of the
$k'_{\perp}$ integration.
In fact, since
effective vertices are not necessarily
simple local vertices or, if they are, the propagators
may no longer be elementary, much more discussion
is required to show that there is a contribution that is closely related to
the triangle anomaly. 

The diagram of
Fig.~2 has the minimal complexity needed to generate triangle couplings,
as in Fig.~3, for both $G_L$ and $G_R$. We will find that this is 
necessary to obtain a non-zero contribution in the full amplitude.

\subhead{3.5 Leading Logs, Non-Leading Logs, and the Anomaly }

A priori, as described in Appendix B, we expect
the leading high-energy behavior of Fig.~2 to be $[ln~s]^4$, multiplied by the 
transverse momentum diagram obtained, as illustrated in Fig.~4, by placing
all vertical lines on-shell. 
\begin{center}
\parbox{4in}{\epsfxsize=1.7in
\epsffile{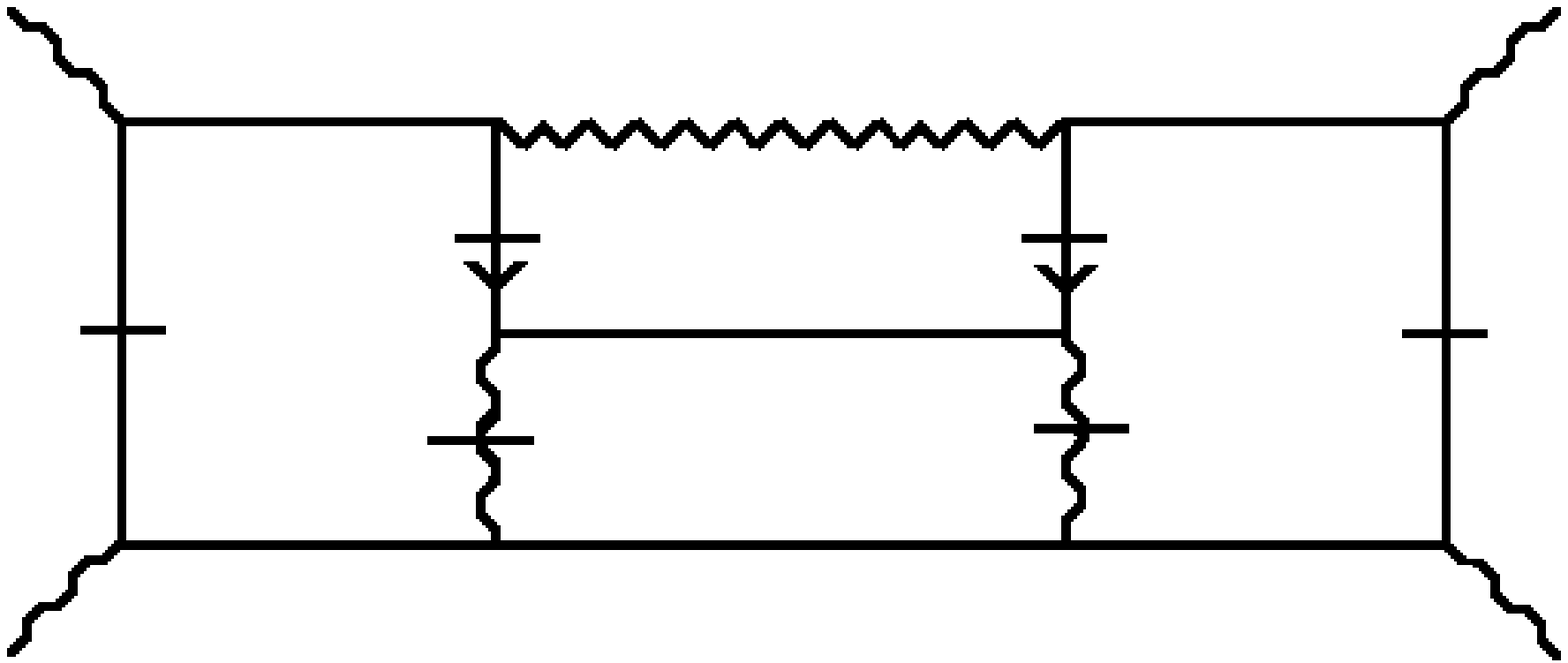}
\epsfxsize=2in
\epsffile{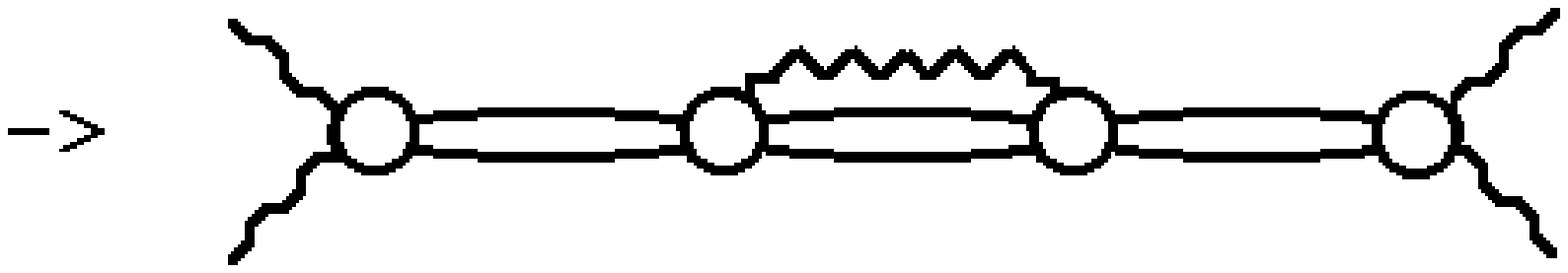}
} 
\parbox{1in}{
$$ \times~ [lns]^4~$$
}

Fig.~4 The Leading-Log Amplitude
\end{center}
(A hatch on a line will always imply that it is on-shell.)
In fact, this transverse momentum diagram  
contains quark loops that are logarithmically divergent and generate 
additional powers of $ln s$, as discussed above.

The transverse momentum diagram of Fig.~1 should appear  
at the next-to-next-to-leading log level
(formally with a factor of $[ln S]^2$).
In this contribution, it would be anticipated that
the dominant internal momenta,
within the $G_L$ and $G_R$ couplings, will be ``close to'' 
that of the corresponding fast external particle (in particular, 
there should be no large internal rapidity difference.) 
 
The expectation would be that large (relative) internal momenta 
within the $G_L$ and $G_R$ couplings are
suppressed because of Ward identity cancelations that are
a consequence of gauge invariance for the gluon
(giving either a finite or, at worst, logarithmically divergent integral). 
As we will see, the anomaly 
contradicts this expectation in that it is
a contribution to the $G_L$ and $G_R$ couplings
from, relatively, large internal momentum  
in which the unhatched vertical quark lines of Fig.~2 (and, correspondingly, the
unhatched vertical line of Fig.~3) are far off-shell. 
This does not, however, imply the failure of a Ward identity.
Rather, as we enlarge on further in Section 5, 
the presence of the anomaly means that large internal momenta
play an important role in the Ward identity. (In Appendix C 
we review the corresponding situation for the vector Ward
identities in the familiar axial-vector/vector/vector triangle diagram in which
the anomaly occurs.) 

In the course of our analysis we will
find that, in the low-order diagrams where the anomaly first occurs,
there are no additional logarithms (multiplying the power enhancement)
associated with lines that are only close to on-shell,
rather than actually on-shell, in the large transverse momentum divergence.

\newpage

\mainhead{4. THE ANOMALY ENHANCEMENT}

Is this Section we study, in detail, the occurrence of the anomaly enhancement
in the high-energy behavior of the feynman diagram of Fig.~2. This diagram
is shown again in Fig.~5, together with the momentum notation that we will use.
\begin{center}
\epsfxsize=3in
\epsffile{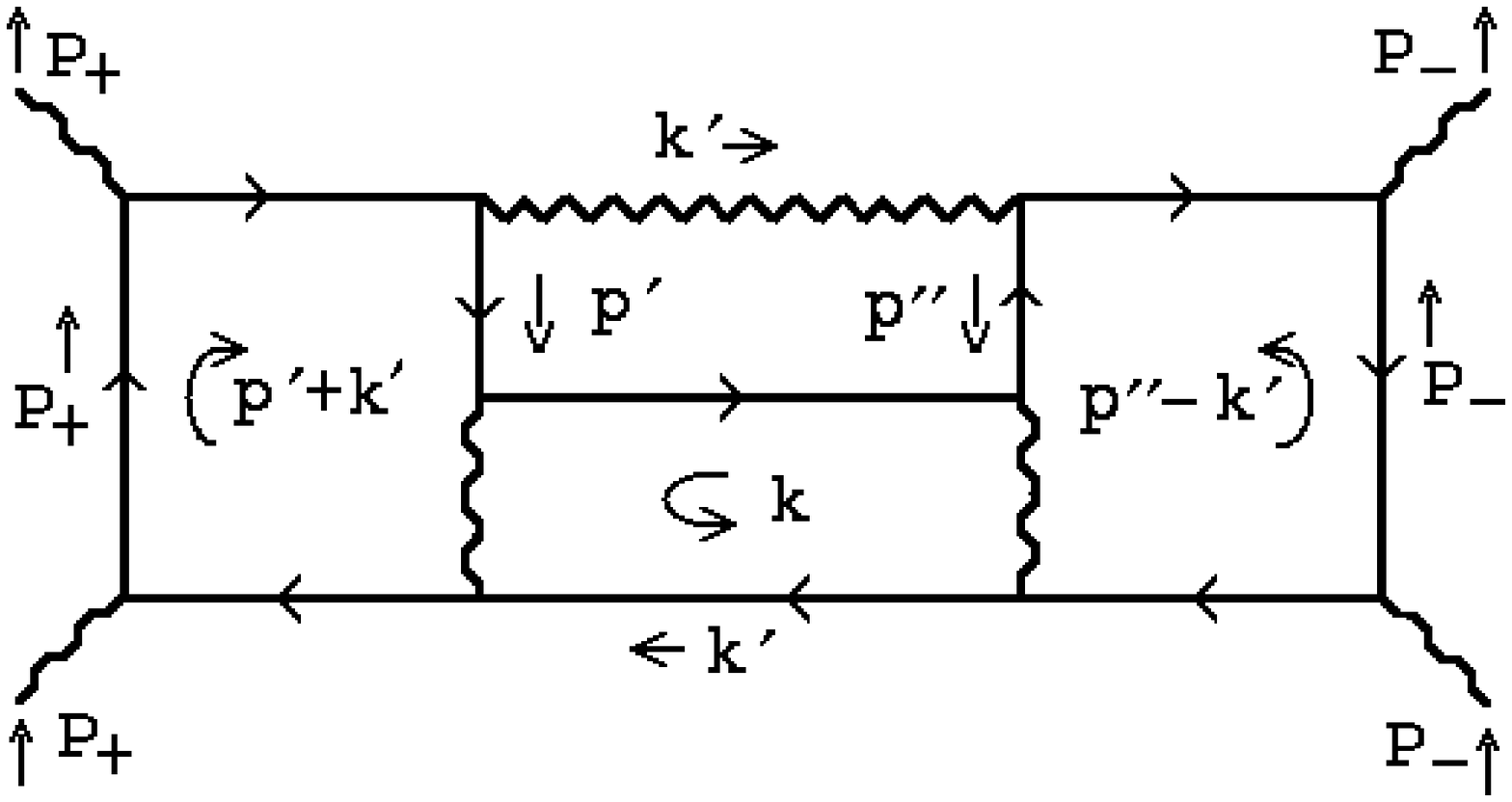}

Fig.~5 Momentum Notation for the Diagram of Fig.~2
\end{center} 
We will consider the limit
$$
P_+ ~ \to ~\bigl( \frac{\sqrt{S}}{2},\frac{\sqrt{S}}{2},0,0\bigr) 
~,~~~~~~~
P_- ~ \to ~\bigl(\frac{\sqrt{S}}{2},\frac{-\sqrt{S}}{2},0,0\bigr)
\auto\label{hel}
$$
and will find that the anomaly 
is a simple pole, of the feynman integral, at $S=\infty$, which results
from the combination of the asymptotic pinching
of mass-shell propagator poles (those hatched in Fig.~2)  
with the large momentum behavior of off-shell propagators. The on-shell 
propagators will be used to carry out longitudinal momentum
integrations and produce a reduction to the transverse momentum integral
of Fig.~1. The large momentum behavior will be a combination of the 
transverse momentum dependence of the exchanged propagators and the internal
loop momentum dependence of the propagators in the left and right side
triangle diagrams corresponding to Fig.~3.

\subhead{4.1 Internal Momenta and the Quark Mass-Shell Conditions }

We consider, first, the left-hand box sub-graph that appears in Fig.~5.
As shown, we direct $P_+$ along the left-most quark line and
use the $k_-'$ integration to put this line on-shell, i.e.
$$
\int d k_-' ~~ \frac{i~ \gamma \cdot (P_+~ +~\cdots)}{ 
(k_-' + p_-')P_+~ + i\epsilon +~\cdots } 
~~~~\centerunder{$\to $}{\raisebox{-4mm}{${\scriptstyle P_+ \to \infty }$}}
~~
\pi~\gamma_- ~+ ~\cdots
\auto\label{k'os}
$$
By using $k_-'$ for this purpose, we keep the $p'$ integration as a
four-dimensional integral 
that we can anticipate will have an anomaly contribution from the large 
momentum region, as
$P_+ \to \infty$.

Using the $k_-'$ integration as in (\ref{k'os}) has, however, the disadvantage
that it introduces additional $p'$ dependence into two of the propagators 
forming the triangle diagram. Consequently the triangle diagram no longer 
has the elementary structure known to generate the anomaly. We will avoid this
problem by considering only a limited part of the $p'$ integration, i.e.
we consider the region where
the components of $p'$ have the order of magnitude 
$$
 |p'_+| ~\sim ~\epsilon~ 
S^{\frac{1}{2}}~<<~ P_+ ~,~~~~
{p'_{\perp}}^2~\sim ~\epsilon~M~ S^{\frac{1}{2}}~
<<~ MP_+ ~,~~~
| p'_- |~\sim ~\epsilon~ M 
\auto\label{intr1}
$$
where $\epsilon$ is small, but finite. As we will see, this will allow us to 
ignore the $k_-'$ dependence of triangle diagram propagators while simultaneously
keeping only the transverse momentum dependence of the exchanged propagators
and also allowing the anomaly spin structure to emerge as a large 
$k_{\perp}$ approximation. 

We will use powers of $\epsilon$  as a 
simple way to impose inequalities amongst momenta 
that we could equally well impose more abstractly. 
We will integrate over a range of 
momenta having the given order of magnitude. Since we are only interested in showing
that an anomaly power enhancement occurs, and will make no attempt to determine the
coefficient multiplying it, the use of 
powers of $\epsilon$ will be sufficient to carry
through our arguments. Note that
as we explicitly discuss later, if we
allowed $p'_-$ to be slightly larger, i.e. $|p'_-| \sim M$, 
a Lorentz transformation on (\ref{intr1}) would give all components the same 
order of magnitude, i.e. $p_i' \sim 
(\epsilon M)^{\frac{1}{2}}S^{\frac{1}{4}}$. In this case, however,  
the approximations we
make in the following would be more marginal, and more discussion of their
justification would be required. For simplicity, therefore, we keep  
$|p'_-| \sim \epsilon M$, although we believe the full anomaly generating region
includes $|p'_-| \sim M$.

If $k'_{\perp}$ is also large, but small
compared to $p'_{\perp}$, say 
$$
{k'_{\perp}}^2 ~~ \sim~ 
 ~ \epsilon^{\frac{3}{2}} ~M ~S^{\frac{1}{2}} ~~<<~ {p'_{\perp}}^2
\auto\label{intr2}
$$
then the mass-shell condition (\ref{k'os}) becomes 
$$
|k_-' + p_-'|~ \sim ~ ~\frac{ {p'_{\perp}}^2}{P_+} 
~~\sim~~\frac{\epsilon~M~S^{\frac{1}{2}}}{S^{\frac{1}{2}}}~~
\sim ~~ \epsilon ~M
\auto\label{msc1}
$$
which, together with (\ref{intr1}), implies that 
$$
|k'_-| \sim ~\epsilon~M 
\auto\label{msc11}
$$

If we similarly direct $P_-$ along the right-most quark line in Fig.~5 and 
consider the analagous region of the $p''$ integration, then using $k'_+$
to put this line on-shell we will obtain a similar constraint on $k'_+$.
Together, these constraints ensure that, in the momentum region we are 
considering, transverse momenta dominate the gluon propagator.
(\ref{intr1}) and (\ref{intr2}), together with the corresponding range for 
$p''$, define a range of internal transverse momenta 
that is growing with the external energy but which is, nevertheless, close
to mass-shell for the hatched quark lines.

\subhead{4.2 Adjacent Quark Numerators and the External Effective Vertices }

Consider, next, the contribution of the quark numerators
that are adjacent to the fast quark line. When combined with (\ref{k'os}),
the $\gamma_-$ components give zero. Therefore, the leading 
contribution, in the momentum region we are considering, 
is given by the transverse numerator components. To discuss this contribution
we use the complex $\gamma$ - matrix formalism\cite{kms} described in Appendix A.  
We can then write both numerators in the form
$$
\st{p}_{\perp}~=~\frac{1}{2}[~( \pt'+\kt')\gam^* ~+~(\pt'+\kt')^*\gam~]
\auto\label{qn1}
$$
We take the vector meson states to be transversely polarized. The 
helicity of each vector meson will be conserved, but the two helicities
can be equal or opposite. 
Using (\ref{prj3}) and (\ref{prj4}), the $(1-\gamma_5)$ vector meson coupling
implies that, as illustrated in Fig.~A1, there is just one combination of 
$\gam$ and $\gam^*$ numerator matrices that can contribute for each helicity. 
For helicity $\lambda=-1$, the resulting coupling is that
illustrated in Fig.~6(a), while $\lambda=+1$ gives that of Fig.~6(b).
\begin{center}
\epsfxsize=4.8in
\epsffile{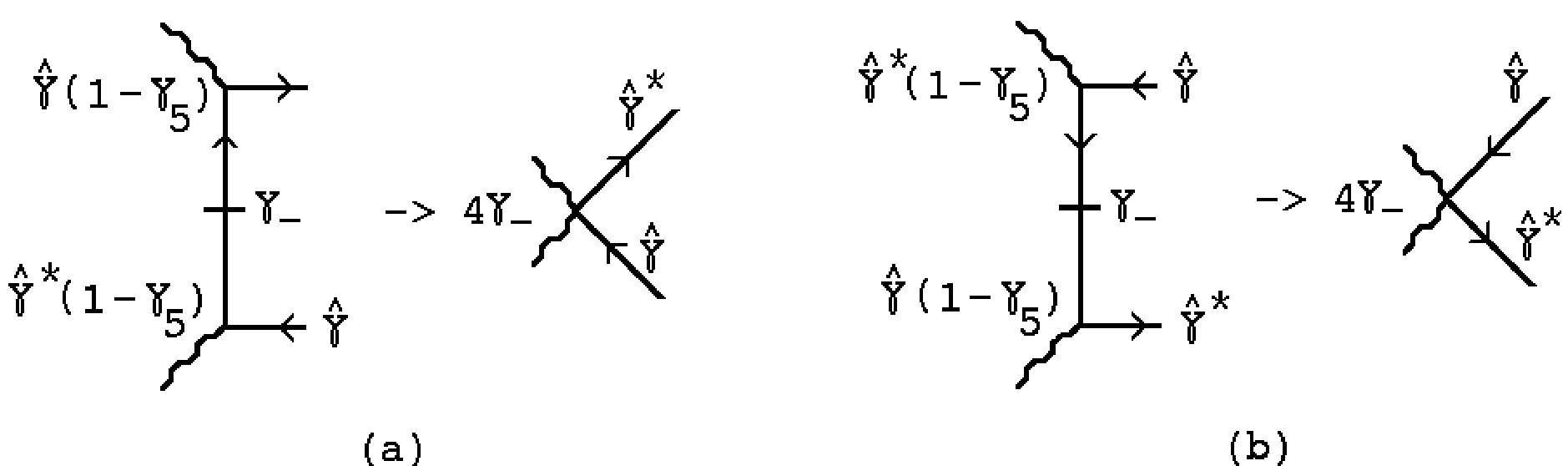}

Fig.~6 Effective Vertices (a)  for Helicity $\Lambda=-1$
(b)  for Helicity $\Lambda=+1$
\end{center}
(In all the figures of this kind we follow the
normal convention and multiply $\gamma$ - matrices 
in the direction of the quark arrow.)

\subhead{4.3 The Internal Vector Mesons}

The on-shell contribution of the internal vector meson propagator is, of course,
gauge independent and can be obtained from the unitary gauge propagator 
$$
\Gamma_{\mu\nu}(p'+k)~=~ \frac{\bigl(~g_{\mu\nu}~-~
(p'+k)_{\mu}(p'+k)_{\nu}/{M^2}~\bigr)}{(p'+k)^2 ~-~M^2}
\auto\label{vpr}
$$ 
It is the second part of the on-shell numerator,
corresponding to longitudinal polarization of the intermediate state, that
produces\cite{arw02} the vector-like (cross-channel)
coupling needed to obtain the anomaly. Since we are looking for a large 
momentum contribution of the $p'$ integration, we note that 
we obtain a factor of $p'_+$ in this part of the numerator
if there is a $\gamma_-$ at one of the vertices. This is possible
at the upper vertex in Fig.~7 (see below), but is not possible 
at the lower vertex, since a $\gamma_-$ at the bottom vertex 
would (anti-)commute through the adjacent transverse numerator and give zero.

As we will see, we obtain the factor of $p'_+$, in combination with the spin
structure needed to obtain the anomaly in the resulting triangle diagram, 
if $k_{\perp}^2~>> ~ {p'_{\perp}}^2~$. Since we have already
imposed (\ref{intr1}), we can achieve this by taking 
$$
k_{\perp}^2 ~~\sim ~~\epsilon^{\frac{1}{2}}~M~S^{\frac{1}{2}}~~
\sim ~\epsilon^{-1/2}~ {p'_{\perp}}^2 
\auto\label{intr3}
$$ 
The dominant contribution of the vector meson numerator is then
as shown in Fig.~7. 
(We remove the $(1-\gamma_5)$ factors in the quark couplings of the internal
vector meson by (anti-)commuting them around the quark loop.)
\begin{center}
\epsfxsize=5.5in
\epsffile{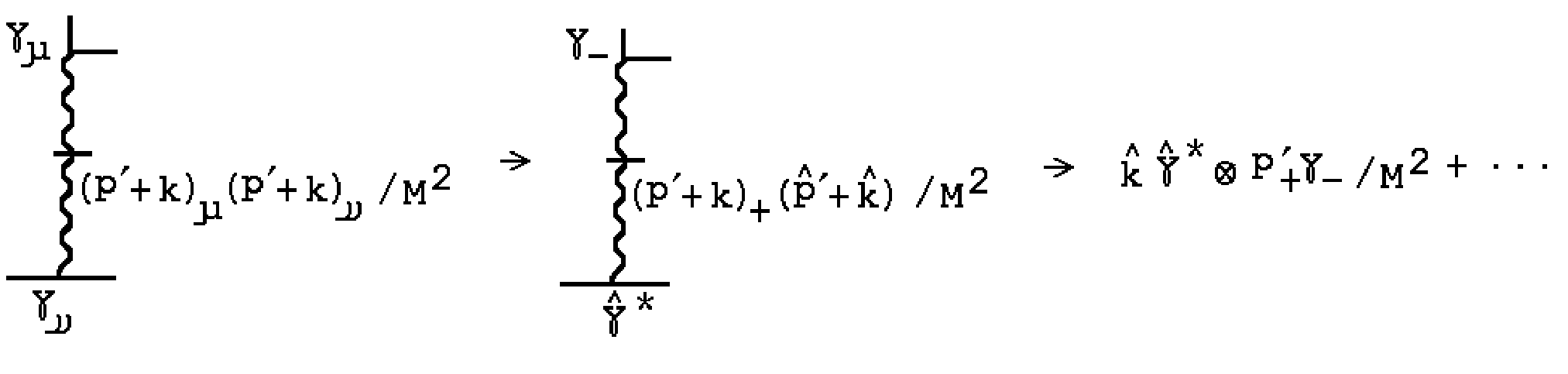}

Fig.~7 The Internal Vector Meson Numerator
\end{center}
The $\otimes$ notation indicates that the two $\gamma$-matrices are
not multiplied.

With $k_{\perp}^2~>> ~ {p'_{\perp}}^2~$, 
putting the vector meson propagator on-shell via the $k_-$ integration gives 
$$
\int d k_- ~~ \frac{(\kt \gam^*~\otimes ~
i~p'_+ \gamma_- ~ +~\cdots)/M^2}{  [(k_- + p_-')p'_+ ~\cdots] } 
~~~~\centerunder{$\sim $}{\raisebox{-4mm}{$p'_+ \to \infty $}}
~~\pi ~\kt \gam^* ~\otimes
 ~\gamma_- /M^2
\auto\label{kos}
$$
and, if (\ref{intr3}) is satisfied, the mass-shell constraint gives
$$
|k_-| ~~ \sim ~~ \frac{k_{\perp}^2}{|p_+'|} ~~
\sim ~~\frac{\epsilon^{\frac{1}{2}}~M~S^{\frac{1}{2}}}
{\epsilon~S^{\frac{1}{2}}}~~\sim~~
 \epsilon^{-1/2}~M
\auto\label{msc2}
$$
The parallel discussion of the on-shell contribution of the right-side internal 
vector meson will give a corresponding constraint on $|k_+|$. 
The two constraints,
taken together with (\ref{intr3}), imply that (as for the gluon)
transverse momenta dominate
the central quark and antiquark propagators. As we noted, 
ensuring that $k_-$ remains
finite, with (\ref{intr3}) satisfied, provides part of the motivation
for the the initial choice of (\ref{intr1}) as the $p'$ integration region.

Note that,
although we consider very large transverse momenta, because the vector mesons 
remain on-shell the high-energy behavior we will find will be scaled by $M^2$. 
Consequently, there is no possibility that it could be canceled by the 
contribution of a Higgs particle. (That is, if the Higgs mechanism were 
used to generate the vector meson mass.)

\subhead{4.4 The Internal Quark Numerator and the Triangle Amplitude}

For the remaining components of Fig.~3 that we have not yet discussed,
the largest contribution (that also gives the $\gamma_-$ vertex as in Fig.~7) 
is obtained by taking the gluon coupling to be $\gamma_+$
and taking the remaining quark numerator to also be transverse.
The chirality then feeds through the propagator as illustrated in Fig.~8. 
\begin{center}
\epsfxsize=2.8in
\epsffile{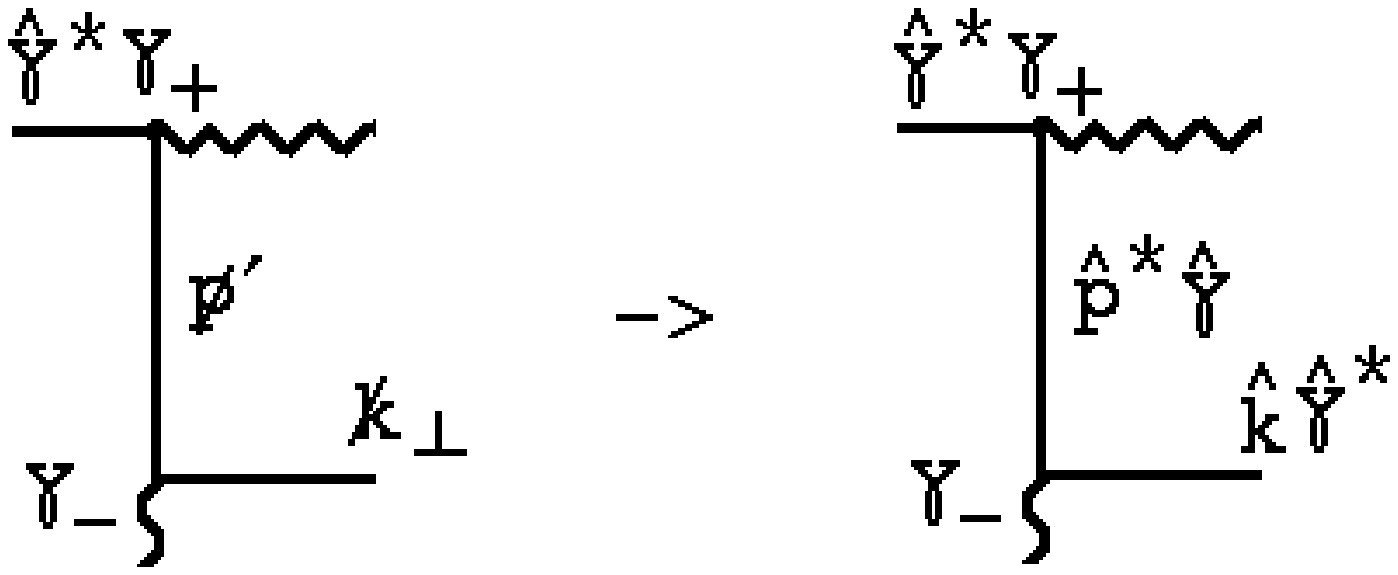}

Fig.~8 The Internal Quark Numerator
\end{center}

Combining Figs.~6, 7, and 8, and using (\ref{kos})
we obtain an effective triangle diagram with the numerators
and vertices shown in Fig.~9. 
In this figure we have also included the transverse quark and antiquark
propagators ($\gam/\kt$ and $\gam^*/\kt^*$) that are external
to the triangle diagram.
\begin{center}
\epsfxsize=1.9in
\epsffile{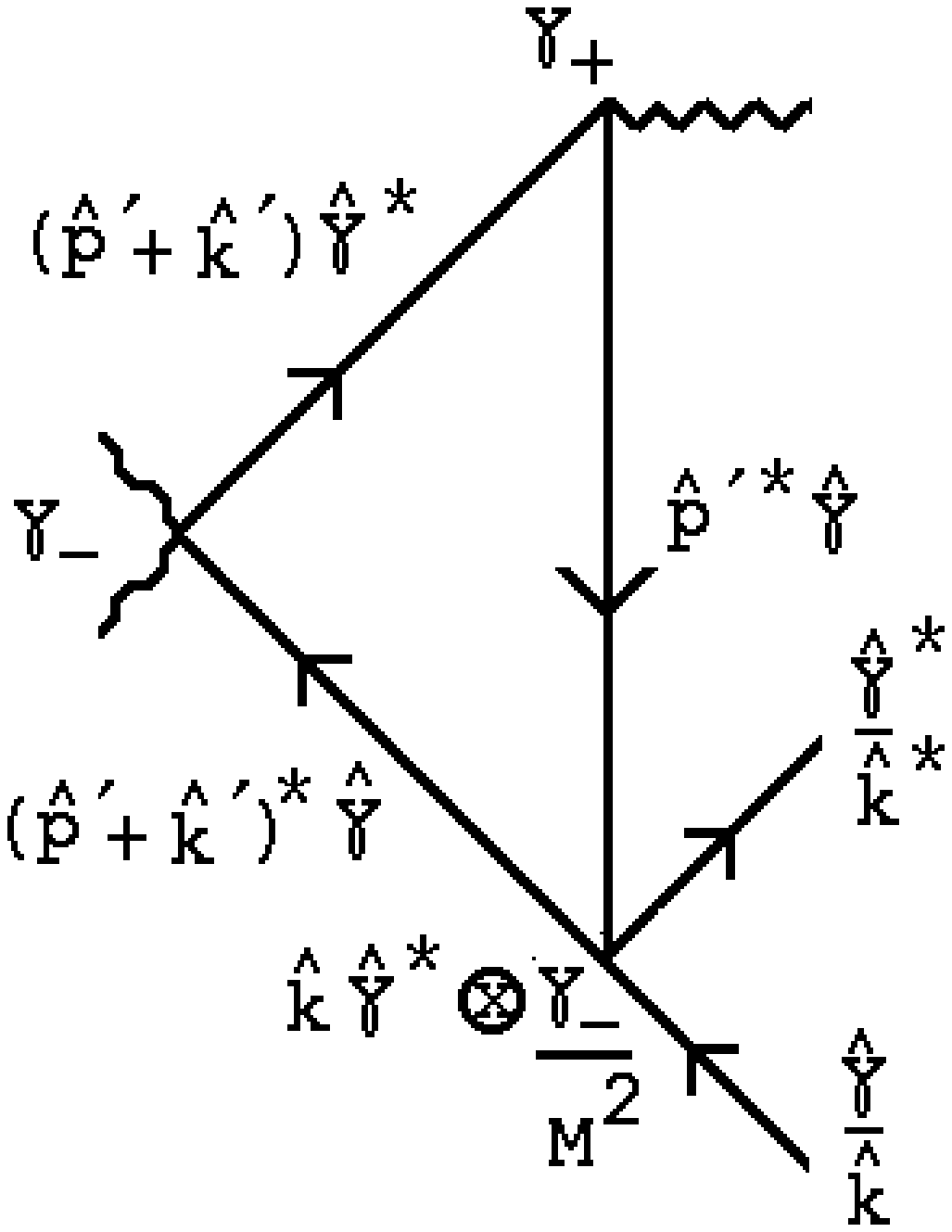}

Fig.~9 The Effective Triangle Diagram
\end{center}
The amplitude obtained from Fig.~9 
is (apart from an overall constant that we neglect - involving a numerical factor, 
factors of $\pi$, and powers of the coupling constants)
$$
\frac{\gam}{\kt}~G_L(k_{\perp},k_{\perp}')~\frac{\gam^*}{\kt^*}~~\sim~~
\frac{\gam}{\kt}~\biggl(~\int d^4p'~
\frac{\kt\gam^*  ~({\pt}'+{\kt}')^*\gam
~\gamma_-~ ({\pt}'+ {\kt}')
~\gam^* ~ \gamma_+[\pt']^*\gam}{
(p'+k')^4~(p')^2}~\frac{ \gamma_- }{M^2}\biggr)~\frac{\gam^*}{\kt^*}
\auto\label{bxi}
$$
with the integration region specified by (\ref{intr1}).

With (\ref{intr1})-(\ref{msc1}) satisfied, we can make the 
approximations 
$$
\eqalign{|k'_+(p'_- +k'_-)| ~~&\sim ~~\epsilon^2~M^2~, \cr
 |p'_+(p'_- +k'_-)| ~~ &\sim
~~{\epsilon}^2~ M~S^{\frac{1}{2}}~,\cr
~~(p'+k')_{\perp}^2~~
&\sim~~\bigl(\epsilon ~+~O(\epsilon^{\frac{5}{4}})\bigr)~M~S^{\frac{1}{2}}
}
\auto\label{asb1}
$$
and so
$$
(p'+k')^2 ~~\sim ~~(p'+k')_{\perp}^2
\auto\label{asb2}
$$
Consequently, we can approximate (\ref{bxi}) as (again ignoring an overall constant)
$$
\frac{\gam}{\kt}~G_L(k_{\perp},k_{\perp}')~\frac{\gam^*}{\kt^*}~~\sim~
~\frac{\gam}{\kt}~~\biggl(~\int~ d^4p' ~\frac{\kt[\pt']^* }{
[p'_{\perp}+k'_{\perp}]^2 ~[2p'_+p'_- - (p'_{\perp})^2 ]}~
\frac{\gamma_-}{M^2} ~\biggr)
 ~~\frac{\gam^*}{\kt^*}
\auto\label{bxi1}
$$
with the integration region still specified by (\ref{intr1}).

Because we have eliminated $k_-'$ and $k_+'$ (which are functions
of $p'$ and $p''$), the amplitude (\ref{bxi1}) is (almost) that of 
a conventional triangle feynman diagram - with local vertices. Achieving the
elimination of $k_-'$ and $k_+'$ from (\ref{bxi1}) was, as we remarked,
part of the motivation
for the initial restriction to the momentum region (\ref{intr1})-(\ref{msc1}).

\subhead{4.5 The Anomaly Contribution}

Using (\ref{cta1}) we can write the numerator momentum factor of (\ref{bxi1}) as
$$
\kt [\pt']^*~=~k_{\perp}\cdot p_{\perp}'~+~i~k_{\perp}\times p_{\perp}'
\auto\label{cta11}
$$
The first term is not special to a vector vertex
fermion triangle diagram and is not related to the anomaly.
We expect it to arise from (and to eventually
be canceled by) a variety of contributions to the complete transverse
momentum couplings of Fig.~1.
It is the second term in (\ref{cta11}) that we expect to give an anomaly
contribution. It's parity properties result directly
from the product of an odd number of quark numerators and so we anticipate
that it can only be canceled by effective triangle diagrams that contain
three quark propagators. 

Keeping just the second term in (\ref{bxi1}) gives,
for the integral within the brackets (apart from the factor of $\gamma_-/M^2$)
$$
\int~dp_+' dp_-' \int d^2p_{\perp}' 
~\frac{i~k_{\perp}\times p_{\perp}' }{
[p_{\perp}'+k'_{\perp}]^2 ~[2p_+'p_-' - (p_{\perp}')^2 ]}
\auto\label{bxi3}
$$
To carry out the angular integration for $p_{\perp}'$
we choose co-ordinates $(p_2',p_3')$ 
such that $k_{\perp}'$ lies along the 2-axis. In this case,  
$$
p_{\perp}'\cdot k_{\perp}' ~=~ |p_{\perp}'|~|k_{\perp}'|~ cos\phi~,
~~~~~p_{\perp}'\times k_{\perp}~=~|p_{\perp}'|~(k_3 cos\phi ~-~ k_2 sin \phi)
\auto\label{bxi4}
$$
where $k_2$ and $k_3$ are projections of $k_{\perp}$ along and perpendicular
to $k'_{\perp}$. We can, therefore, write 
$$
\eqalign{\int~&dp_+' dp_-' \int \frac{d(p_{\perp}')^2} 
{[2p_+'p_-' - (p_{\perp}')^2 ]}
~\frac{k_{\perp}\times p_{\perp}' }{
[p_{\perp}'+k'_{\perp}]^2 } \cr
&= ~ \int~dp_+' dp_-' \int \frac{d(p_{\perp}')^2} 
{[2p_+'p_-' - (p_{\perp}')^2 ]}
\int_0^{2\pi}d\phi~ \frac{|p_{\perp}'|(k_3cos\phi -k_2sin\phi) }
{[(p_{\perp}')^2 + (k_{\perp}')^2+2 |p_{\perp}'||k'_{\perp}|~cos\phi]}}
\auto\label{bxi30}
$$

Using
$$
\int_0^{2\pi}d\phi ~\frac{sin\phi}{a+b~cos\phi} ~
=~-~ \frac{1}{b}~  \bigl[~a+b~cos \phi ~\bigr]_0^{2\pi}
~=~0
\auto\label{bxi5}
$$
and
$$
\eqalign{\int_0^{2\pi}d\phi ~\frac{cos\phi}{a+b~cos\phi} ~~
&\centerunder{$\sim$}{\raisebox{-4mm}{${\scriptstyle b~ <<~ a}$}}
 ~~\frac{1}{a}~\int_0^{2\pi}d\phi ~cos\phi~-~
\frac{b}{a^2}\int_0^{2\pi}d\phi ~cos^2\phi ~+ 
~O\bigl(~\bigl[\frac{b^2}{a^3}\bigr]~\bigr)\cr
&\sim 0~-~\frac{\pi b}{a^2}~+ ~O\bigl(~\bigl[\frac{b^2}{a^3}\bigr]~\bigr) }
\auto\label{bxi6}
$$
we obtain
$$
\int d\phi 
~\frac{k_{\perp}\times p_{\perp}' }{
[p_{\perp}'+k'_{\perp}]^2 }
~~~\centerunder{$\sim$}{\raisebox{-4mm}{${\scriptstyle {k_{\perp}'}^2 <<
~ {p_{\perp}'}^2}$}}~~
\frac{k_3~|k'_{\perp}|}{{p_{\perp}'}^2} 
~~\sim~~\frac {k_{\perp}\times k_{\perp}'}{{p_{\perp}'}^2 } 
\auto\label{bxi7}
$$
(\ref{bxi30}) gives, therefore, 
$$
k_{\perp}\times k_{\perp}' ~~
\centerunder{\hbox{\huge $\int$}}{\raisebox{-4mm}{${\scriptstyle
{p_{\perp}'}^2 \sim~ \epsilon M \sqrt{S}}$}}  
~~\frac {d({p_{\perp}'}^2)} {({p_{\perp}'}^2)} ~~
\centerunder{\hbox{\huge $\int$}}{\raisebox{-4mm}{${\scriptstyle
|p_+'| \sim \epsilon \sqrt{S}~,~|p_-'| \sim ~\epsilon~ M}$}}
~~\frac{ dp_+' 
 dp_-'}{[2p_+'p_-' - (p_{\perp}')^2 ]}
\auto\label{bxi8}
$$
If we change variables to 
$$
\sqrt{S}~x~=~{p_{\perp}'}^2~, ~~~~~~~ \sqrt{S}~y~=~p_+'
\auto\label{scS}
$$
we obtain
$$
\eqalign{ \centerunder{\hbox{\huge $\int$}}{\raisebox{-4mm}{${\scriptstyle
{p_{\perp}'}^2 \sim~ \epsilon M \sqrt{S}}$}}  
~~&\frac {d({p_{\perp}'}^2)} {({p_{\perp}'}^2)} ~~
\centerunder{\hbox{\huge $\int$}}{\raisebox{-4mm}{${\scriptstyle
|p_+'| \sim~  \epsilon \sqrt{S},~|p_-'| \sim~ \epsilon M 
}$}}  
~~\frac{ dp_+' 
 dp_-'}{[2p_+'p_-' - (p_{\perp}')^2 ]} \cr
&=~\centerunder{\hbox{\huge $\int$}}{\raisebox{-4mm}{${\scriptstyle
x \sim~ \epsilon M }$}}  
~~\frac {dx} {x} ~~
\centerunder{\hbox{\huge $\int$}}{\raisebox{-4mm}{${\scriptstyle
|y| \sim~ \epsilon ,~|p_-'| \sim~\epsilon  M }$}}  
~~\frac{ dy 
 dp_-'}{[2yp_-' - x ]} }
\auto\label{sc1}
$$
which is clearly a constant - 
that we do not need to evaluate. 

(\ref{bxi8}) is sufficient
to conclude that the integration region on which we have focussed gives, 
for (\ref{bxi1}), the behavior
$$
\sim ~~\frac{\gam^*}{\kt^*}~~\biggl(~\frac {(k_{\perp}\times k_{\perp}')~
\gamma_-}{M^2} ~\biggr)
 ~~\frac{\gam}{\kt}
\auto\label{bxi11}
$$

\subhead{4.6 Behavior of the Full Amplitude}

It is straightforward to obtain the behavior of 
the full amplitude that results from combining (\ref{bxi11})
with the corresponding contribution from the right-side box graph in Fig.~5.
As we have discussed above, the internal 
mass-shell conditions determine that
the longitudinal momenta in the central propagators of Fig.~5 can be neglected.
As a result the $k'_{\perp}$ and $k_{\perp}$ loop 
integrations produce, as anticipated, a
transverse momentum integral of the form of (\ref{tqqg}) which we write,
in complex $\gamma$ - matrix notation, as 
$$
\eqalign{\int& ~\frac{d^2k_{\perp}'}{{k_{\perp}'}^2}~\int d^2k_{\perp} ~
Tr\bigl\{~\frac{\gam}{\kt}~G_L(k_{\perp},k_{\perp}')~\frac{\gam^*}{\kt^*}~
G_R(k_{\perp},k_{\perp}')~\bigr\}\cr
&\sim ~\int ~\frac{d^2k_{\perp}'}{{k_{\perp}'}^2}~\int d^2k_{\perp} ~
\biggl(~\frac{k_{\perp}\times k'_{\perp}}{M^2}
~ \biggr)^2~ 
\frac{ Tr\{ \gamma_- \gam^* \gamma_+ \gam\} }
{\kt \kt^*} \cr
&\sim ~
\int ~\frac{d^2k_{\perp}' d^2k_{\perp}}{{k_{\perp}'}^2 k_{\perp}^2} 
~\biggl(~\frac{k_2k'_3 -k_3k'_2}{M^2}~\biggr)^2\cr
}
\auto\label{fam0}
$$
Since the foregoing analysis assumes that both 
(\ref{intr2}) and (\ref{intr3}) hold, it follows that both ${k'_{\perp}}^2$
and $k_{\perp}^2$ can be integrated over a range of values,
that are $O(MS^{\frac{1}{2}})$  without either the  
approximations that we have made breaking down or the transverse momentum
approximation to the gluon and
quark propagators being invalidated. Therefore, we obtain 
a contribution from (\ref{fam0}) of the form
$$
\eqalign{\int& ~\frac{d^2k_{\perp}'}{{k_{\perp}'}^2}~\int d^2k_{\perp} ~
Tr\bigl\{~\frac{\gam}{\kt}~G_L(k_{\perp},k_{\perp}')~\frac{\gam^*}{\kt^*}~
G_R(k_{\perp},k_{\perp}')~\bigr\}\cr
&\sim ~\frac{1}{M^4}~
\int_{O(M S^{\frac{1}{2}})}d({k_{\perp}'}^2)~\int_{O(M S^{\frac{1}{2}})}
d(k_{\perp}^2)~~~\sim ~\frac{S}{M^2}
}
\auto\label{fam}
$$
(Note that, because the anomaly contribution to $G_L(k_{\perp}, k_{\perp}')$
is linear in $k_{\perp}'$, if it is combined with a $G_R(k_{\perp}, k_{\perp}')$
that does not have this contribution then integration over  $k_{\perp}'$
will give a cancelation of the enhancement effect. This is why we have considered 
a diagram which gives anomaly contributions to both $G_L$ and $G_R$.)

As we noted above, because two fermion exchange is involved, we 
would have expected the amplitude to increase only as some power of $ln S$. 
However, we have now shown 
that the kinematic region of Fig.~5 that we have isolated actually
produces a power enhancement of the expected high energy behavior. As we will
see in the next Section, there are no accompanying logarithms in this lowest-order
appearance of the anomaly.

Clearly, if a fixed transverse momentum cut-off is imposed, i.e.
$$
{k'_{\perp}}^2,~k_{\perp}^2 < \lambda_{\perp}
\auto\label{fco}
$$ 
then there will be no 
contribution of the form (\ref{fam}) when $S$ is sufficiently large. Therefore,
a transverse cut-off eliminates the enhancement effect and restores
the normal behavior expected for two fermion exchange. However, as we discuss
at greater length in Section 7, a transverse cut-off violates gauge invariance
Ward identities in a way that replaces the anomaly enhancement by transverse
momentum infra-red divergences.

\subhead{4.7 Comparison With the Axial Vector Vertex Anomaly}

To see the relationship between the anomaly amplitude (\ref{bxi11}) that we 
have found and the familiar axial vector anomaly we proceed as follows.
First we change variables from $p'$ to $q$ where
$$
q_+ ~=~\frac{p'_+}{\Lambda}~, ~~
q_- ~=~ \Lambda~p'_-~,~~ 
q_{\perp}~=~p'_{\perp}~, ~~~~~~~\Lambda~=~
\bigl(\frac{\epsilon}{M}\bigr)^{\frac{1}{2}}~S^{\frac{1}{4}}
\auto\label{cv1}
$$
If we also extend the integration region for $p_-'$ 
to $p_-' \sim M$ (which, as we have already noted, would not
significantly alter the above analysis) then 
(\ref{bxi1}) becomes (moving the $\gamma_-/M^2 $ outside of the brackets)
$$
~\frac{\gam}{\kt}~~
\biggl(~
\centerunder{\hbox{\huge $\int$}}{\raisebox{-4mm}{${\scriptstyle
|q_j|~ \sim ~(\epsilon M)^{\frac{1}{2}}S^{\frac{1}{4}} }$}}
~ dq_j ~\frac{\kt\qt^* }{
[q_{\perp}+k'_{\perp}]^2 ~[2q_+q_- - q_{\perp}^2 ]}~ \biggr)
~ \frac{\gamma_-}{M^2}~\frac{\gam^*}{\kt^*}
\auto\label{bxi2}
$$
where, as indicated the  
range of integration is now the same for all components of $q$. 
In the limit $S \to \infty$, the integration is over a 
four-dimensional large momentum region and, formally, the integral
is linearly divergent. Also, a product 
of three othogonal $\gamma$ matrices is present 
- although there is no trace involved. Consequently, it is natural to expect
a large momentum contribution of the form
associated with the triangle anomaly. 

The strictly infinite momentum region contribution to (\ref{bxi2}) is
$$
\sim ~~\frac{\gam}{\kt}~\biggl(~ \int~dq_+ dq_- \int d^2q_{\perp} ~
\biggl[~\frac{k_{\perp}\times q_{\perp}}{q_{\perp}^2}
\biggr]~\frac{1}{[2q_+q_- - q_{\perp}^2]}~ \biggr)~\frac{ \gamma_-}{M^2}
~\frac{\gam^*}{\kt^*}
\auto\label{lmr}
$$
which, integrating by parts with respect to $q_{\perp}^2$, we can rewrite
as 
$$
\sim~~ \frac{\gam}{\kt}~\biggl(~ 
\int~dq_+ dq_- \int d^2q_{\perp} ~\frac{k_{\perp}\times q_{\perp}}
{\bigl[2q_+q_- - q_{\perp}^2\bigr]^2} ~ \biggr)~ \frac{\gamma_-}{M^2}
~\frac{\gam^*}{\kt^*}
\auto\label{lmr1}
$$
and, with a further integration by parts, as
$$
\sim ~~ \frac{\gam}{\kt}~\biggl(~ 
\int~dq_+ dq_- \int d^2q_{\perp} ~\frac{q_{\perp}^2
[~k_{\perp}\times q_{\perp}~]}
{\bigl[2q_+q_- - q_{\perp}^2\bigr]^3} ~ \biggr)~ \frac{\gamma_-}{M^2}
~\frac{\gam^*}{\kt^*}
\auto\label{lmr2}
$$
Undoing the $\gamma$ - matrix removal involved 
in going from (\ref{bxi}) to (\ref{bxi1}) 
(or, equivalently, inserting $\gamma$ - matrices using  
$2 = \gamma_+\gamma_- + \gamma_-\gamma_+ = \gam\gam^* + \gam^*\gam$~) we 
can rewrite (\ref{lmr2}) as 
$$
\eqalign{&\sim ~~ \gam ~\biggl(~\int_{|q_i|~> >~ O(M) }~ d^4q ~\frac{  ~
\gam^* ~[\qt^*\gam] ~ \gamma_-
 ~ [\qt \gam^*]~ \gamma_+  
~[\qt^* \gam]}{[~q^2~]^3 } \biggr)
~~\frac{\gamma_-}{M^2}~ \frac{\gam^*}{\kt^*}\cr
& \sim ~~\gam \biggl(~\int_{|q_i|~> >~ O(M) }~ d^4q ~\frac{  ~
\gam^*~ \st{q}_{\perp} ~(1- \gamma_5) \gamma_-
 ~ \st{q}_{\perp}~ \gamma_+  
~\st{q}_{\perp} }{[~q^2~]^3 } \biggr)
~~\frac{\gamma_-}{M^2} ~\frac{\gam^*}{\kt^*} }
\auto\label{bxi40}
$$

We recognize the integral, within the brackets, of (\ref{bxi40})
as a left-handed transverse propagator contribution to a tensor component of 
the standard large momentum anomaly integral (apart from the feature that
there is no trace of the $\gamma$ -matrices involved). Therefore, we could
anticipate (\ref{bxi11}) directly from the familiar anomaly contribution to
a three current vertex $T_{\mu \alpha \beta}(k_1,k_2)$ - the notation is that
of Fig.~C1. In our case, 
$$
k_1~=~ -k_2~= ~k'_{\perp}
\auto\label{k1k2}
$$
and if we consider the decomposition into invariant
amplitudes (\ref{inde}), (\ref{bxi11}) corresponds to the contribution of the first
two terms, which are linear in $k_1$ and $k_2$. 

It is very well known that in the three-current vertex the ambiguity
of the ultra-violet anomaly contribution 
is determined by vector current Ward identities that relate
the anomaly contribution to infra-red triangle diagram contributions. 
In our case, we anticipate that 
there will be a (reggeized) gluon
Ward identity which similarly determines the coefficient of the anomaly
contribution we have found. We discuss this point further in Section 5.
Note, however, that, in the special
momentum configuration (\ref{k1k2}), all the other terms in (\ref{inde}) vanish
- if there are no infra-red divergences to consider. Therefore, in the lowest-order
graphs we are discussing, the ultra-violet anomaly contributions we are discussing
can not be canceled by the contribution of infra-red transverse momentum regions.

\newpage

\mainhead{5. NON-CANCELATION OF THE ANOMALY}

In this Section we consider other diagrams that are also $O(\alpha_s)$
and similarly have anomaly enhancements that might
produce an overall cancelation.

\subhead{5.1 Reality of the Anomaly Amplitude}

It is significant that the anomaly amplitude we have found, although
calculated with internal lines on-shell, is real. Indeed there is no evidence, in
the amplitude, of either the 
$s$ - channel or the $t$ - channel intermediate states 
that are present in the diagram from which it was calculated.
At first sight this seems paradoxical since it 
would appear that the analysis of Fig.~2, in the previous Section,
can be viewed as the
calculation of an $s$ - channel discontinuity - via the unitarity cut corresponding
to the dashed line in Fig.~10(a).
\begin{center}
$~~~~$\epsfxsize=4.5in \epsffile{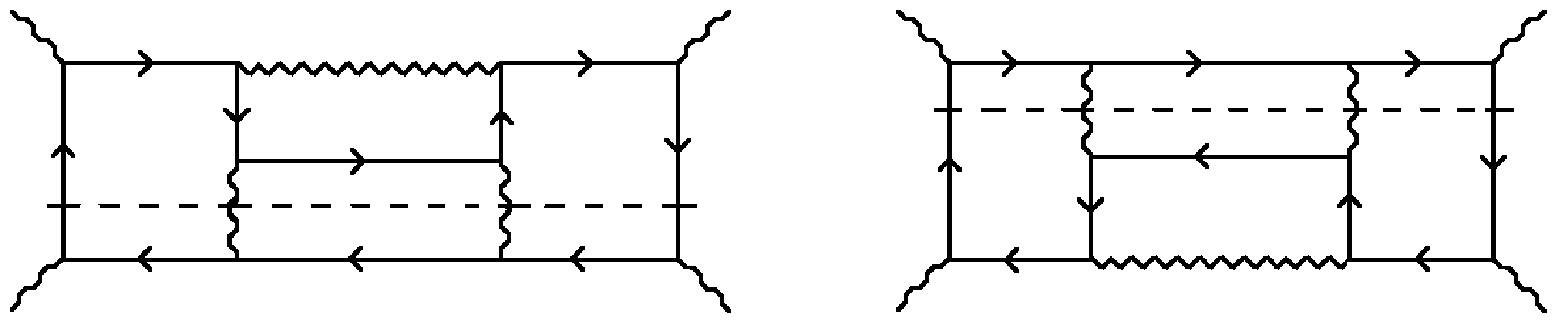}
\newline \centerline{(a)\hspace{2.4in}(b)}

Fig.~10 Unitarity Cuts of (a) the Diagram of Fig.~2 (b) a Related Diagram
\end{center}
The tree amplitude that appears below the cut
is integrated with the one loop amplitude that appears above the cut. 
That the calculation can be related to 
the evaluation of a discontinuity immediately justifies, in fact,
our choice of lines to place on-shell. 

That there is ultimately no discontinuity
associated with the anomaly is due to a second discontinuity 
contribution from the unitarity cut of a closely related 
graph shown in Fig.~10(b).
Clearly Fig.~10(b) is so 
similar to Fig.~10(a) that our analysis carries over directly.
In both cases, the intermediate
state integration over $k_{\perp}$ produces an imaginary
contribution of $i|S|^{\frac{1}{2}}$. However, 
the $k'_{\perp}$ integration, that also gives a factor
of  $i|S|^{\frac{1}{2}}$, is part of the integration
within the loop amplitude.

If we formulate the above analysis as a 
unitarity calculation
then the amplitude on one side of the cut must be complex conjugated.
As a result, the loop amplitude will have the opposite sign in the 
contributions
from Figs.~10(a) and 10(b) and since all other parts of the diagrams
contribute identically, adding the two will give a factor of 
$$
(2\pi )^4~\bigl(~[i~|S|^{\frac{1}{2}}~]~[~i~|S|^{\frac{1}{2}}~]
~+~[~i~|S|^{\frac{1}{2}}~]~[- i~|S|^{\frac{1}{2}}~]~\bigr)
~~=~0
\auto\label{disc}
$$
Alternatively, if we calculate the contribution of the two diagrams as that
of amplitudes then the loop amplitude will have the same sign in both cases 
and (\ref{disc}) will be replaced by
$$
(\pi )^4~\bigl(~[~i~|S|^{\frac{1}{2}}~]~[~i~|S|^{\frac{1}{2}}~]
~+~[~i~|S|^{\frac{1}{2}}~]~[~i~|S|^{\frac{1}{2}}~]~\bigr)
~~=~2~\pi^4~S
\auto\label{ampl}
$$ 
which will, indeed, give a real amplitude. The contribution of Fig.~10(b) simply
doubles that of Fig.~10(a). The absence of a discontinuity implies that, as we
anticipated earlier, there are no additional logarithms accompanying the power
enhancement due to the anomaly. As we stated in the last Section, the anomaly 
is a simple pole at $S=\infty$ which results
from the combination of the asymptotic pinching
of the mass-shell poles of the hatched propagators
with the large momentum behavior of the unhatched propagators.

\subhead{5.2 Another Anomaly Generating Diagram}

The diagram of Fig.~10(b) is obtained from that of Fig.~10(a) by simultaneously
``twisting'' both the left and right-side box diagrams.
For much of our discussion (including the addition of extra gluons in the 
next Section) we will keep the right-side of the diagrams we 
consider, and therefore the corresponding $G_R$, fixed 
and discuss anomaly amplitudes entirely in terms 
of possible left-side contributions to $G_L$.
In the simple case of the one gluon diagrams that we are presently discussing, 
the right-side
coupling will be that of Fig.~2 (or Fig.~10(a) ) and it will be clear that, as in
the above discussion, diagrams with the right-side coupling of
Fig.~10(b) simply give parallel contributions. However, when we consider infra-red 
cancelations in the next Section, 
it will be essential to also consider all contributions to $G_R$.

Consider, next, the diagram shown in Fig.~11 that is 
obtained from that of Fig.~2 by twisting the left half of the diagram
relative to the right half.
\begin{center}
\epsfxsize=2.5in
\epsffile{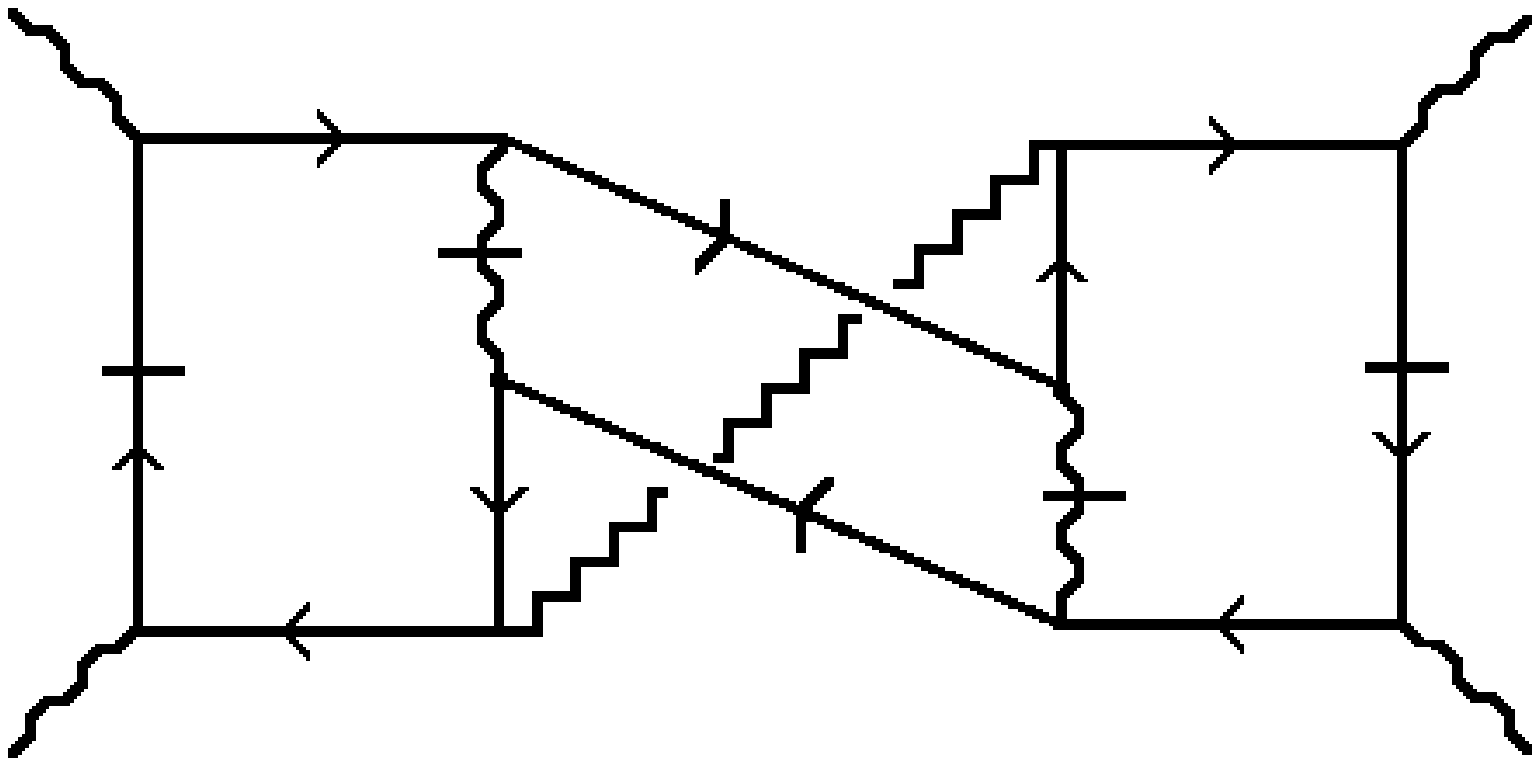}

Fig.~11 A Twisted Diagram 
\end{center}
By a similar application of the above analysis, which puts on-shell
the hatched lines shown in Fig.~11, 
the transverse momentum diagram of Fig.~1 will again be
generated. The $G_L$ shown in Fig.~12(a), 
obtained from the left part of Fig.~11, 
contains the effective triangle diagram shown in Fig.~12(b).
\begin{center}
\epsfxsize=5in
\epsffile{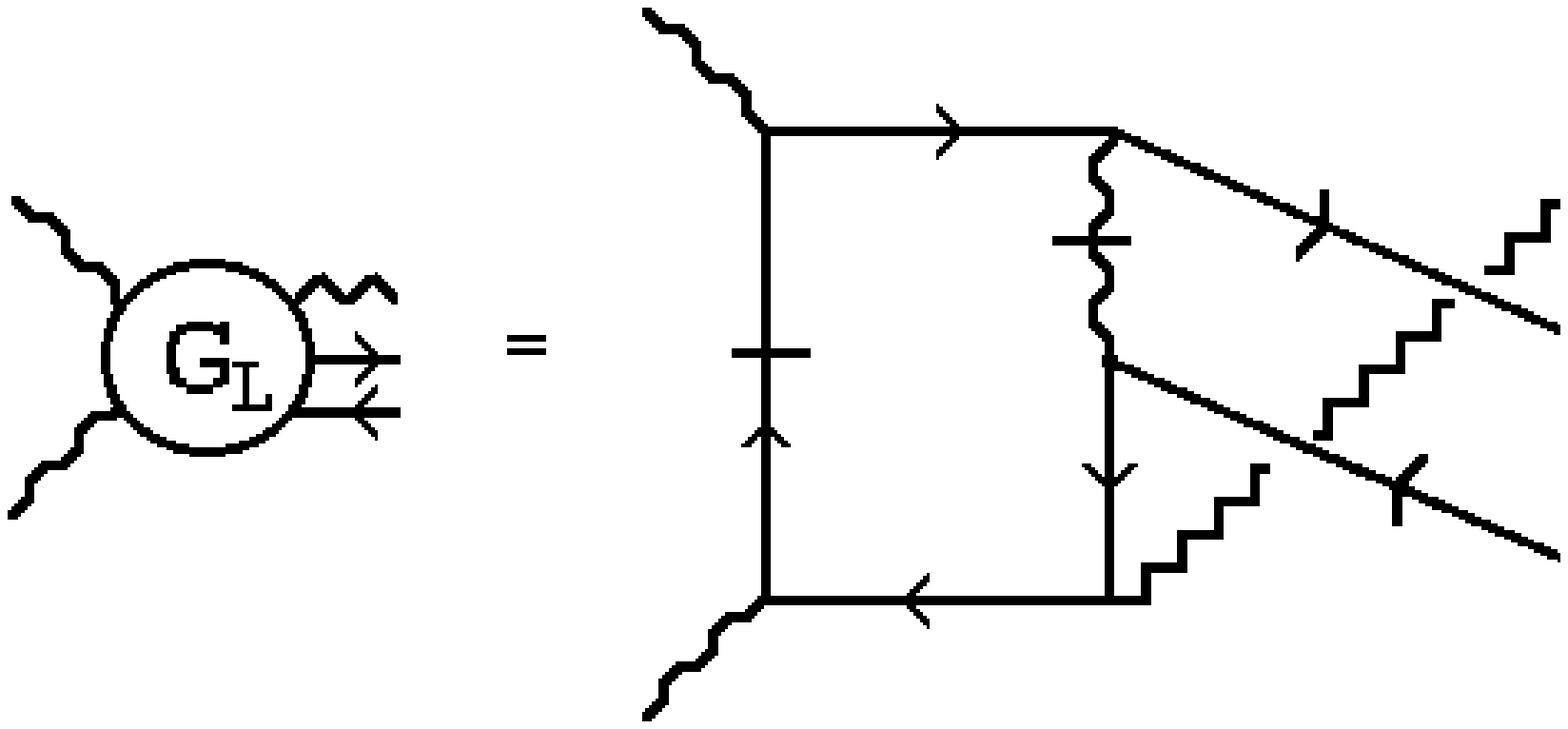}

(a)\hspace{3in}(b)

Fig.~12 (a) The $G_L$ Generated by Fig.~11 (b) The 
Effective Triangle Diagram
\end{center}
A very similar expression to (\ref{bxi11}) will clearly be obtained.
The differences in the analysis can be summarized as follows. 

\noindent (i) ~~~ The analogue of (\ref{bxi}) gives (\ref{bxi1}) 
but with $\kt [\pt']^* \to 
\kt^*\pt'$ which leads to $k_{\perp}\times p'_{\perp} \to
-~ k_{\perp}\times p'_{\perp}$ in (\ref{bxi3}) and the following. 

\noindent (ii)~ ~ A second change of sign arises from $k'_{\perp}
\to - k'_{\perp}$ in (\ref{bxi1}). 

\noindent The net result is that an identical anomaly contribution, 
to that obtained from Fig.~2, is obtained from the diagram of Fig.~11. 

The lines placed on-shell, asymptotically, 
in Fig.~11 do not correspond to a simple cut of the
diagram, as was the case for Fig.~2. However, Fig.~11 can also 
be represented as in Fig.~13(b), i.e. as an exchanged gluon attached to the
off-shell lines of the cut amplitude of Fig.~13(a).
The exchanged gluon has transverse momentum much less than 
the off-shell quark or antiquark
to which it couples in the large momentum $p'$ and $p''$
regions which generate the anomaly. Consequently, it does
not interfere (kinematically) with either the quark/antiquark scattering process,
or the asymptotic placing on-shell of the left side fast quark and the right side
antiquark. Therefore, the justification for the choice of lines placed on-shell
is closely related to 
the existence of the asymptotic physical region discontinuity of Fig.~13(a).
\newline
\parbox{2.6in}{
\begin{center}
\epsfxsize=1.8in
\epsffile{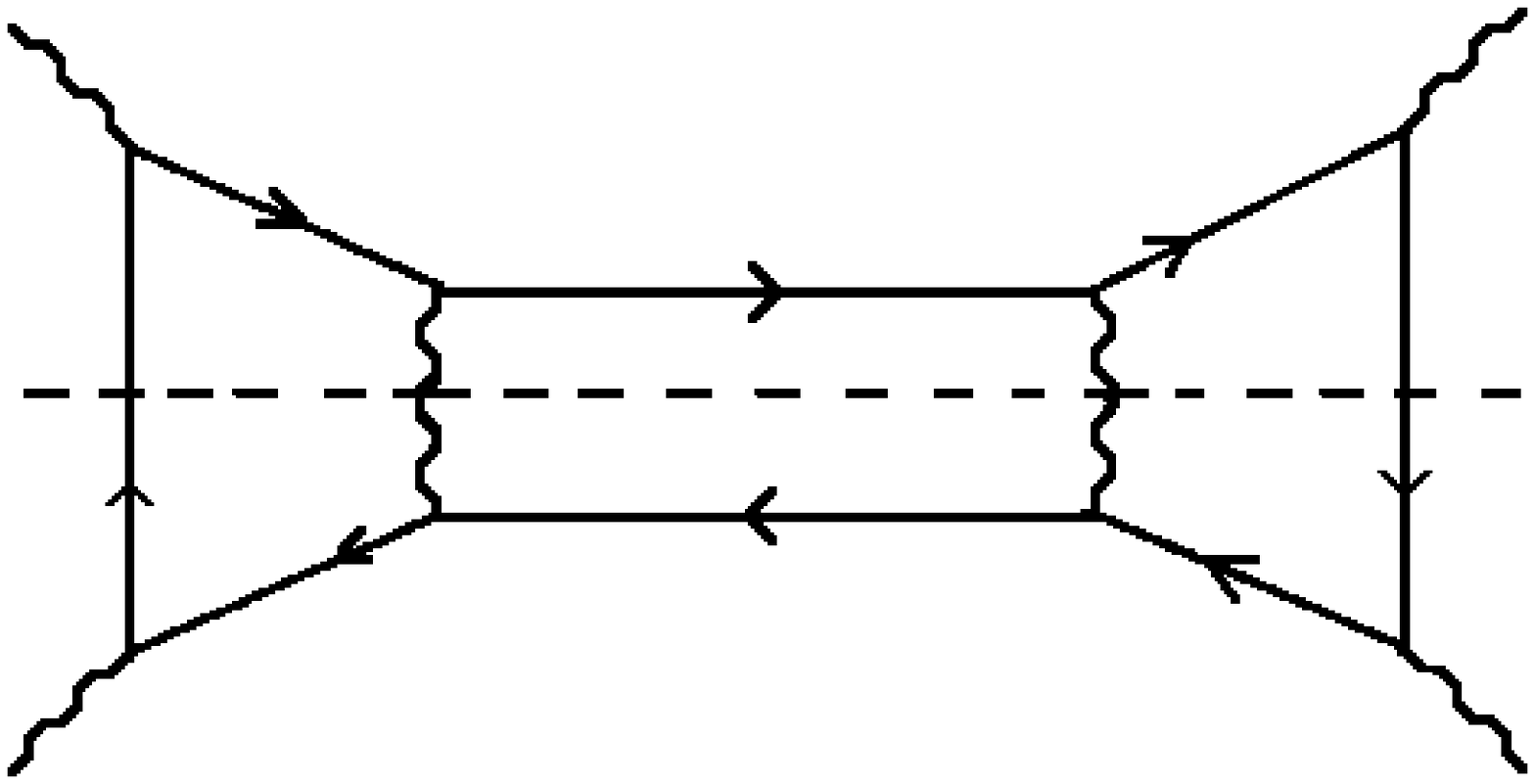}
\end{center}}
\parbox{0.6in}{$\to$}
\parbox{2.6in}{
\begin{center}
\epsfxsize=1.8in
\epsffile{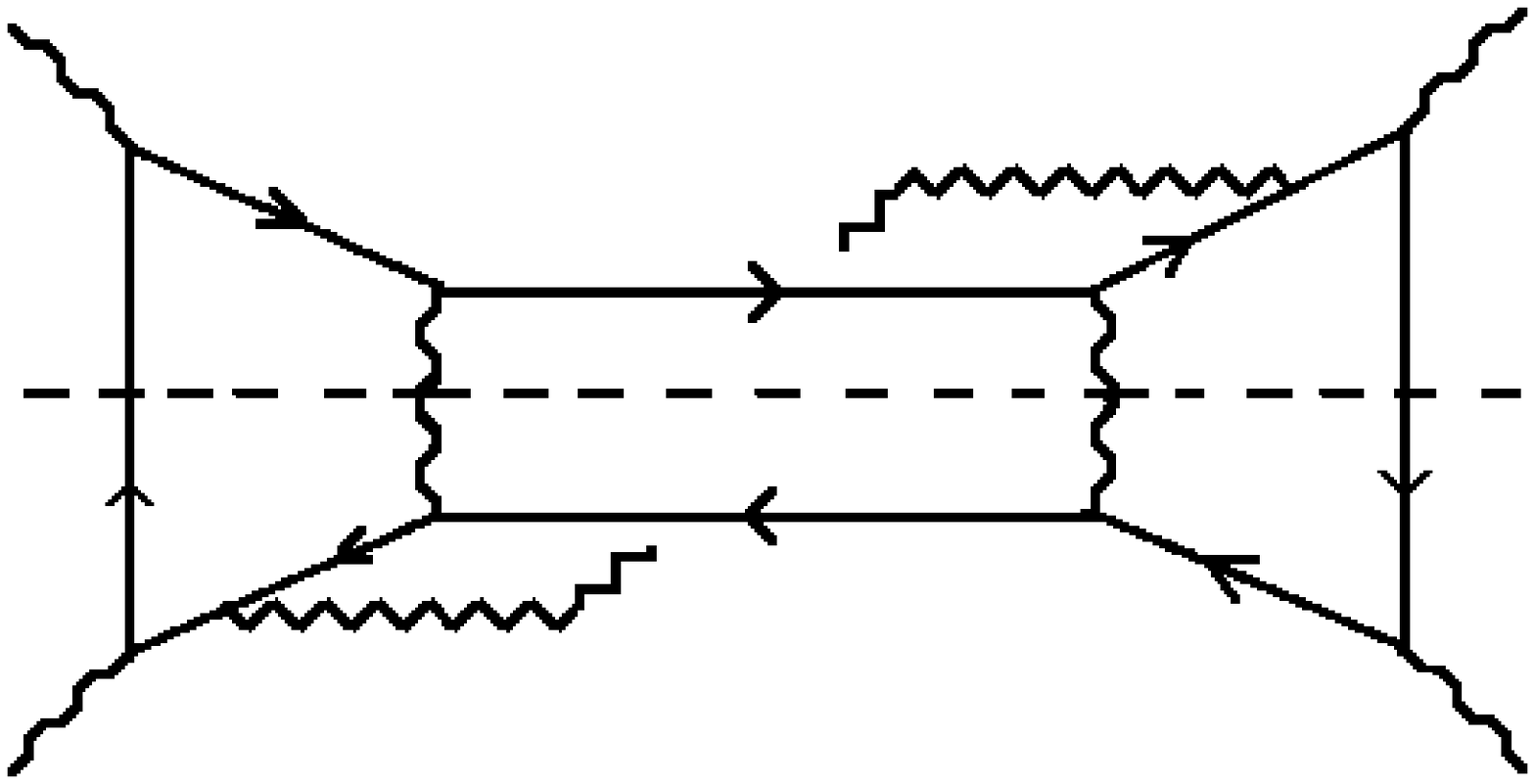}
\end{center}
}
\newline \centerline{(a)\hspace{3in}(b)}
\begin{center}
Fig.~13 Addition of an Exchanged Gluon to a Cut Amplitude
\end{center}
The asymptotic pinching of the particle poles that gives the discontinuity in 
Fig.~13(a), together with the large momentum behavior of the uncut propagators,
is responsible for the pole at infinite momentum in Fig.~13(b)
that corresponds to the anomaly.

A second, essential, point related to the choice of on-shell lines 
is the following. According to multi-regge theory,
the coupling $G_L$ can be evaluated by a double dispersion relation, represented
schematically in Fig.~14 - where the cuts represent the discontinuities involved.
As a consequence, $G_L$ can be expressed as a 
sum over dispersion integrals which give 
amplitudes corresponding to all possible double discontinuities plus, 
possibly but not necessarily, (generalized) 
subtraction terms containing just single discontinuities. The anomaly
contributions we have found are, in fact, generalized subtraction terms and
the contributions of Fig.~3 and Fig.~12, respectively, correspond to the 
two single discontinuity terms shown explicitly in Fig.~14.
\begin{center}
\epsfxsize=5in
\epsffile{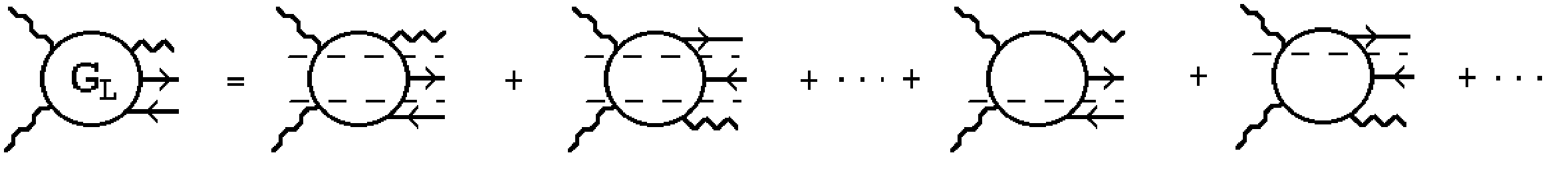}

Fig.~14 Representation of the Double Dispersion Relation for $G_L$
\end{center}
However, since we are
evaluating an amplitude, and not a discontinuity, to have a contribution
with on-shell lines corresponding to particular single discontinuities
of $G_L$ and $G_R$,
these discontinuities must be present in the asymptotic
kinematic region we are considering.
In fact, the discontinuity line in Fig.~13(b) can be regarded as 
representing the combination of the relevant discontinuities of $G_L$ and $G_R$.

\subhead{5.3 Possible Cancelation Mechanisms}

If we consider just contributions to
the transverse momentum diagram of Fig.~1, then 
Fig.~11 is the only diagram which contributes (via the coupling of Fig.~6(a))
to the same helicity amplitude as Fig.~2 and which 
generates an appropriate effective triangle diagram,
apart from the diagram obtained by similarly twisting Fig.~10(b). 
We can not twist just the quark-antiquark state since this would reverse the 
direction of the quark arrow along the fast quark line, requiring a change of the
external helicity to obtain a coupling.  
We conclude, therefore, that the full anomaly contribution to $G_L$ is
obtained by adding the two effective triangle diagrams of Fig.~15.

\begin{center}
\parbox{3.2in}{\epsfxsize=3in
\epsffile{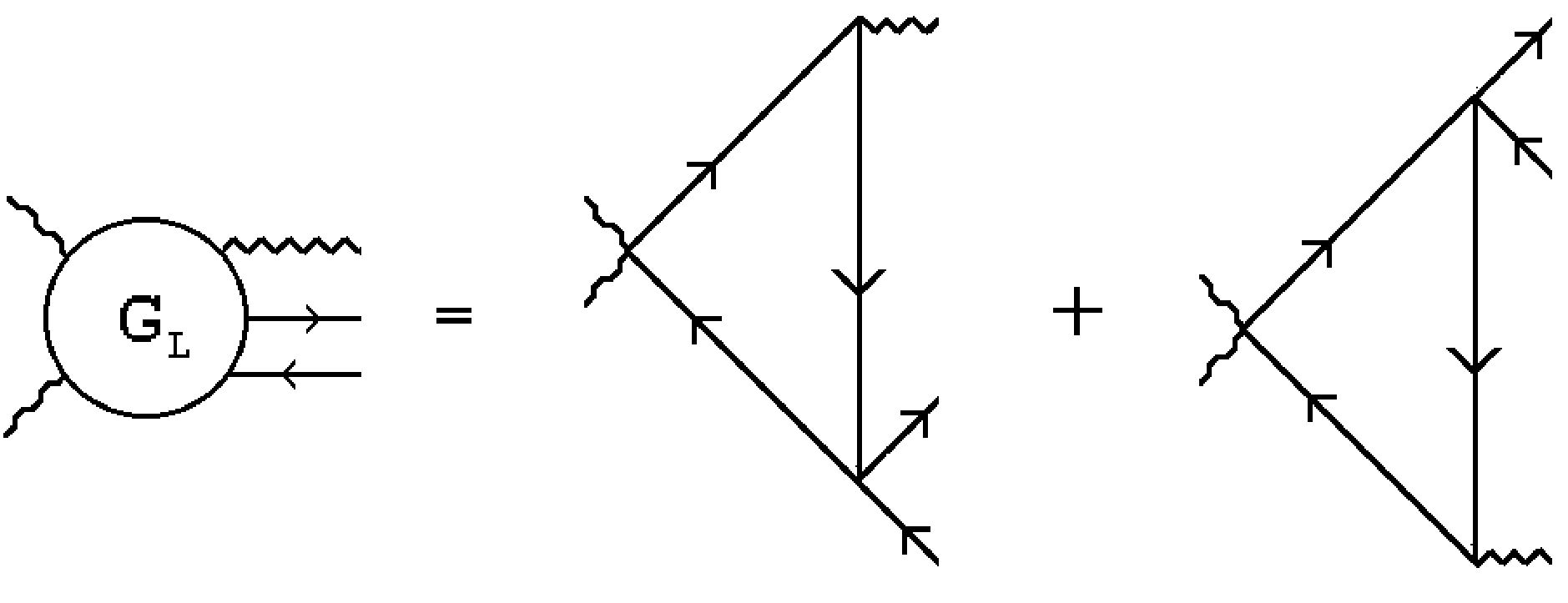}
}

Fig.~15 The Two Effective Triangle Diagrams Generating the Anomaly 
\end{center}
Therefore, within the transverse momentum diagram of Fig.~1, the anomaly 
enhancement does not cancel.

It is natural, however, to expect that there will be further cancelations.
As we emphasized in the Introduction,
because longitudinal vector meson contributions are
involved, it is important to look for all possible 
cancelation mechanisms that could be associated with an underlying 
gauge invariance. In particular,
because the left-side quark and right-side antiquark are asymptotically
on-shell, we must consider whether asymptotic electroweak Ward identities 
could lead to the cancelation of the vector meson 
numerator contributions
that are producing the anomaly enhancement. 

There are two obvious Ward identity related cancelations that we should consider.
First, we consider the tree diagram that appears in the lower half of
Fig.~10(a). At finite momentum, if the intermediate state
quark and antiquark are strictly on-shell,
there will be Ward identities involving this
diagram and all other diagrams obtained by attaching the internal vector meson 
lines at all possible points. Examples of such diagrams, together with the initial
diagram, are shown in Fig.~16.  
\begin{center}
\epsfxsize=4.5in
\epsffile{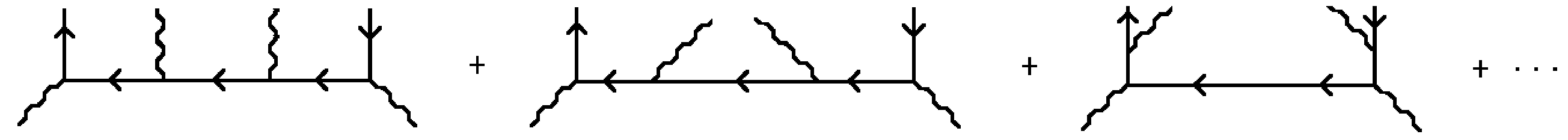}

Fig.~16 Tree Diagrams Contributing to an Electroweak Ward Identities
\end{center}
In fact, because the intermediate state quark and antiquark are 
only asymptotically on-shell, we might expect that only
the vector meson numerator components that are parallel to the 
asymptotic light-cone quark and antiquark momenta must decouple.
This decoupling has
already appeared in the analysis of sub-section {\bf 4.3}.

In Appendix D we study in detail the Ward identity cancelations associated with
the tree diagrams of Fig.~16. The essential
part of the first diagram in Fig.~16 is, indeed, 
directly canceled by the contribution
of the second diagram, which corresponds to the 
feynman diagram shown in Fig.~17. 
\begin{center}
\epsfxsize=1.8in
\epsffile{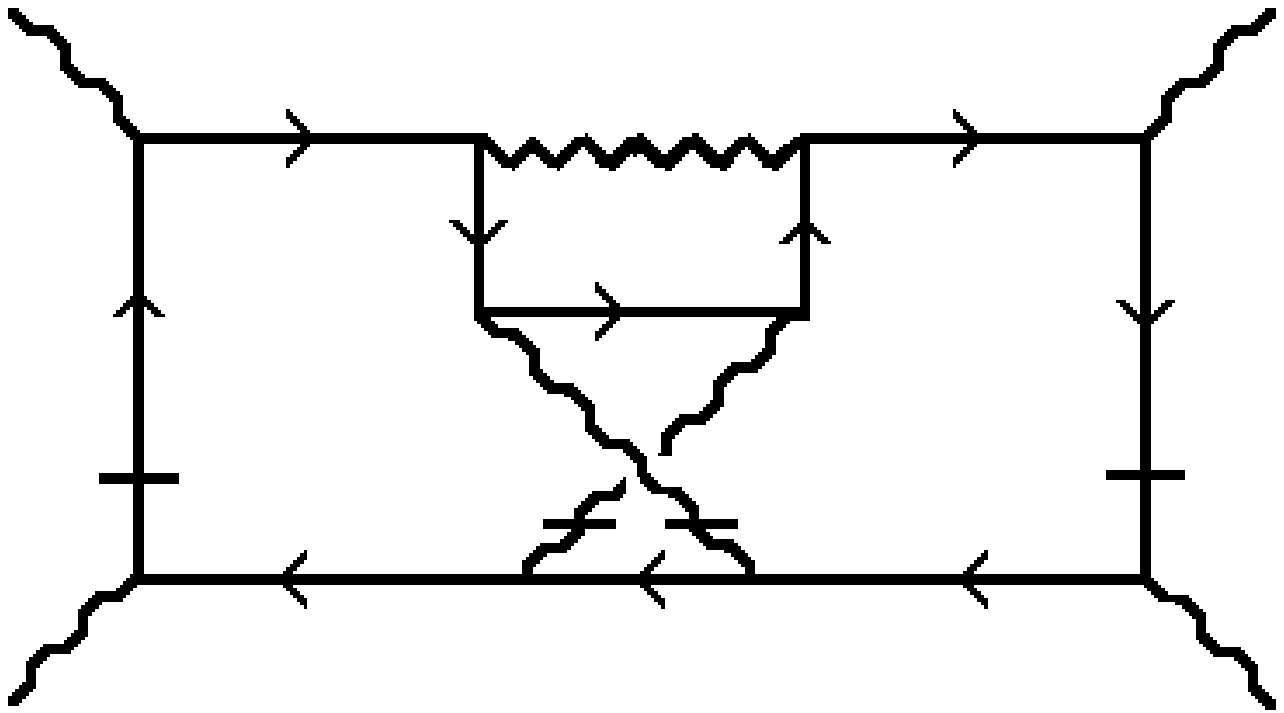}

Fig.~17 A Feynman Diagram With a Canceling Anomaly Contribution.
\end{center}
However, the anomaly enhanced amplitude produced 
by Fig.~17 appears to not be representable as a transverse momentum 
diagram divergence.

More surprisingly, perhaps, essentially the same anomaly enhanced
amplitude then reappears via  
the contribution of the third diagram in Fig.~16, which corresponds to the
feynman diagram shown in Fig.~18.
This is a diagram that would normally be neglected because off-shell
propagators are carrying large light-cone momenta.
\begin{center}
\epsfxsize=1.8in  
\epsffile{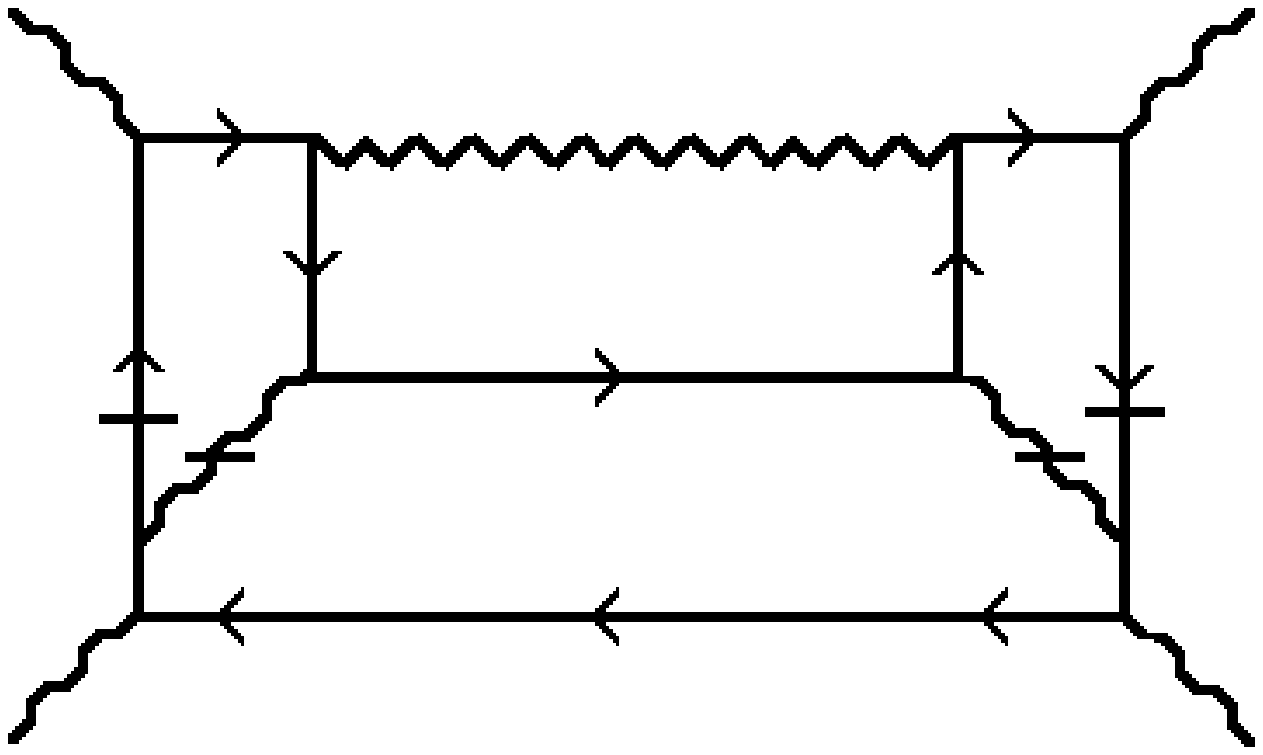}

Fig.~18 The Feynman Diagram Corresponding to the Third Tree Diagram in Fig.~16
\end{center}

In a sense, therefore, 
nothing is gained by implementing the Ward identity cancelations.
However, after this implementation it is apparent that  
the lack of anomaly cancelation is
entirely due to the asymptotic nature of the placing on-shell
of the quark and antiquark lines. 
Also, when the Ward identity cancelations are carried out several diagrams are 
included, in addition to Fig.~18, that would normally be considered non-leading.  
This makes it clear that there is a general phenomenon of superficially
non-leading high-energy behavior contributing to the leading behavior
because of large transverse momentum divergences. 

There is also 
a second Ward identity, involving the top part of Fig.~10(a) and other
loop diagrams, some of which are shown in Fig.~19, that might be expected to
lead to the decoupling of the top $\gamma_- p_+'$ vertex in Fig.~7, together with
the corresponding $\gamma_+p_-''$ right-side vertex. These vertices  
are crucially important for our analysis.
\begin{center}
\epsfxsize=4.5in
\epsffile{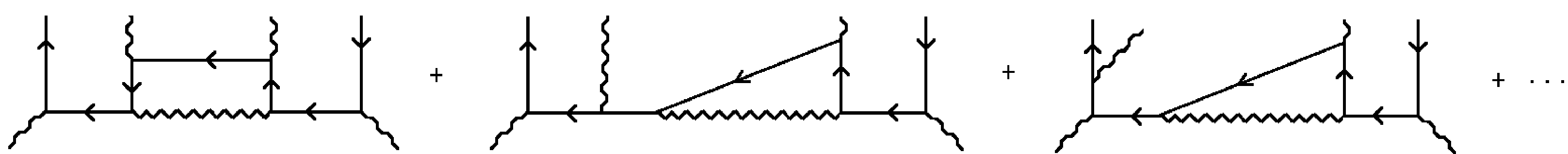}

Fig.~19 Loop Diagrams Contributing to Electroweak Ward Identities
\end{center}
However, in this case it is 
the intervention of the asymptotic anomaly that invalidates the potential
asymptotic Ward identities and, self-consistently, prevents 
the decoupling of the vector meson vertices that are involved. 

In conclusion, we can say that there is no cancelation 
of the transverse momentum coupling
effective triangle diagram anomaly by another diagram with a similar
anomaly. There may very well be a cancelation
outside of the transverse momentum diagram formalism. However, 
in this paper at least, we will not pursue this possibility any further.

\subhead{5.4 The Same Helicity Scattering Amplitude }

To obtain an anomaly amplitude for the scattering of 
vector mesons which both have helicity $\lambda = +1$ 
we include the left side coupling of Fig.~6(b) within a diagram that
otherwise is the same as Fig.~2 or Fig.~11. 
The result is the two diagrams shown in Fig.~20.
When displayed in the first form, it is clear that 
the only difference between these diagrams and, respectively, Figs.~2 and 11
is that along the left-most vertical line
$P_+ \to - P_+$. Therefore, if we evaluate the diagrams with the sign of $P_+$
reversed, corresponding to a cross-channel physical region, the appropriate
on-shell configurations will be present. The diagrams 
will be kinematically identical, respectively, to Figs.~2 and 11
and will give identical anomaly contributions, but with $S \to -S$.

The second form for the diagrams displayed in Fig.~20 is
more transparent for discussing symmetry properties of the intermediate state. In
particular, in this form, it is
clear that Fig.~20(b) can be obtained from Fig.~2 by twisting the quark-antiquark
intermediate state (together with the necessary redirection of the quark arrow
in the left part of the diagram). Fig.~20(a) can similarly be obtained from 
Fig.~11.
\begin{center}
\epsfxsize=4.5in
\epsffile{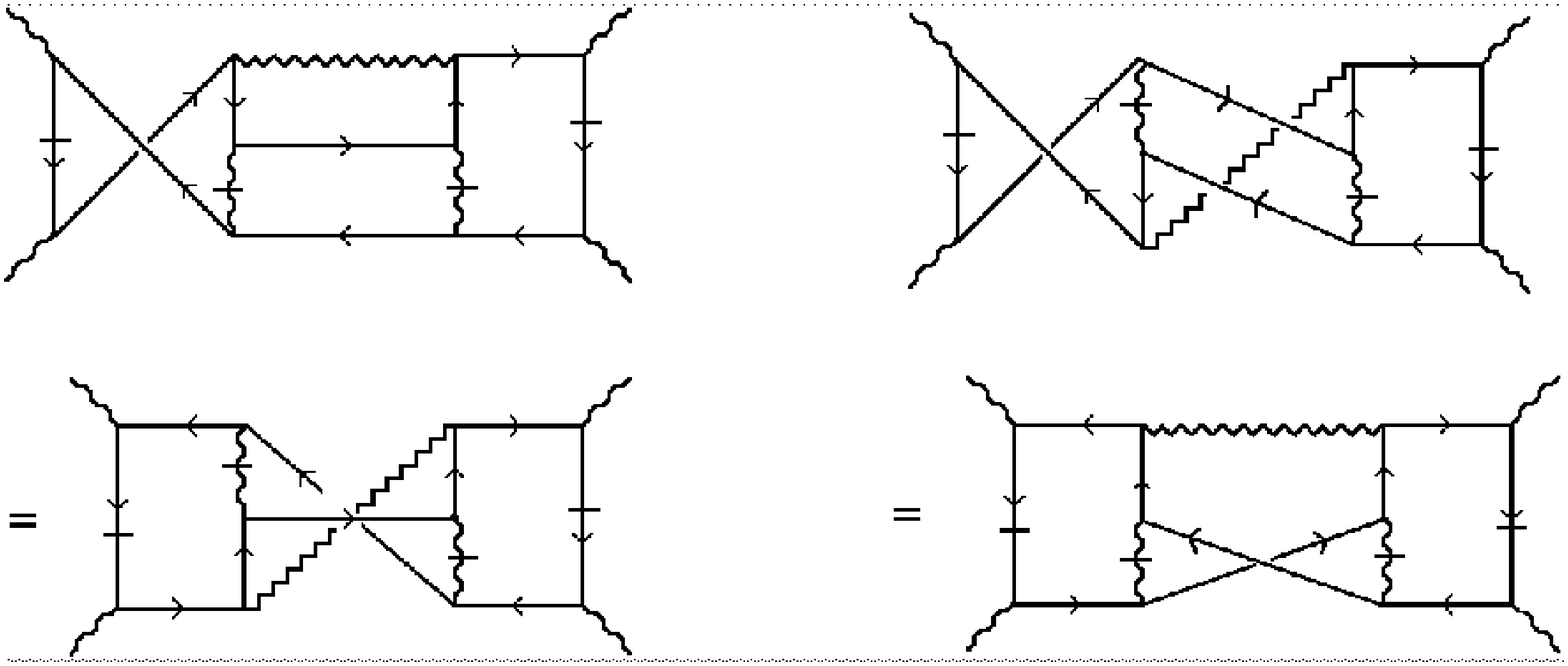}

(a)\hspace{2.5in}(b)

Fig.~20 The Scattering of Same Helicity Vector Mesons
\end{center}

\subhead{5.5 Cancelation in a Vector Theory}

If the vector mesons we are considering had a vector coupling, rather than a 
left-handed coupling, to quarks, then the diagrams of Fig.~16 would appear also
in the opposite sign helicity amplitude but with $(1+\gamma_5)$ couplings
replacing the $(1-\gamma_5)$ couplings in Fig.~6(b). In this case, after the use of
(\ref{prj5}), the relative
minus sign discussed in the previous sub-section 
(resulting from $S\to -S$) produces a cancelation between the
anomaly contributions from Fig.~2 and Fig.~20(a). Similarly, the contributions from 
Fig.~11 and Fig.~20(b) cancel. In a vector theory,
this cancelation of right and left-handed 
coupling contributions would persist, even as we add more gluons as discussed in
the next Section.

\subhead{5.6 The Even Signature Amplitude}

To form signatured scattering amplitudes we should add to, or subtract from,
a particular helicity amplitude the amplitude 
obtained by a CPT transformation of one scattering state relative to the 
other. Therefore, if $A_{-+}$ and $A_{++}$ are the opposite sign and
same sign amplitudes we have discussed, 
$$
A^{\pm}(P_+,P_-)~=~A_{-+}(P_+,P_-)~\pm~ A_{++}(-P_+,P_-)
\auto\label{esa}
$$
is an even/odd signature amplitude. This implies that the anomaly amplitudes
arising from Fig.~2 and Fig.~20(a) are added in the even signature amplitude
and subtracted in the odd signature amplitude (as are, also,
the anomaly amplitudes arising from Fig.~11 and Fig.~20(b). Therefore,
the anomaly cancels in the odd signature amplitude and
is present only in the even signature amplitude. This will continue to be the 
case as we add more gluons in the next Section. It is directly related, via a 
generalization of the discussion of the previous two sub-sections, to the
cancelation in a vector theory. 

\subhead{5.7 $C$ and $P$ Properties of the Transverse Momentum State}

Since the intermediate state in Fig.~1 is completely transverse (or,
equivalently, is a $t$ - channel intermediate state) the $T$ part of the 
$CPT$ transformation, defining the signature of an amplitude, has no effect on it.
Therefore, we should be able to relate signature directly to the $CP$ properties
of the transverse momentum state.

The parity transformation reverses the transverse momentum of the gluon
and so, because of the coupling (\ref{bxi11}), 
simply gives a minus sign. Without a color factor, the charge conjugation
transformation also gives just a minus sign. Therefore, 
the gluon component of the intermediate state is even under $CP$.
For quarks the left-handed coupling violates both $P$ and $C$.
As a result, the quark-antiquark
intermediate state only has simple 
transformation properties under the combined $CP$ transformation.
Charge conjugation transforms
a quark (antiquark) to the 
corresponding antiquark (quark), with the same helicity (opposite chirality). 
The parity part of the $CP$ transformation then 
reverses the helicities. In our case, the quark and antiquark have opposite 
helicities and so they will be simply
interchanged by the $CP$ transformation. Individually, the diagrams we 
are discussing do not have simple symmetry properties with respect to 
quark/antiquark interchange. Not surprisingly, however, the full set of anomaly 
contributions in the even signature amplitude does have such a property. 

With the four diagrams, Figs.~2, 11, 20(a) and (b),
added in the even signature amplitude, it is clear
(using the second display form in Fig.~20) that the left-side
coupling is symmetric, diagrammatically, 
with respect to the interchange of the quark and antiquark. 
The interchange relates Fig.~2 to Fig.~20(b) and Fig.~11 to Fig.~20(a).
In addition to the 
reversal of the quark line, the contribution of
Fig.~20(b) to the even signature amplitude differs kinematically from that  
of Fig.~2 in two ways that produce canceling sign changes. Firstly,
$k_{\perp} \to -k_{\perp}$, and, secondly, the effect in the
$k_-'$ integration of $P_+ \to -P_+$ resulting from the definition of 
the even signature amplitude. Therefore, in this
amplitude, the quark/antiquark intermediate state is even under $CP$.
Since the gluon state is also even under $CP$, 
the full transverse momentum state is indeed even under $CP$, as
it should be.

Note that the full even signature 
amplitude will contain, in addition to the four diagrams 
of Figs.~2, 11, 20(a) and 20(b), the four related diagrams obtained by 
substituting the right side of Fig.~10(b) for that of Fig.~10(a). In effect,
in this second set of diagrams the twists are made on $G_R$ that are made on
$G_L$ in the first set of diagrams. Each set of twists is sufficient to give
an intermediate state with the appropriate $CP$ property. As a result, the
discussion of each set of four diagrams can be made separately and
is directly parallel. In higher orders it will sometimes 
be necessary to consider both sets of twists 
together to obtain an intermediate state with the right $CP$ property.

\newpage

\mainhead{6. COLOR FACTORS AND MORE GLUONS}

We begin with a discussion of SU(2) flavor that will, essentially, allow us to 
ignore it in the following.

\subhead{6.1 SU(2) Flavor}

The SU(2) flavor symmetry will play only a minimal role in our discussion
and we will introduce it in a very elementary manner.
We consider the exchange of a quark-antiquark  
$\{I=1, I_z=0\}$ state that, in the standard model, would
carry the quantum numbers of the $\pi^0$. Identifying $W^{\pm},W^0$ with the 
$\{I=1, I_z = \pm,0\}$ vector mesons and identifying 
$u,d$ with the $\{I= \frac{1}{2}, I_z= \frac{\pm1}{2}\}$ 
quarks, we can add flavor quantum numbers to the discussion
of the previous Section by using the vector meson / quark vertices of Fig.~21.
\begin{center}
\epsfxsize=3.5in
\epsffile{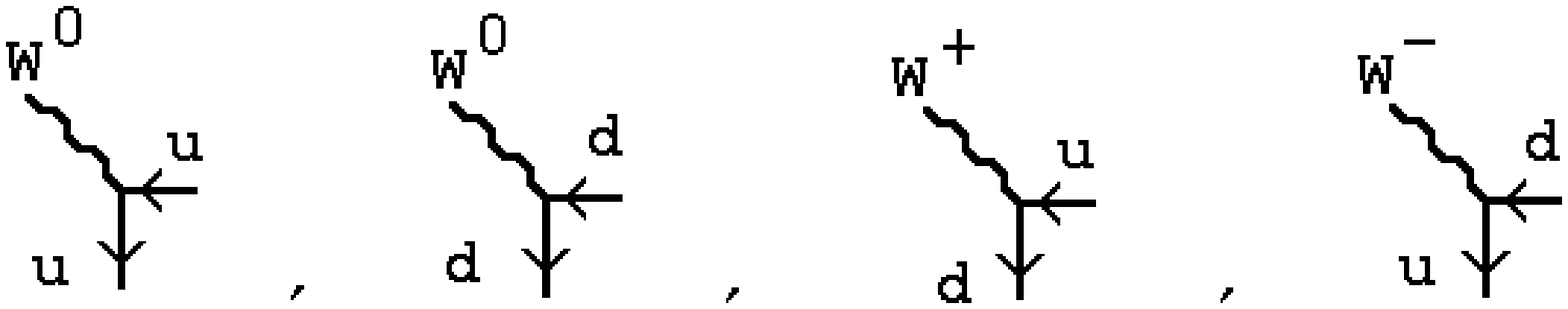}

Fig.~21  Vector Meson / Quark Vertices
\end{center}
The flavorless couplings of Fig.~6 are then replaced by 
the sums of couplings shown in Figs.~22(a) and (b) and the internal vector meson
on-shell contributions are replaced by a similar sum. 
\begin{center}
\epsfxsize=2.5in
\epsffile{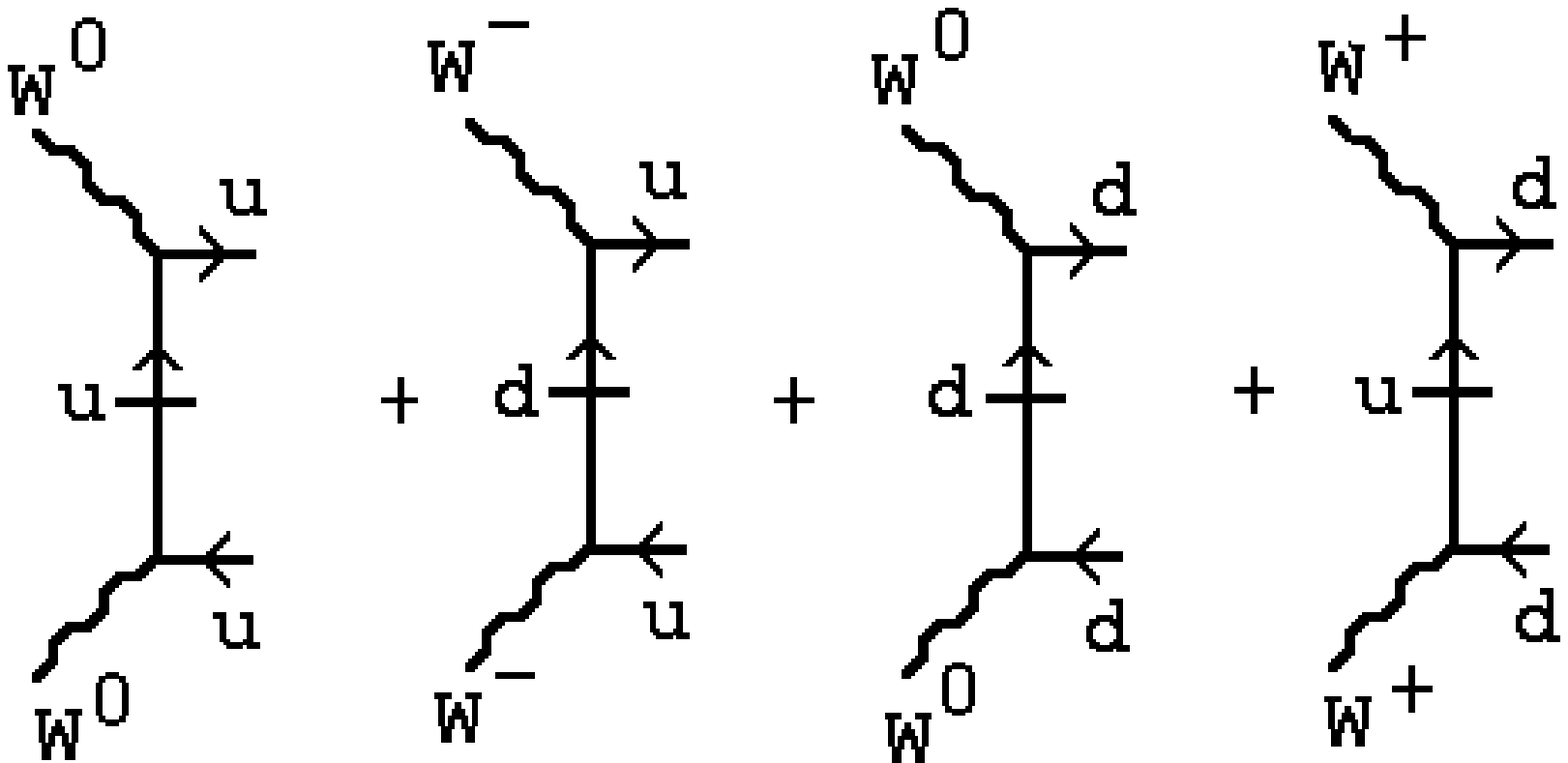}
\hspace{0.5in}
\epsfxsize=2.5in
\epsffile{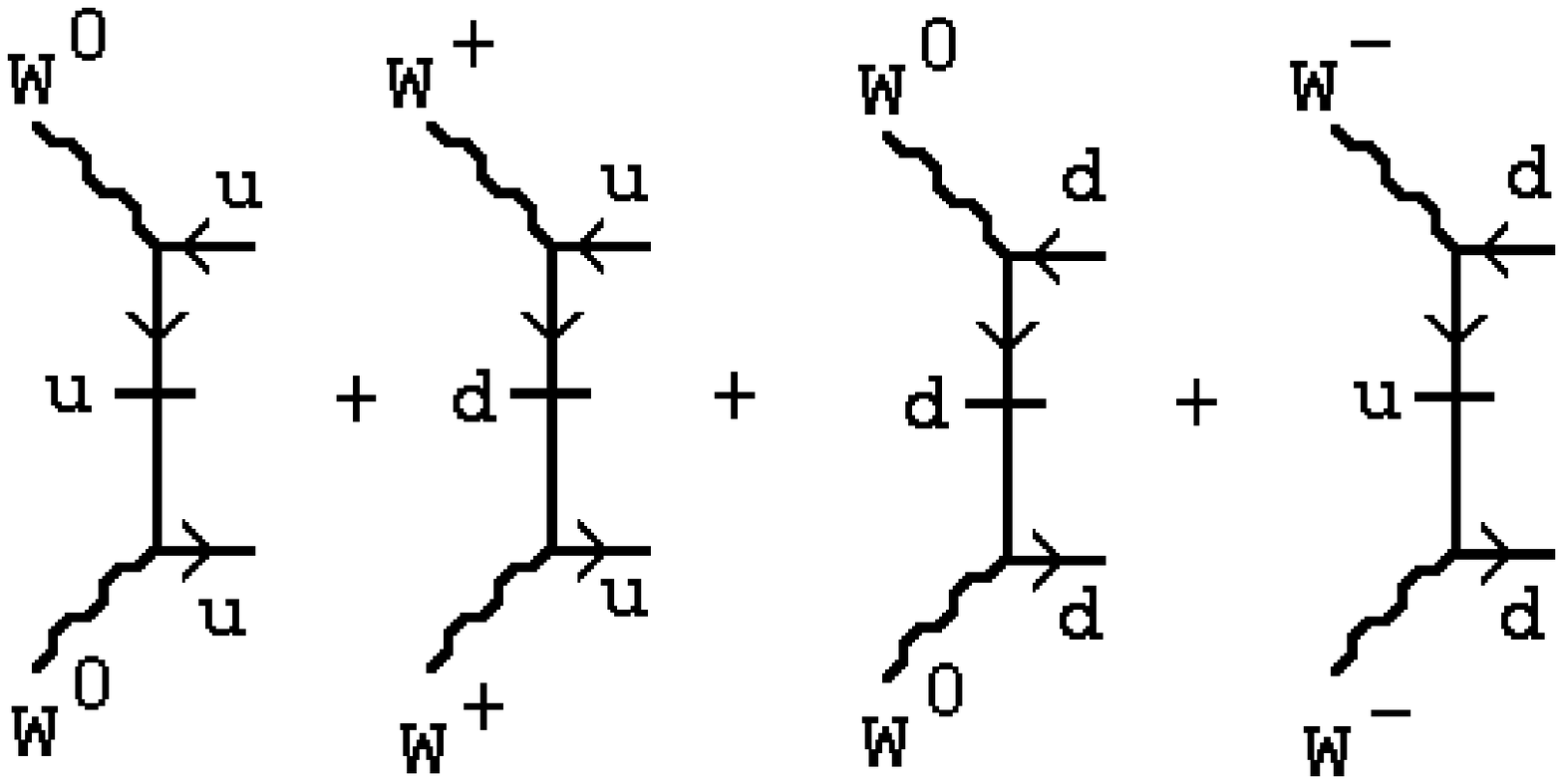}

(a)\hspace{2.8in}(b)

Fig.~22 Couplings with SU(2) Flavor Quantum Numbers
\end{center}
We then add all the diagrams obtained with this set of couplings. 
The most important feature of these couplings is that they are 
symmetric with respect to $u \leftrightarrow d$ and that this symmetry is preserved
by the internal vector meson exchange interactions. 
(It is also important for the $CP$ properties of the diagrams we discuss
that, in going from Fig.~22(a) to Fig.~22(b), left-handed quarks and right-handed
antiquarks are interchanged, as was the case for Figs.~6(a) and (b).) 
Consequently, the addition
of SU(2) flavor factors will not produce any diagram cancelations and we can
leave, as implicit, the replacement of the couplings of Fig.~6 by those of Fig.~22.

\subhead{6.2 Color Factor Diagrams and the One Gluon Color Factor}

SU(3) color factors will also be relatively
simple. In all the diagrams we discuss,
there will be only one quark loop. There is
no external color and so color is introduced into the quark loop only by the 
couplings to the internal gluons. Also, for the diagrams we consider, gluons
will appear only as part of the exchanged transverse momentum state and will
be attached within the corresponding $G_L$ and $G_R$ transverse momentum couplings.
As a result, we can use a simple notation
to describe color factors. We represent the quark loop as a rectangle,
and attach gluons only to the vertical lines.
The attachment of gluons to the left-side vertical line represents the order
of attachment to the quark loop within the left-side transverse momentum 
coupling $G_L$, while the right-side vertical line similarly represents the
order of attachment within $G_R$. For each gluon
there is a color matrix  $\lambda_i$ at each attachment point. 
The full color factor is the trace of 
the product of the $\lambda$ - matrices taken around the loop, 
and then summed over $i=1,..,8$ for each gluon.
The notation is illustrated for various numbers of exchanged gluons in Fig.~23.
\begin{center}
\epsfxsize=3.5in
\epsffile{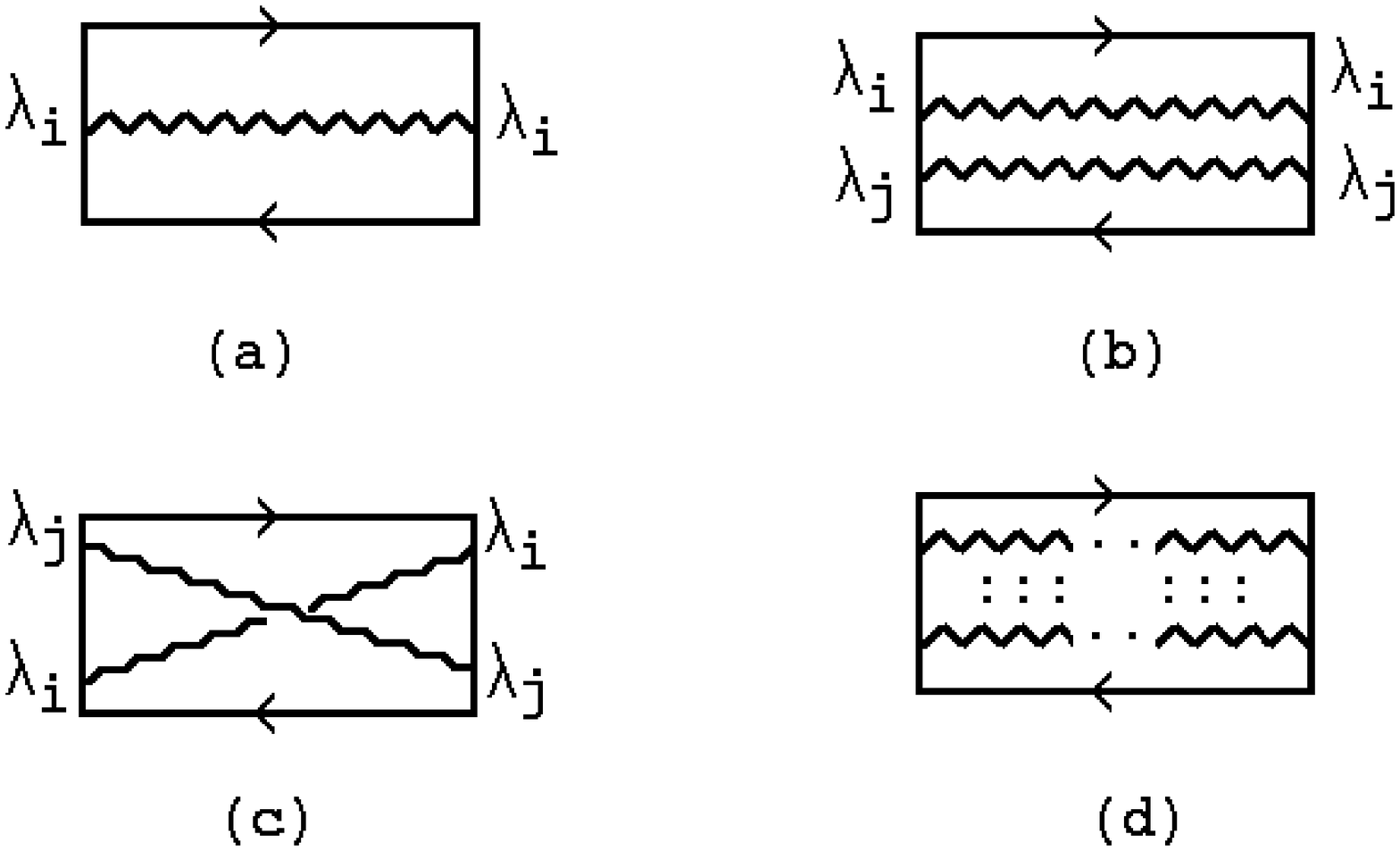}

Fig.~23 Color Factor Diagrams
\end{center}

All of the diagrams discussed in the previous Section contain just one 
gluon and have the same color factor. This is represented by Fig.~23(a) and is 
simply
$$
\centerunder{$\hbox{\Large $\Sigma$}$}{\raisebox{-3mm}{$
{\scriptstyle i}$}}
~~Tr\{\lambda_i^2\}
\auto\label{cf1}
$$
Since all diagrams have the same color factor (and flavor factor) 
all of the discussion in the previous Section is essentially unchanged, apart from 
the discussion of charge conjugation, which now has to include color charge
conjugation.

For a general gluon field with color matrix $M_{ab}$, color charge conjugation 
is defined as 
$$
M_{ab}~~\to~~ \bigl[-M_{ab}~\bigr]^T ~~= ~~-M_{ba}
\auto\label{gcc}
$$
For the hermitian color matrix vertices $\lambda_i$
$$
\bigl[-\lambda_i~\bigr]^T ~~= ~-~\bigl[\lambda_i ~\bigr]^* 
\auto\label{gcc1}
$$
where $[~]^*$ denotes complex conjugation. Therefore, in addition to the charge 
conjugation minus sign discussed in the last section, the coupling of the gluon
to the quark line (within $G_L$, say) is complex conjugated. Correspondingly,  
for the quark-antiquark pair, in addition to the
charge conjugation discussed in the previous Section, 
quark/antiquark interchange gives 
$$
\bigl[\lambda_i ~\bigr]^* ~\to ~ \bigl[\lambda_i ~\bigr]
\auto\label{qcc}
$$
in (\ref{gcc1}). Since the parity transformation is unchanged, 
the full gluon plus quark/antiquark transverse
momentum state remains even under $CP$ when color charge
conjugation is included.
 
\subhead{6.3 The Addition of a Soft Gluon }

Next, we look for feynman diagrams that contain two gluons and that also, 
potentially, contain the anomaly enhancement. We will 
assume that only one gluon is involved in the transverse momentum divergence
and will consider two
possibilities for the scale of the transverse momentum carried by the second
gluon. It can either be ``soft'', 
i.e. it carries a very small momentum $k''$, with 
$$
|k_{\perp}''|~<< ~M~<<~|k_{\perp}|, |k'_{\perp}| ~~~~
\leftrightarrow ~~~~``soft~''
\auto\label{soft}
$$ 
or ``finite'', i.e. 
$$
|k_{\perp}''|~\sim ~M~<<~|k_{\perp}|, |k'_{\perp}| ~~~~
\leftrightarrow~~~~``finite~''
\auto\label{fnte}
$$ 
As we will see,
in some diagrams soft gluon exchange is possible, in addition to the anomaly
generation, while in others only finite gluon exchange is possible. In both 
cases, the second gluon will provide an important color factor.  
A soft gluon, however, will also produce an infra-red divergence.  
Since the full transverse momentum
state carries zero color, such divergences must cancel. This will 
help us to locate other diagrams generating the anomaly.

We consider the ``soft'' gluon case first and
look for diagrams that contribute to the
transverse momentum diagram of Fig.~24 which, as discussed further in Appendix B,
would again be expected to
contribute (formally) only at next-to-next-to-leading log. 
(Note that, when the anomaly is not present, this diagram again,
potentially, includes a logarithmic transverse momentum divergence generating
an additional energy logarithm.)
\begin{center}
\epsfxsize=2in
\epsffile{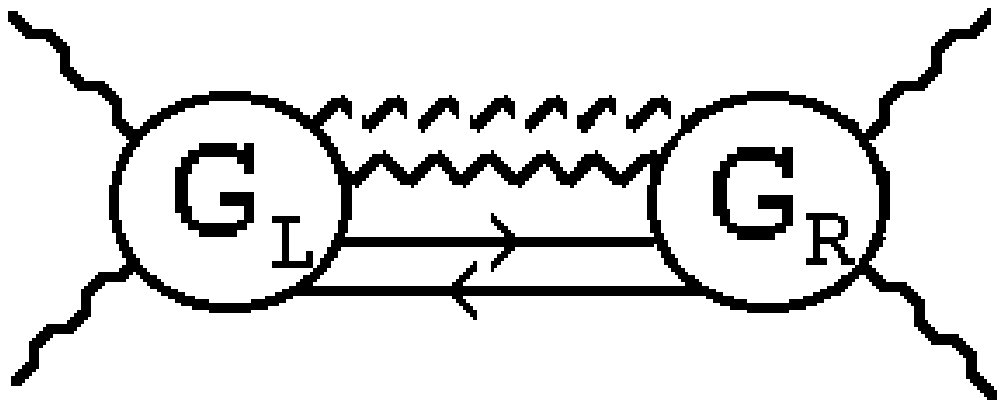}

Fig.~24 The Two Gluon Transverse Momentum Diagram - the Broken Line Denotes
a Soft Gluon
\end{center}
 
We begin with the addition of a soft gluon to the diagram of Fig.~2.
The anomaly will appear in the same manner as before if, in the high-energy limit, 
an effective triangle diagram is generated as in Fig.~3, but with the 
additional gluon attached, via a point coupling, 
to one of the three vertices of the diagram. 
The required local coupling could appear, 
in principle, if 
$k''$ can be directed through an adjacent quark line which can
be put on-shell by  the $k_-''$  integration. If this line carries 
(predominantly) a large light-cone momentum then, in analogy with
(\ref{k'os}), the integration will produce
couplings that are independent of $k_{\perp}''$. 
In Figs.~25(a), (b), and (c) we show how the extra gluon could be added 
to Figs.~6,7, and 8, respectively, with 
the final $\gamma$ - matrices remaining the same as in Fig.~9.
\begin{center}
\epsfxsize=4.5in
\epsffile{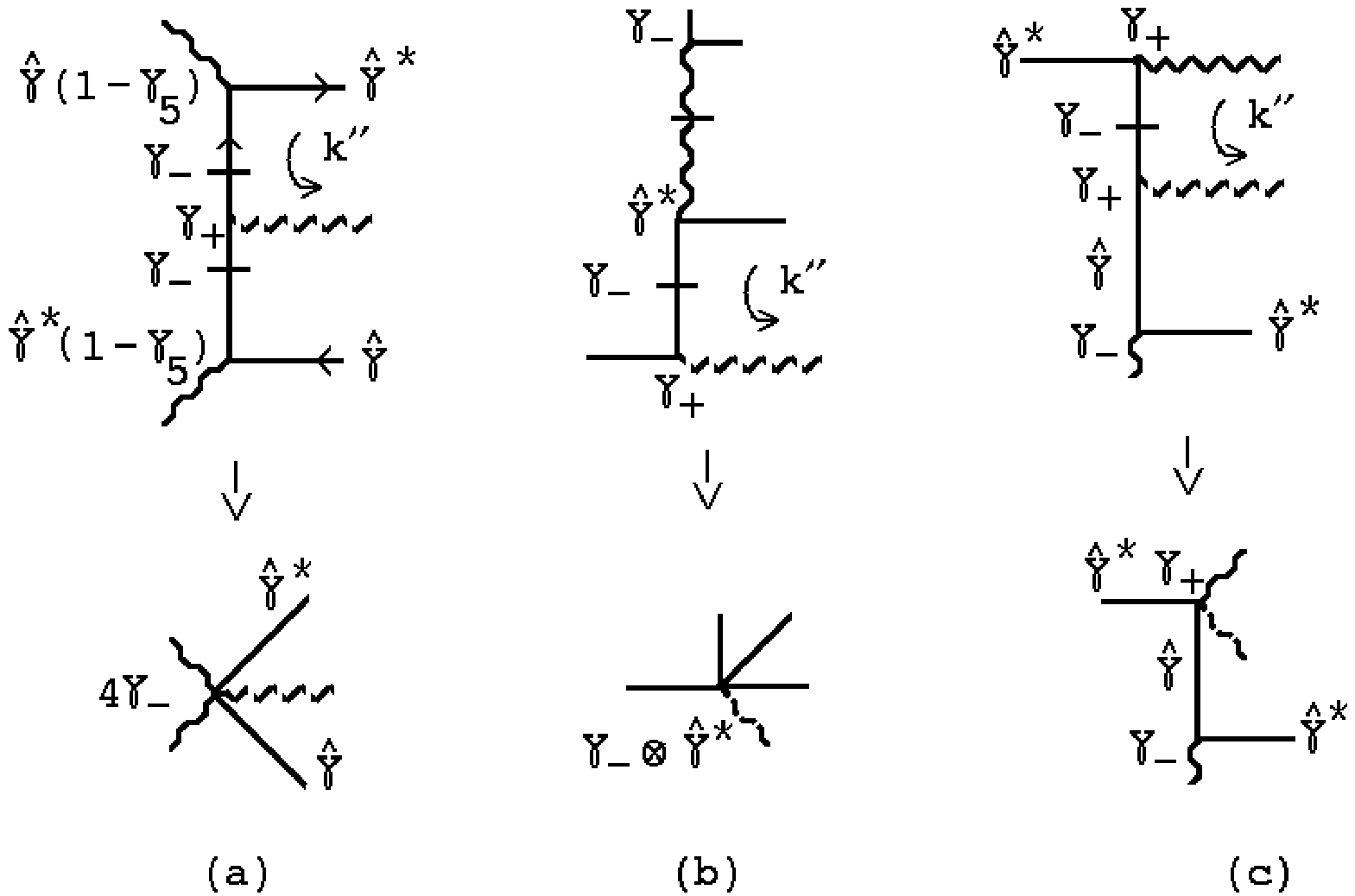}

Fig.~25 Adding a Gluon to the Vertices of (a) Fig.~6~, (b) Fig.~7~, and
(c) Fig.~8
\end{center}
In Fig.~25(c) the soft and hard gluon can be interchanged, whereas in Figs.~25(a)
and (b) there is no ambiguity as to where the soft gluon has to be attached,
if the $\gamma$ - matrix structure is to remain the same.

For Fig.~25(a) we can, essentially, apply (\ref{k'os}) directly.
For the couplings of Figs.~25(b) and (c) there is, however, a problem
if the extra gluon is soft and so 
carries only very small transverse momentum. In these cases,
the propagator that has to be placed on-shell by the $k_-''$ - integration 
is adjacent to an off-shell propagator
that, in the anomaly configuration, 
is carrying very large transverse momentum ($ p_{\perp}'$). 
In this case the mass-shell condition is 
$$
k_-''~\sim ~ \frac{(p_{\perp}'~+~ k_{\perp}'')^2}{p_+'}~\sim~
\frac{\epsilon M \sqrt{S}}{\epsilon  \sqrt{S}} ~\sim~M
\auto\label{k''os}
$$
which can not be satisfied with $k_-''~<<~ |k_{\perp}''| 
~<<~M$. Therefore, if the vertex for the extra gluon is of the form of
Fig.~25(b) or (c), in both the $G_L$ and $G_R$ couplings, then it can not carry
$|k_{\perp}''|^2 ~<<~M^2$. 

Later, we will discuss potential contributions from 
vertices of the form of Fig.~25(b) and (c) when $k''$ is ``finite'', i.e.
$|k_{\perp}''|^2 ~\sim~M^2$. For the moment,
we consider only the vertex of Fig.~25(a).
Generation of the corresponding triangle diagram 
is shown in Fig.~26
\begin{center}
\epsfxsize=4.5in
\epsffile{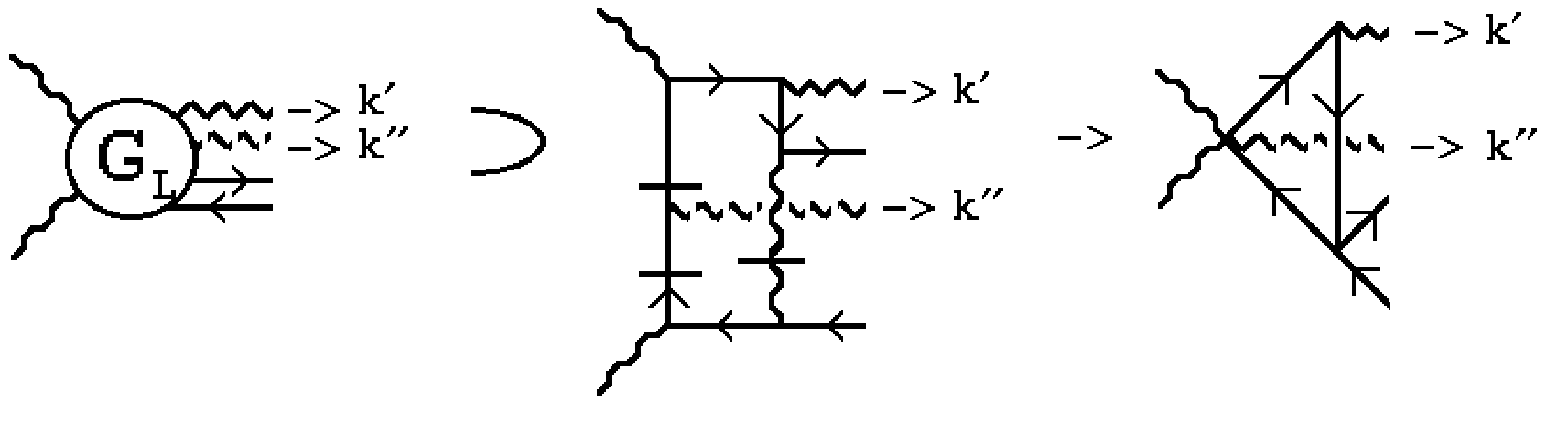}

Fig.~26 A Two Gluon Effective Triangle Diagram
\end{center}
and the full feynman diagram, with the extra gluon attached in the same 
manner to both sides of Fig.~5, is shown in Fig.~27.
\begin{center}
\parbox{2.5in}{\epsfxsize=2.1in
\epsffile{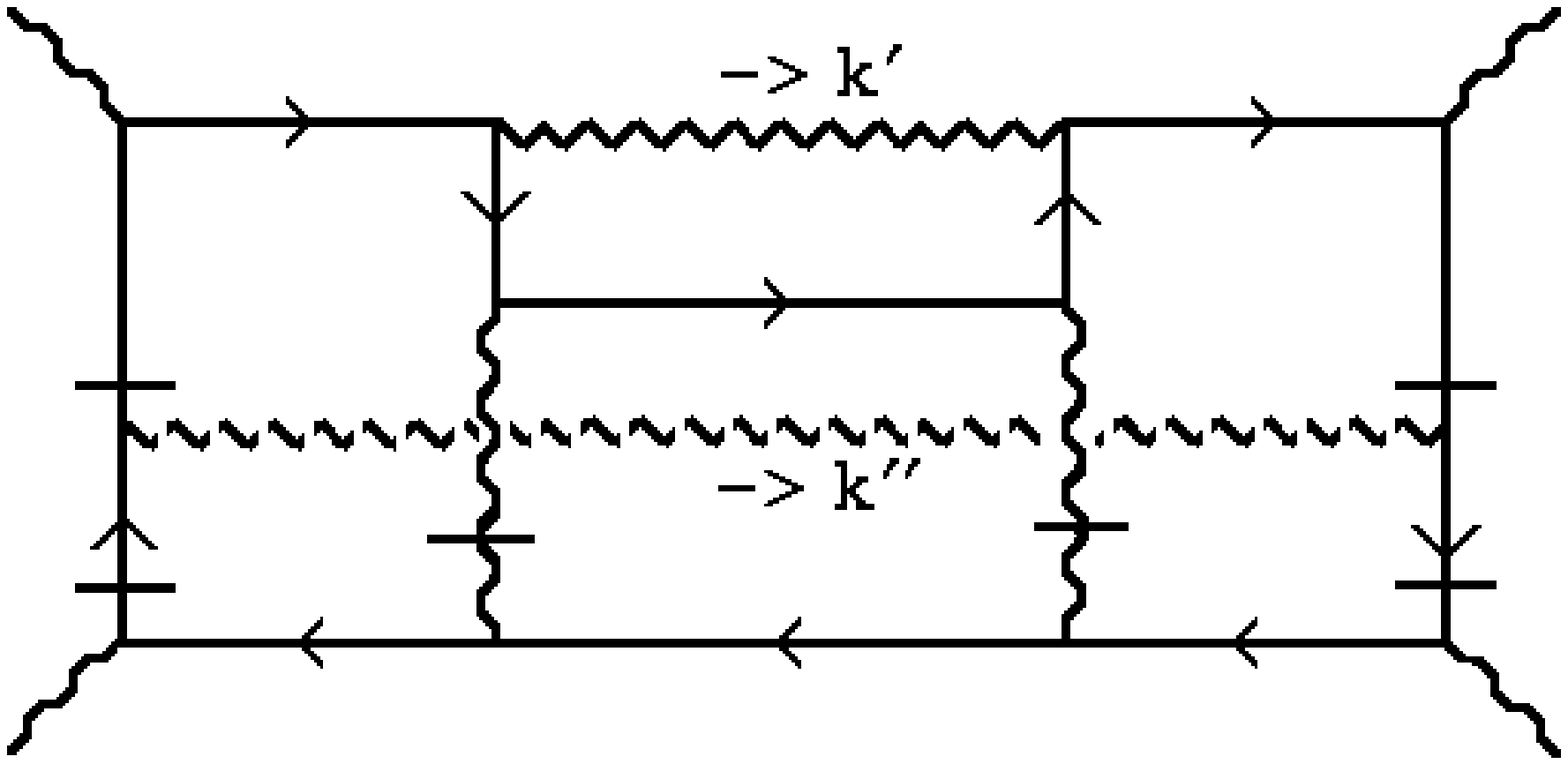}}
\parbox{0.5in}{$\leftrightarrow$}
\parbox{2.5in}{\epsfxsize=2.1in
\epsffile{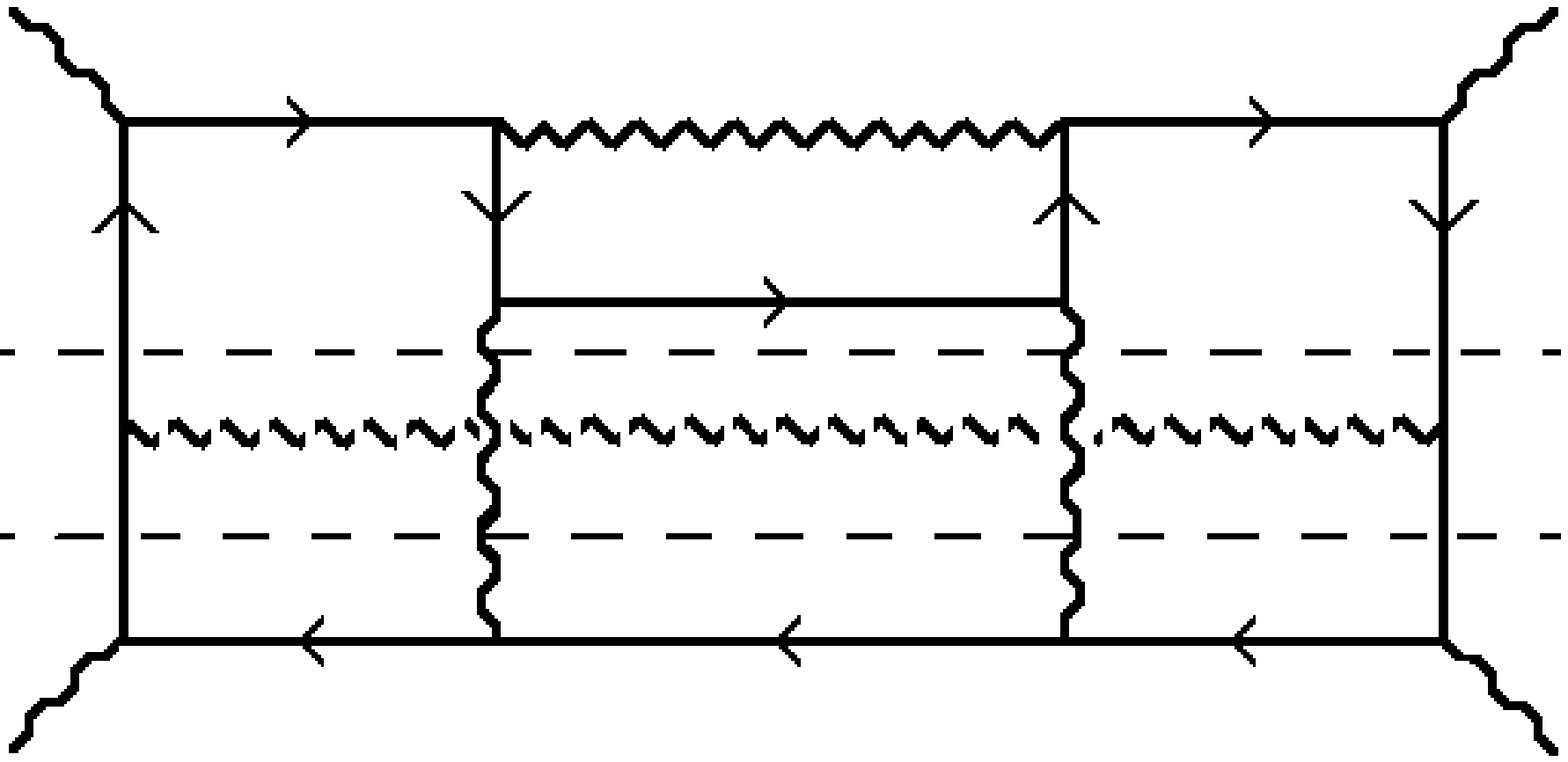}}

Fig.~27 A Two Gluon Feynman Diagram With Two Cuts
\end{center}
As illustrated, the lines put on-shell correspond to making a double cut
of the diagram. This corresponds to double discontinuity contributions to
the $G_L$ and $G_R$ couplings. (These contributions are again generalized
subtractions in that the full dispersion relations for $G_L$ and $G_R$  
contain triple discontinuities.)

\subhead{6.4 The Two Gluon Anomaly Amplitude}

If $k_{\perp}''$ is much smaller than any other transverse momentum in the 
diagram of Fig.~23, the only significant
$k_{\perp}''$ dependence will be in the $k''$ propagator. 
Hence, the $k_{\perp}''$ integration can be factored out from the 
remaining integrations and,
before the inclusion of any color factor, the diagram of Fig.~23 gives,
via the reduction of Fig.~25(a) and Fig.~26, 
a high-energy anomaly enhanced amplitude of the form
$$
\eqalign{ \int^{|k_{\perp}''|<< M }&
 ~\frac{d^2k_{\perp}''}{{k_{\perp}''}^2}~
\int ~\frac{d^2k_{\perp}'}{{k_{\perp}'}^2}~\int d^2k_{\perp} ~
~~\biggl(~\frac{k_{\perp}\times k'_{\perp}}{M^2}
~ \biggr)^2~ 
\frac{ Tr\{ \gamma_- \gam^* \gamma_+ \gam\} }
{\kt \kt^*} \cr
&\sim ~\biggl(~\int^{|k_{\perp}''|<< M } ~\frac{d^2k_{\perp}''}{{k_{\perp}''}^2}~
\biggr)~\int ~\frac{d^2k_{\perp}' d^2k_{\perp}}{{k_{\perp}'}^2 k_{\perp}^2} 
~\biggl(~\frac{k_2k'_3 -k_3k'_2}{M^2}\biggr)^2~\cr
&\sim ~\biggl(~\int^{|k_{\perp}''|<< M } ~\frac{d^2k_{\perp}''}{{k_{\perp}''}^2}~
\biggr)~\int_{O(MS^{\frac{1}{2}})}~\frac{d({k_{\perp}'}^2)}{M^2}~
\int_{O(MS^{\frac{1}{2}})}~ \frac{d(k_{\perp}^2)}{M^2}~\cr
&\sim ~S~/~M^2
}
\auto\label{amp2}
$$
We can similarly add a soft gluon to each of the one gluon diagrams
discussed in the last Section and generate a high energy amplitude
of the same form. There is, however, clearly a divergence at $|k_{\perp}''|^2 
=0$ that we must discuss. First, however, we discuss the relevant
color factors.

\subhead{6.5 Two Gluon Color Factors}

There are two possible color factors for the transverse momentum diagram
of Fig.~24. They are shown in Figs.~23(b) and (c) and have the form
$$
(b)~~ \centerunder{$\hbox{\Large $\Sigma$}$}{\raisebox{-3mm}{$
{\scriptstyle i,j}$}}
~~Tr\{\lambda_j\lambda_i\lambda_i\lambda_j\}~~~~~~~~
(c)~~\centerunder{$\hbox{\Large $\Sigma$}$}{\raisebox{-3mm}{${
\scriptstyle i,j}$}}
~~Tr\{\lambda_i\lambda_j\lambda_i\lambda_j\} 
\auto\label{cf2}
$$
We discuss these two factors in a manner that will generalize when 
further gluons are added. The essential formula we need is
$$
\lambda_i\lambda_j ~=~ \frac{2}{3}~\delta_{ij}~
+~\centerunder{$\hbox{\Large $\Sigma$}$}{\raisebox{-3mm}{$
{\scriptstyle k}$}}~
(if_{ijk} + 
d_{ijk})\lambda_k 
\auto\label{lij}
$$
which, since $f_{ijk}$ is antisymetric and $d_{ijk}$ is symmetric, 
implies that 
$$
\lambda_j\lambda_i ~+~ \lambda_i\lambda_j ~=~ \frac{4}{3}~\delta_{ij}~
 ~+~2~ \centerunder{$\hbox{\Large $\Sigma$}$}{\raisebox{-3mm}{$
{\scriptstyle k}$}}~d_{ijk}\lambda_k 
\auto\label{lji+}
$$
and 
$$
\lambda_j\lambda_i ~-~ \lambda_i\lambda_j ~=~
-~2i ~\centerunder{$\hbox{\Large $\Sigma$}$}{\raisebox{-3mm}{$
{\scriptstyle k}$}}~f_{ijk}\lambda_k ~~~~~~~
\auto\label{lji-}
$$
Therefore, the sum of the two color factors (\ref{cf2}) is given by 
$$
\eqalign{ \centerunder{$\hbox{\Large $\Sigma$}$}{\raisebox{-3mm}{$
{\scriptstyle i,j}$}}
~~Tr\{\lambda_j\lambda_i\lambda_i\lambda_j ~+~
\lambda_i\lambda_j\lambda_i\lambda_j\} 
&=~  \centerunder{$\hbox{\Large $\Sigma$}$}{\raisebox{-3mm}{$
{\scriptstyle i,j}$}}
~~Tr\{[~\frac{4}{3}~\delta_{ij} ~+ ~2~
\centerunder{$\hbox{\Large $\Sigma$}$}{\raisebox{-3mm}{$
{\scriptstyle k}$}}~
d_{ijk}\lambda_k]~\lambda_i\lambda_j~\}\cr
&= ~ \centerunder{$\hbox{\Large $\Sigma$}$}{\raisebox{-3mm}{$
{\scriptstyle i,j}$}}
~~Tr\{[~\frac{4}{3}~\delta_{ij}~ 
+~2~
\centerunder{$\hbox{\Large $\Sigma$}$}{\raisebox{-3mm}{$
{\scriptstyle k}$}}
d_{ijk}\lambda_k]~[~\frac{2}{3}~\delta_{ij} ~+~
\centerunder{$\hbox{\Large $\Sigma$}$}{\raisebox{-3mm}{$
{\scriptstyle l}$}}~d_{ijl}\lambda_l]~
\}\cr
&= ~ \centerunder{$\hbox{\Large $\Sigma$}$}{\raisebox{-3mm}{$
{\scriptstyle i,j}$}} ~[~\frac{8}{9}~\delta_{ij}~+\frac{4}{3} 
\centerunder{$\hbox{\Large $\Sigma$}$}{\raisebox{-3mm}{$
{\scriptstyle k}$}}~ d_{ijk}^2]~=~20\frac{4}{9} 
}
\auto\label{cf3}
$$
and the difference of the two gives 
$$
\eqalign{ \centerunder{$\hbox{\Large $\Sigma$}$}{\raisebox{-3mm}{$
{\scriptstyle i,j}$}}
~~Tr\{\lambda_j\lambda_i\lambda_i\lambda_j ~-~
\lambda_i\lambda_j\lambda_i\lambda_j\} 
&=  \centerunder{$\hbox{\Large $\Sigma$}$}{\raisebox{-3mm}{$
{\scriptstyle i,j,k,l}$}}
~~Tr\{[-2i~f_{ijk}\lambda_k]~\lambda_i\lambda_j\}\cr
&=  \centerunder{$\hbox{\Large $\Sigma$}$}{\raisebox{-3mm}{$
{\scriptstyle i,j,k}$}}
~~Tr\{[2i~f_{jik}\lambda_k]~[i~f_{ijl}\lambda_l]\}\cr
&=  \centerunder{$\hbox{\Large $\Sigma$}$}{\raisebox{-3mm}{$
{\scriptstyle i,j,k}$}} ~\frac{4}{3} f_{ijk}^2~=~12
}
\auto\label{cf4}
$$

Both (\ref{cf3}) and (\ref{cf4}) are expressed as a sum of squares of color
factors where each individual term corresponds to a particular color for
the gluon intermediate state. The states that contribute can be found by writing
the left-side color factor as a sum over distinct intermediate states, as
is effectively done in 
(\ref{lji+}) and (\ref{lji-}). The full color factor can then be written, 
relying on the orthogonality of the intermediate states, as a sum of squares of the
left-side factors.
The quark-antiquark intermediate state can
only carry zero or octet color. Correspondingly, since
the total color of the intermediate state is zero, the 
gluon sums must also contribute either zero or octet color. This will continue to
be the case when more gluons are present in the 
transverse momentum state. 
For the two gluon case, as illustrated in Fig.~28,
\begin{center}
\epsfxsize=4in
\epsffile{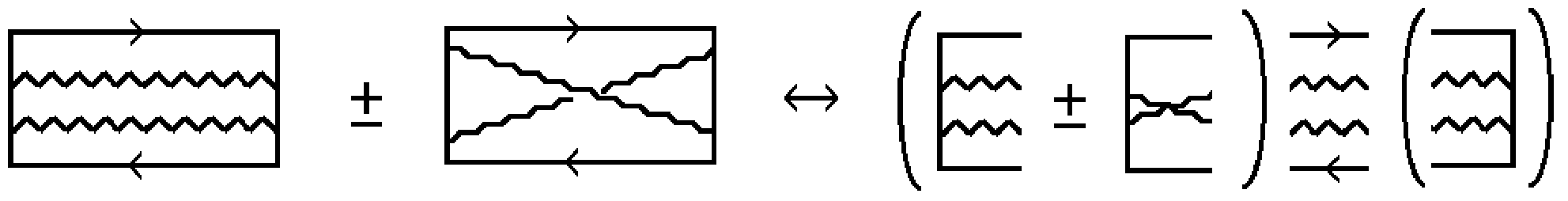}

\epsfxsize=3.5in
\epsffile{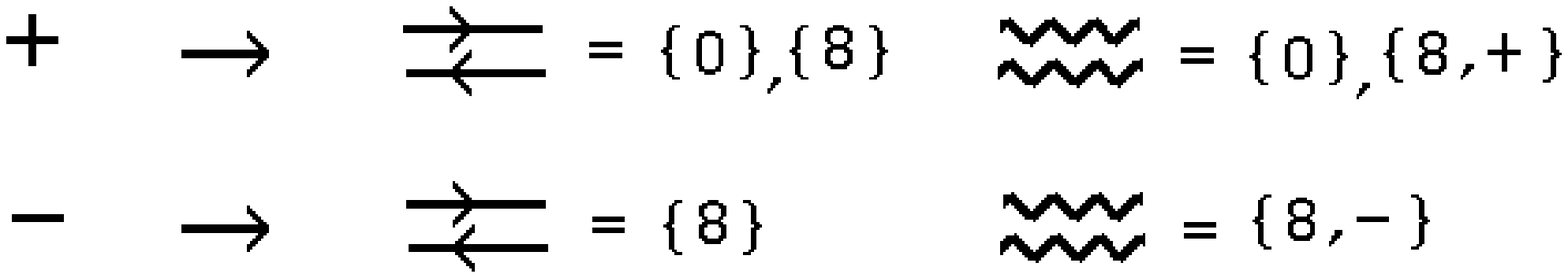}

Fig.~28 Color Breakdown of the Intermediate State
\end{center}
in the symmetric color 
factor (\ref{cf3}) the gluon sums give zero and
octet color contributions, which both have even color parity. In the antisymmetric
factor (\ref{cf4}) only the color octet with odd color parity contributes.

\subhead{6.6 The Color Factor for the Anomaly}

Consider again the effective triangle diagram of Fig.~22. 
(As in the last Section, we will initially discuss only the various
anomaly contributions to $G_L$, with $G_R$ kept fixed.)
Comparing with 
the diagrams of Fig.~3 and Fig.~12(a), the twisted triangle
diagram that should give an anomaly contribution to
add to that of Fig.~26  
is that shown in Fig.~29(a) and the corresponding full diagram is that of
Fig.~29(b). 
\begin{center}
\epsfxsize=3in
\epsffile{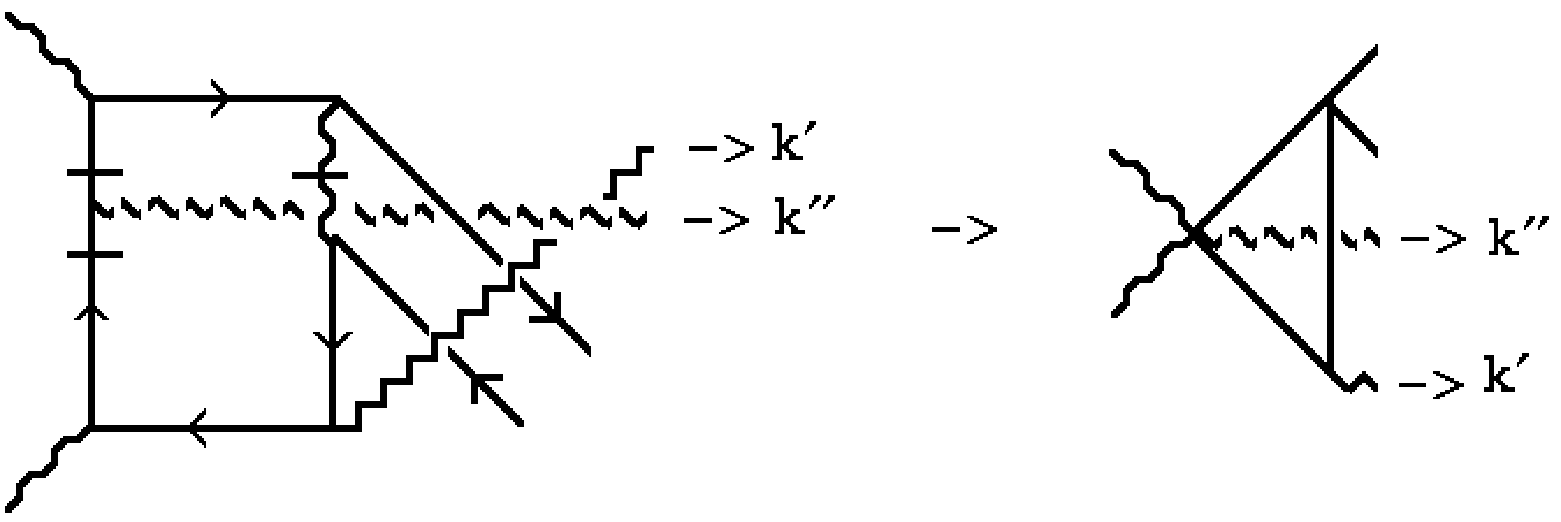}
\hspace{0.6in}
\epsfxsize=2in
\epsffile{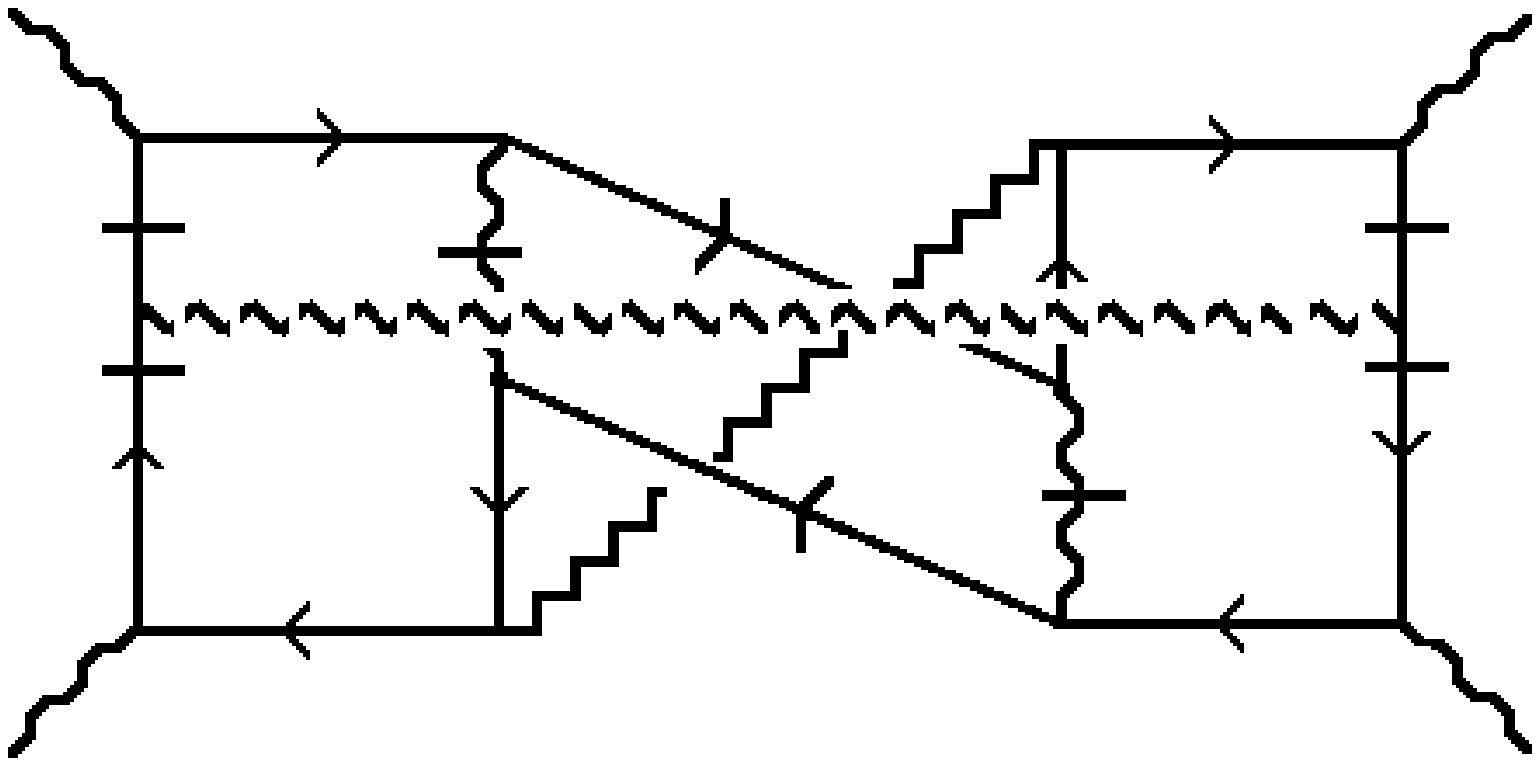}

(a)\hspace{3in}(b)

Fig.~29 (a) A Twisted Effective Diagram (b) The Full Twisted Diagram
\end{center}
As illustrated in Fig.~30
and in analogy with Fig.~13, Fig.~29(b) can be obtained 
by adding an exchanged gluon to the two cut amplitude of Fig.~30(a).
\begin{center}
\parbox{2.6in}{
\begin{center}
\epsfxsize=2in
\epsffile{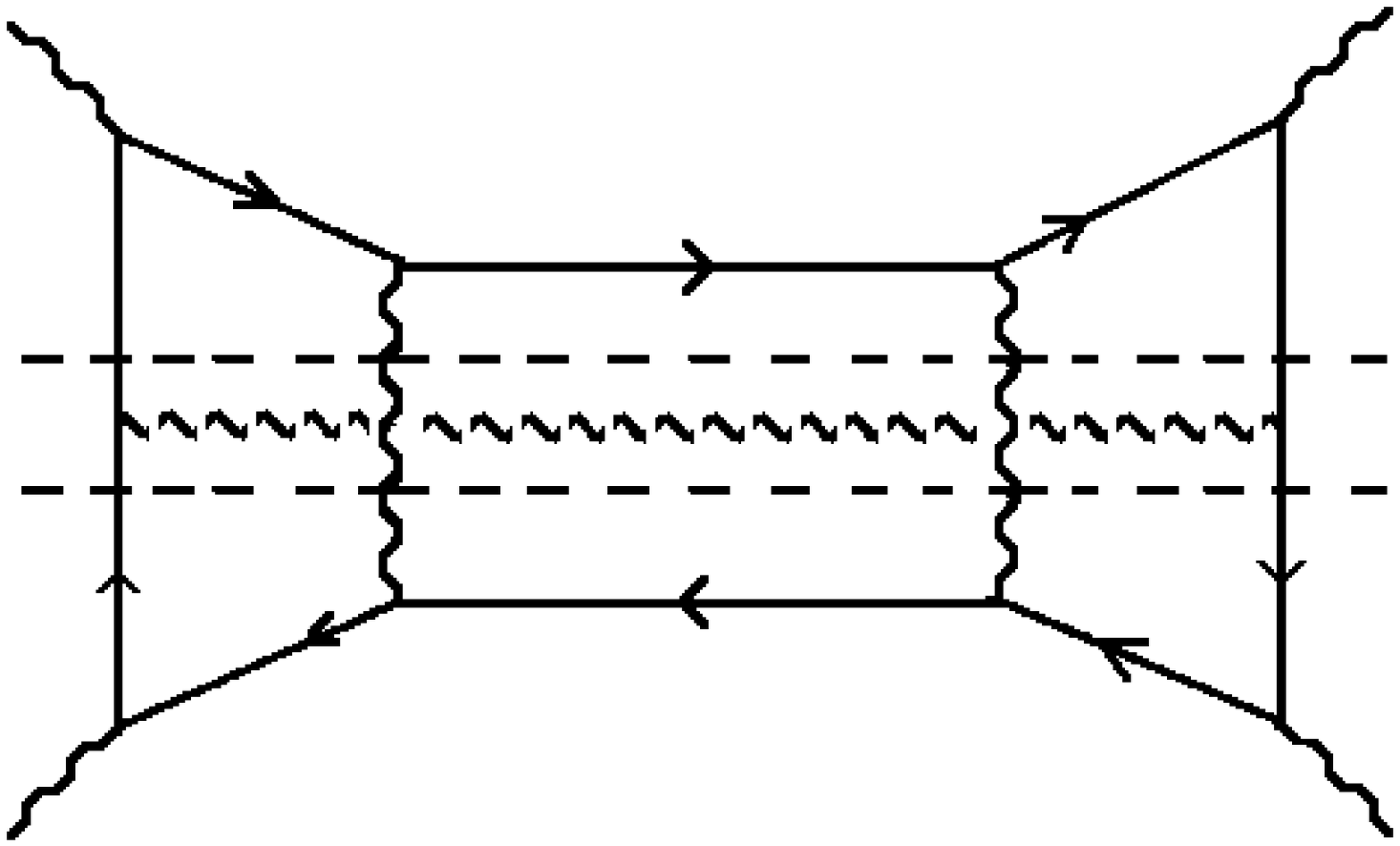}
\end{center}}
\parbox{0.6in}{$\to$}
\parbox{2.6in}{
\begin{center}
\epsfxsize=2in
\epsffile{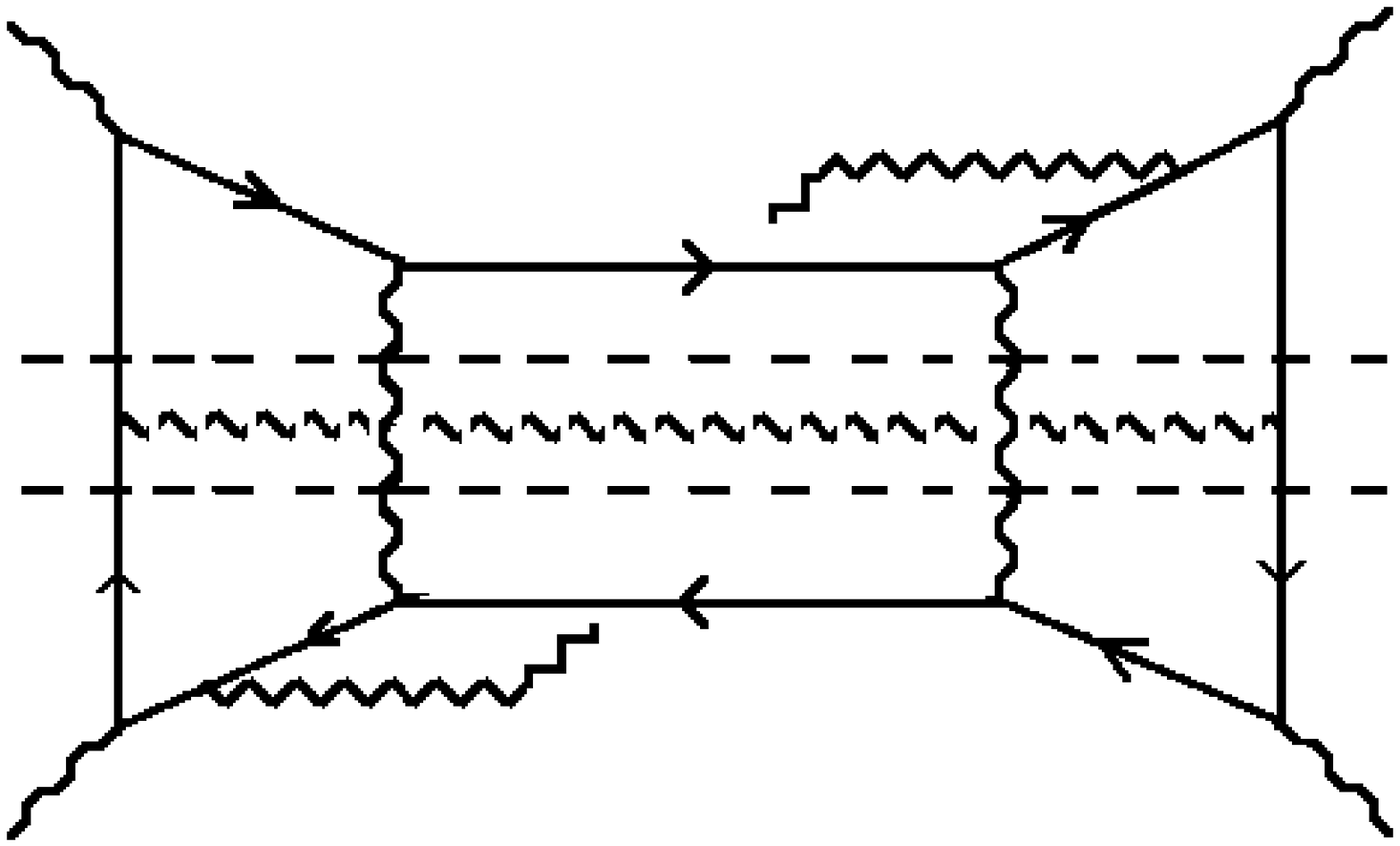}
\end{center}
}

(a)\hspace{3in}(b)

Fig.~30 Addition of an Exchanged Gluon to a Two Cut Amplitude
\end{center}
Again, in the anomaly region, with $|k'_{\perp}|^2 ~<<~
|p'_{\perp}|^2,~ |p''_{\perp}|^2$, the exchanged gluon does not interfere,
significantly, with the kinematics of taking the double discontinuity of
Fig.~30(a). It is also clear from  
Fig.~30, that the on-shell lines of Fig.~29(b)
correspond to physical double discontinuities of the $G_L$ and $G_R$.

If we consider the order of $\lambda$ matrix 
multiplication (following the quark arrow) we see that 
it is reversed for the two
gluons in Fig.~29(a) compared to those in Fig.~26. As a result,
if we add the two diagrams
the anomaly is multiplied by the sum of the two color factors (\ref{cf3}). 
As we have discussed, in this case 
there are separate contributions corresponding to whether
the quark-antiquark pair is
in a color octet or a color zero state. 

\subhead{6.7 Signature and Color Charge Parity}

To form signatured amplitudes we consider, with the diagrams of Fig.~27 
and Fig.~29(b), the corresponding diagrams for same helicity vector scattering.
These are shown in Figs.~31(a) and (b), respectively. With flavor included, 
the left-hand couplings in these diagrams should be replaced 
by the full, flavor symmetric, couplings of Fig.~22(b), while the right-hand
couplings should continue to be the analogue of Fig.~22(a).
As in Fig.~18, we display the diagrams in two different ways, each of which will
be simpler for particular arguments. 
\begin{center}
\epsfxsize=2.5in
\epsffile{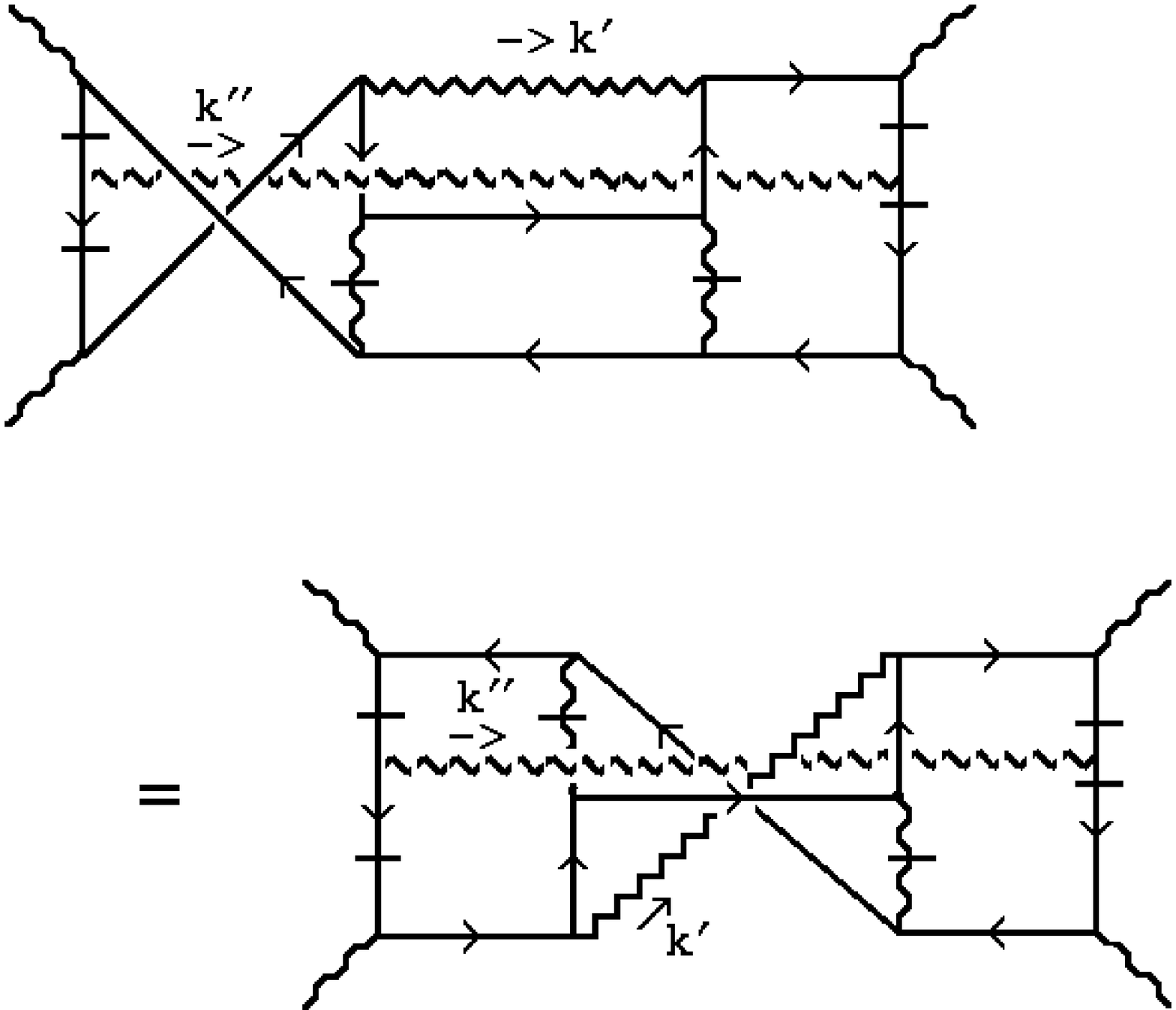}
\hspace{0.6in}
\epsfxsize=2.5in
\epsffile{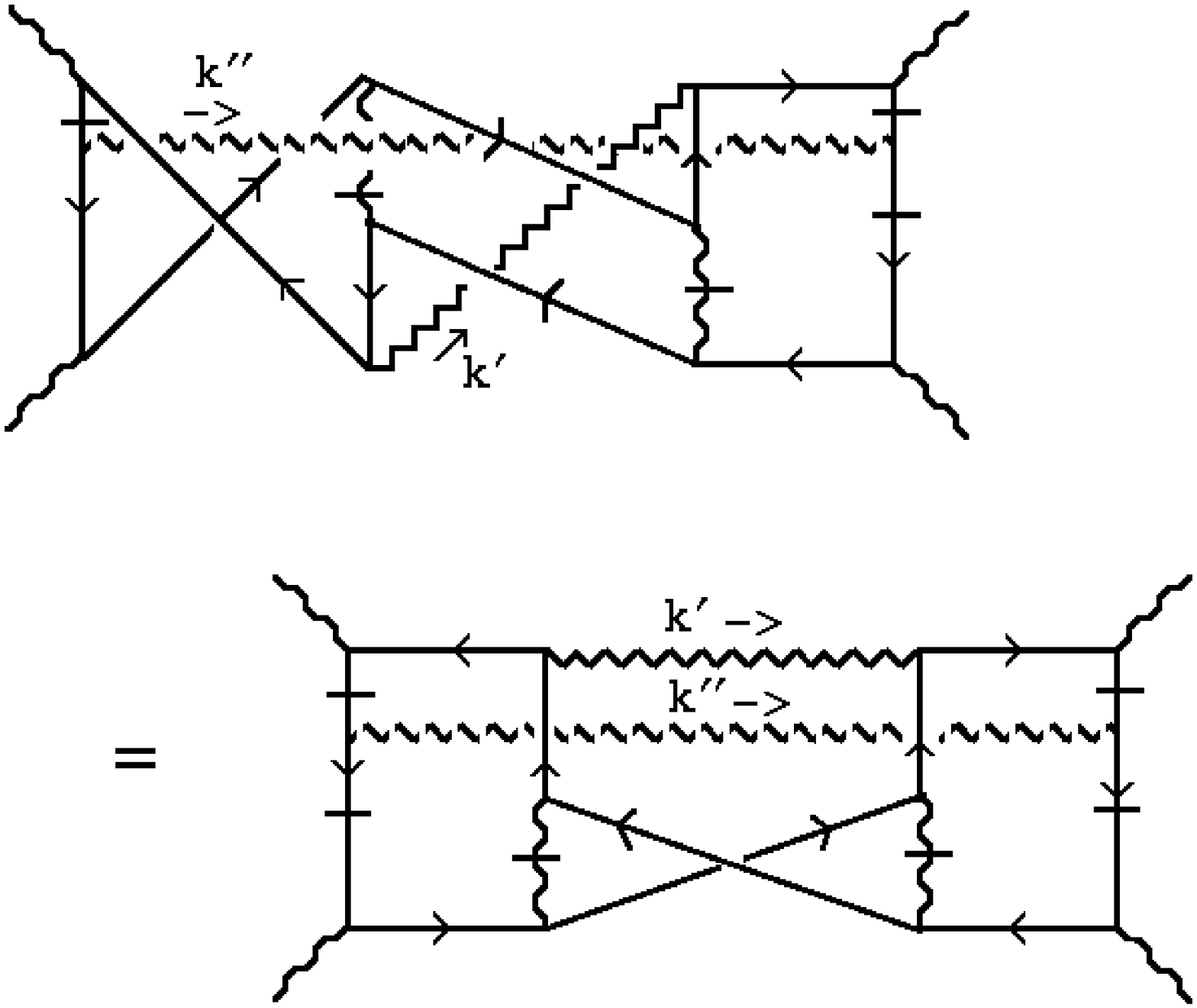}

(a)\hspace{3in}(b)

Fig.~31 Same Helicity Scattering Diagrams With One Soft Gluon
\end{center}

To discuss signature 
we focus on the first forms displayed.
Comparing with our discussion of the diagrams of 
Fig.~20, we note that Figs.~31(a)and(b) again
differ, kinematically, from Fig.~27 and Fig.~29(b) (respectively)
only in that $P_+ \to -P_+$ 
along the left-most vertical line. Therefore, if we 
evaluate the diagrams of Fig.~31 in the cross-channel physical
region - with the sign of $P_+$
reversed, they will be kinematically identical, respectively, to Figs.~27 and 29(b)
and will give identical anomaly contributions, but with $S \to -S$.
Therefore, these diagrams have anomaly 
contributions with the opposite sign to those of Figs.~26 and 29(b)
and in a vector theory would provide a cancelation.
In the present case, since the color factors are the same, the anomaly
contributions of the two diagrams of Fig.~31 add
to those  of Figs.~27 and 29(b) in the even signature amplitude and produce a 
cancelation in the odd signature amplitude.

As in our discussion of the one gluon diagrams, 
we can also obtain the signature from 
the $CP$ symmetry properties of the intermediate transverse momentum state.
We consider, first, properties of the gluon component.
Since the gluons have only QCD vertices, they can have simple 
transformation properties under $C$ and $P$ separately.
Applying color charge conjugation to (\ref{lji+}) gives
$$
\eqalign{ [\lambda_j\lambda_i ~+~ \lambda_i\lambda_j]
~&\to~ (-1)^2 [\lambda_j^*\lambda_i^* ~+~ \lambda_i^*\lambda_j^*] \cr
&=~ [\lambda_j\lambda_i ~+~ \lambda_i\lambda_j]^*
~=~\frac{2}{3} \delta_{ij}~
+~ \centerunder{$\hbox{\Large $\Sigma$}$}{\raisebox{-3mm}{$
{\scriptstyle k}$}}~d_{ijk}[\lambda_k]^*}
\auto\label{lji-c}
$$
The parity transformation reverses the transverse momentum of the gluons
and so, 
because of the coupling (\ref{bxi11}) for the large transverse momentum gluon, 
again gives a minus sign. Therefore, the full effect of the $CP$ 
transformation of the gluon component of the intermediate state is given by
$$
~\frac{2}{3} \delta_{ij}~
+~ \centerunder{$\hbox{\Large $\Sigma$}$}{\raisebox{-3mm}{$
{\scriptstyle k}$}}~d_{ijk}[\lambda_k]~~\to ~~
-~\frac{2}{3} \delta_{ij}~
-~ \centerunder{$\hbox{\Large $\Sigma$}$}{\raisebox{-3mm}{$
{\scriptstyle k}$}}~d_{ijk}[\lambda_k]^*
\auto\label{cpg}
$$
Comparing (\ref{lji-c}) and (\ref{cpg}) with (\ref{lji+}) we observe that
the color zero two gluon state carries  negative $CP$ and is 
separately even under $C$ and odd under $P$.
It therefore has ``normal'' color charge parity (equal to the number of gluons
in the state) but has ``anomalous'' negative parity, producing an ``anomalous''
negative signature. The color parity of the octet gluon state is also 
well-defined if we ignore $[\lambda_k] \to [\lambda_k]^*$ (which is, of course,
compensated for by the quark/antiquark interchange discussed below) 
and is similarly even.

To discuss the $CP$ transformation of the quark/antiquark pair
we compare with our discussion in sub-section {\bf 3.15} and note, first, that
the left-side coupling of Fig.~31(b) can be obtained, diagrammatically, 
from that of Fig.~27 by quark/antiquark interchange. 
The left-side coupling of Fig.~31(a) can similarly be obtained 
from that of Fig.~29(b). In the even signature
amplitude we must evaluate the diagrams of 
Fig.~31 with $P_+ \to -P_+$ compared to the other diagrams. 
Therefore, quark/antiquark interchange now gives three kinematic changes of sign
$$
\eqalign{& (i) ~~~~~~~~ k_{\perp} \to -k_{\perp} \cr
&(ii)~~~~~ ~~ k_-' \to - k_-' ~~~\hbox{from}~~~ P_+ \to -P_+ \cr
&(iii) ~ ~~~~~k_-'' \to - k_-''~~~ \hbox{from}~~~ P_+ \to -P_+}
\auto\label{sc2}
$$
When all four diagrams are
added the amplitude is, kinematically,  
antisymmetric under quark/antiquark interchange. As a result,
when the quark/antiquark pair has color zero it is, straightforwardly,
negative under $CP$.
Combined with the negative $CP$ of the color zero two gluon state this gives
no change under the full
$CP$ transformation, as is necessary to obtain an even signature amplitude.

When octet color is involved, the color
effect of interchanging the quark-antiquark pair will, as we already noted
above, again be
$[\lambda_h]~ \to~ [\lambda_h]^*$. Therefore, 
with the negative sign coming from the kinematic interchange, the
complete $CP$ transformation on the quark-antiquark pair again 
combines with the octet part of (\ref{cpg}) to produce an overall positive 
$CP$ result for the full two gluon quark-antiquark state.  

\subhead{6.8 Infra-Red Cancelations}

Since (\ref{amp2}) contains an infra-red divergence (at
${k_{\perp}''}^2 \sim 0$), and there is no external color, 
there must be other anomaly contributions that
cancel this divergence. Before discussing the possible diagrams
that could be involved, it will be useful to first discuss the lower limit on the
${k''_{\perp}}^2$ integration in Fig.~27.

The momentum flow through the two lines that are put on-shell by the 
$k_-'$ and $k_-''$ integrations is shown in Fig.~32.
\begin{center}
\epsfxsize=2.5in
\epsffile{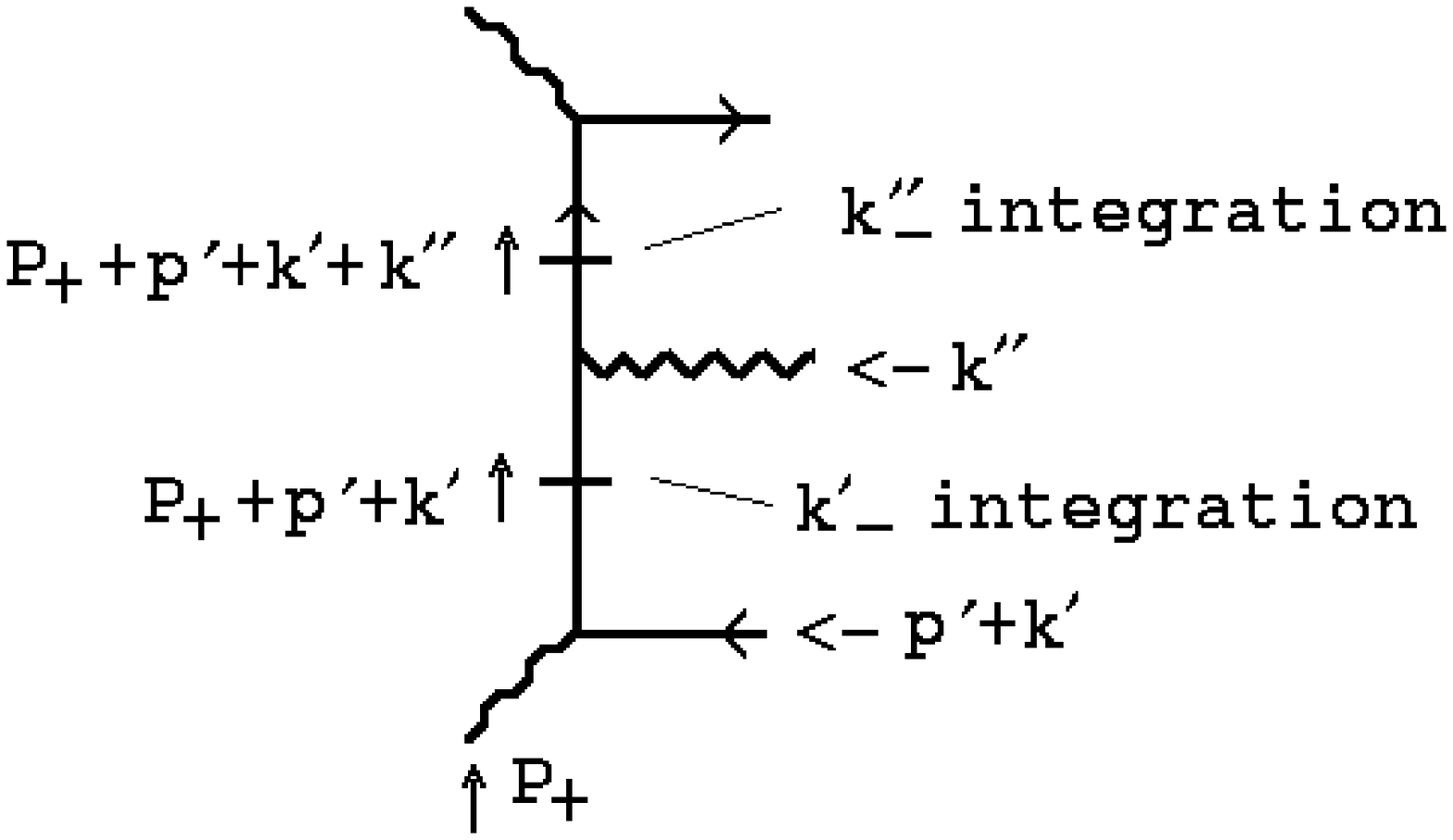}

Fig.~32 Momentum Flow for the $k_-'$ and $k_-''$ Integrations
\end{center}
Since $k'_{\perp}$ is large and $k''$ is small, 
all the large momenta flow through both lines. Therefore, the large momenta are
significantly constrained by the $k'$ mass-shell condition before the 
$k''$ mass-shell condition is imposed. If we, temporarily, introduce a quark 
mass $m$ then 
the $k_-'$ mass-shell condition is $(P_+ + p'+k')^2 =m^2$ and, with this 
constraint, the mass-shell condition for $k_-''$ becomes
$$
\eqalign{ k_-''~&\sim~ \frac{- 2k_{\perp}''(p'+k')_{\perp} + m^2}{P_+} \cr
&\centerunder{$\sim$}{\raisebox{-4mm}{${\scriptstyle k_{\perp}'' \to 0}$}}
~~~~\frac{m^2}{S^{\frac{1}{2}}}}
\auto\label{llmt}
$$
If $k_+''$ is similarly constrained, and (to justify
the reduction to a transverse momentum diagram) we require
${k_{\perp}''}^2 
~\centerunder{$>$}{\raisebox{-2mm}{$\sim $}}~|k_-''k_+''|$, 
the lower limit for the $k_{\perp}''$ - integration is 
$$
|k_{\perp}''|^2 ~~\sim ~~\frac{m^4}{ S}
\auto\label{llmt1}
$$
and so an infra-red divergence appears at $|k_{\perp}''|^2=0$,
as $S\to \infty$, of the form
$$
~\int_{|k_{\perp}''|^2\sim \frac{m^4}{S}} ~\frac{d^2k_{\perp}''}{
{k_{\perp}''}^2}~~\sim ~~ln~S ~-~ln~ m^4
\auto\label{ird1}
$$

As we discuss briefly in the next Section, we expect
the cancelation of the infra-red divergence (\ref{ird1}) to be a consequence 
of a Ward identity that results from attaching the soft gluon at every possible
point around the effective triangle diagram.
Consider, therefore, the attachment of the soft gluon, in $G_L$, according to the
other possibilities illustrated in Fig.~25. With the attachment
shown in Fig.~25(b) we obtain the full diagrams shown in Fig.~33.
\begin{center}
\epsfxsize=2in
\epsffile{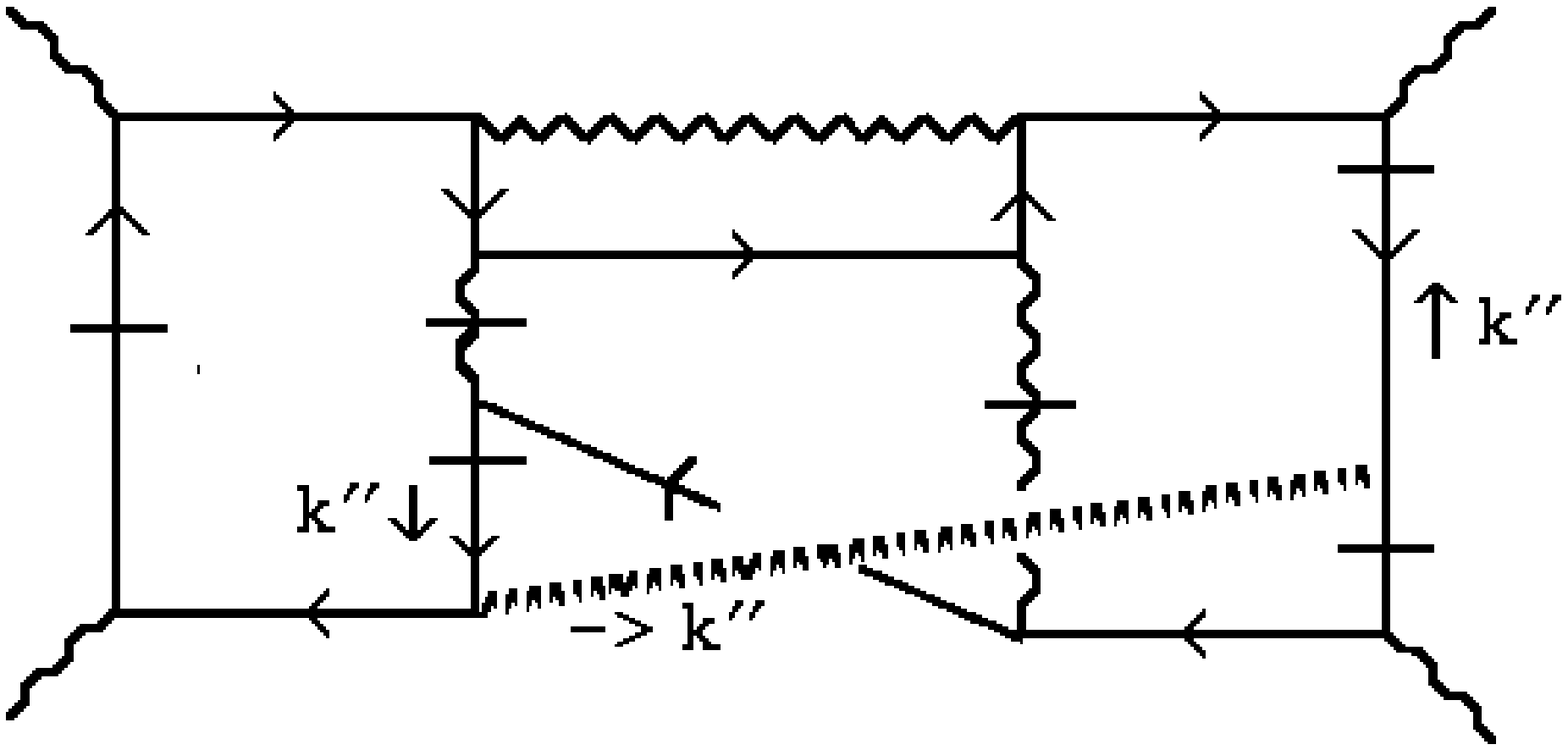}
\hspace{0.8in}
\epsfxsize=2in
\epsffile{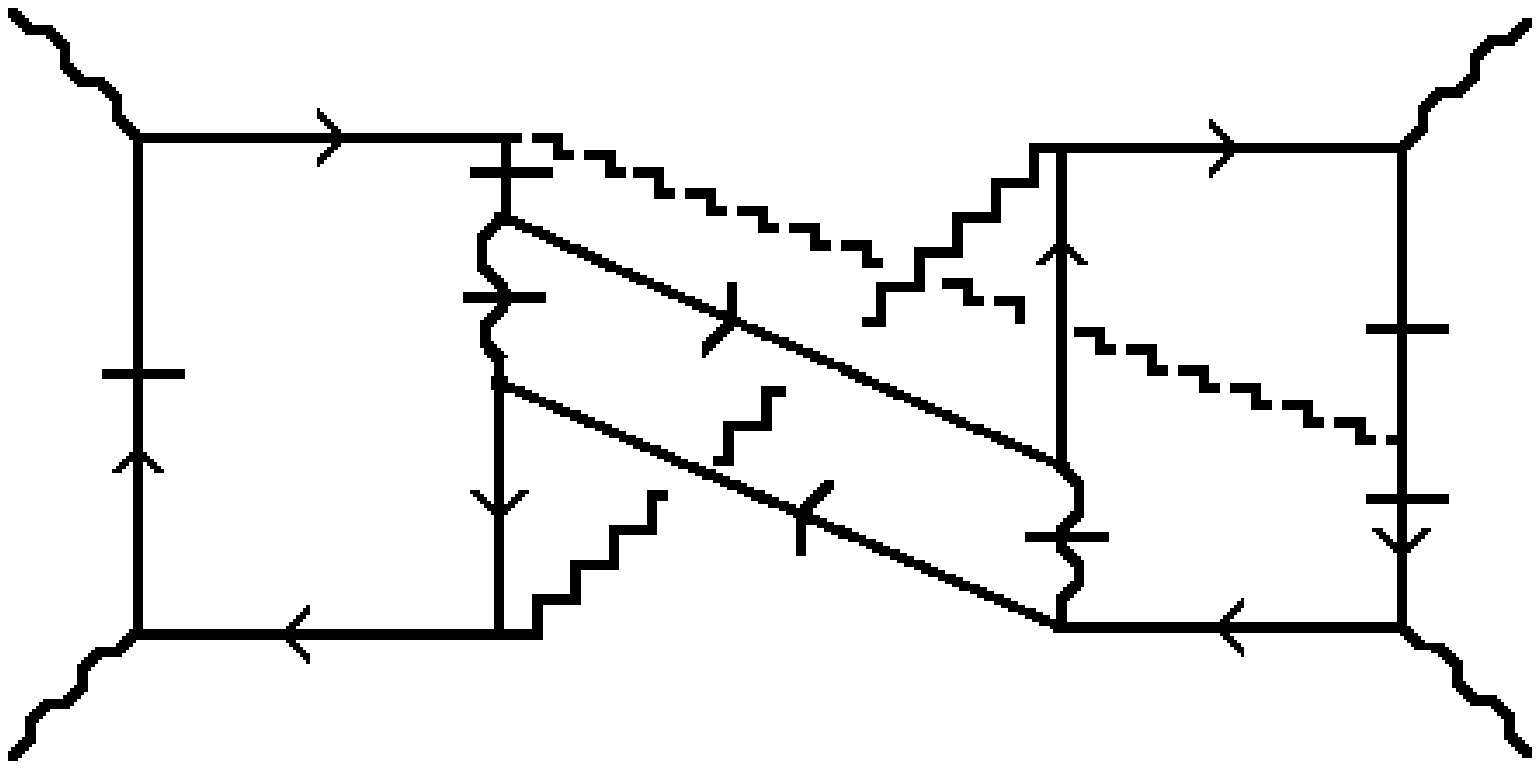}

(a)\hspace{2.8in}(b)

Fig.~33 Diagrams Obtained with the Soft Gluon Attachment of Fig.~25(b) 
\end{center}
The mass-shell condition now has the form (\ref{k''os}) and so,
if the lower limit for the $k_+'$ integration still has
the form (\ref{llmt}), requiring
$|k_{\perp}'|^2 > k_+'k_-'$ gives the lower limit
$$
|k_{\perp}'|^2 ~~\sim ~~\frac{m^2~ M}{S^{\frac{1}{2}}}
\auto\label{llmt3}
$$
Therefore, if the diagrams of Fig.~33 give anomaly contributions they will have,
as a factor, an infra-red divergence of the form
$$
~ \int_{|k_{\perp}''|^2\sim m^2M /S^{\frac{1}{2}} }
~\frac{d^2k_{\perp}''}{{k_{\perp}''}^2}~~
\sim ~~~\frac{1}{2}~ln~S ~-~ln~ m^2~-~ln~M
\auto\label{ird2}
$$
If, instead, one attachment of the soft gluon is that 
of Fig.~25(c) rather than that of Fig.~25(b), as we have just discussed, 
there will clearly be a similar infra-red divergence.

There is, however, a major reason why the diagrams of Fig.~33, ultimately,
do not give an anomaly contribution. Because of the location
of the soft gluon attachment, the sign change $(iii)$ in
(\ref{sc2}) will not be present when $P_+ \to -P_+$ in the diagrams appropriately 
related to those of Fig.~33 in the same sign helicity amplitude. As a result,
in the even signature anomaly amplitude,
the quark-antiquark intermediate state must be positive under $CP$.
Alternatively, if we add all the diagrams related to
those of Fig.~33 that give all possible contributions 
to each $G_R$, we will obtain $G_R$ couplings to the
quark-antiquark intermediate state that have negative $CP$. Therefore,
when all diagrams related to those of Fig.~33 are added in the even signature
amplitude, the resulting $G_L$ requires positive $CP$ for the quark-antiquark
state, while the $G_R$ requires negative $CP$ 
Consequently, if there is an anomaly
contribution from any of the diagrams, it must cancel in the sum.

This last problem similarly applies 
to all diagrams in which one soft gluon attachment is as 
in Fig.~25(c) while the other attachment is that of Fig.~25(a).
As we will shortly discuss there will, nevertheless, 
be important anomaly contributions from
diagrams in which the attachments of the second gluon, in both $G_L$ and $G_R$,
are either of the form of Fig.~25(b) or Fig.~25(c). In this case, however, 
the second gluon necessarily has ``finite'' transverse momentum and so
can not produce an infra-red divergence.

To look further for a divergence that could cancel that due to Fig.~27,
we must consider
whether there are any new kinematic configurations, generating the anomaly
and involving a soft gluon, 
that can not be viewed as a soft gluon accompanying the one gluon enhancement
diagrams. In fact, if we are considering the 
attachment of the soft gluon at every possible
point around the effective triangle diagram, there is
one possibility that we have not
yet included or discussed. This is to 
interchange the momenta of the two gluons involved in Fig.~26.  
In Fig.~34 we show the full diagram obtained from Fig.~27 by 
interchanging $k'$ and $k''$ in one side of the diagram relative to the other.
\begin{center}
\epsfxsize=2.5in
\epsffile{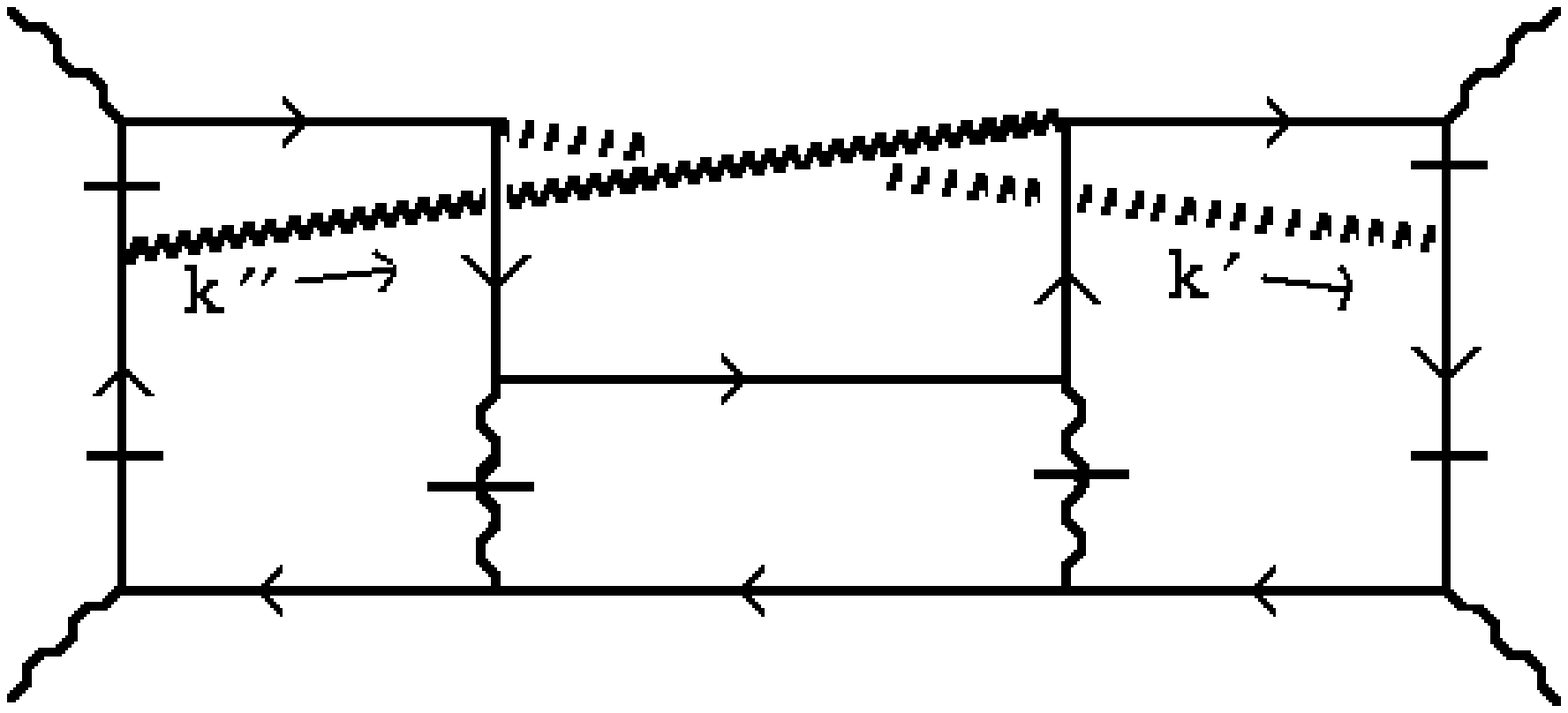}

Fig.~34 The Diagrams Obtained by Interchanging $k'$ and $k''$ in Fig.~27
\end{center}
The soft gluon is again indicated by a broken line. (Note that Fig.~34 is 
symmetric with respect to $k'$ and $k''$, if we interchange the roles
of $G_L$ and $G_R$.)

If we interchange $k'$ and $k''$ in Fig.~26, then  
the large transverse momentum flows into the triangle diagram at the left-most 
vertex while the single gluon vertex carries only small transverse momentum.
This does indeed give an anomaly contribution. This contribution has, however,
some different properties compared to those we have so far discussed.
In Fig.~35 we compare the
large momentum route for $k'$ with the possible routes for $k''$, 
around the same triangle.
\begin{center}
\epsfxsize=4in
\epsffile{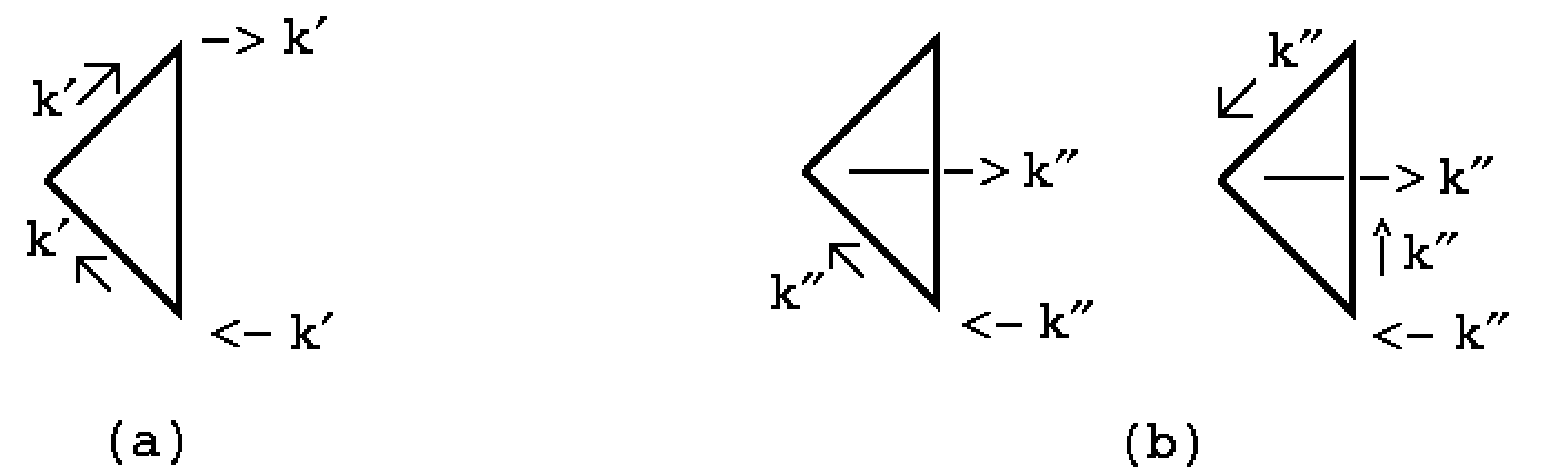}

Fig.~35 (a) Momentum Route for $k_{\perp}'$   (b) 
Possible Routes for $k_{\perp}''$  
\end{center}
We observe, first, that while the route for $k'$ large was determined by 
the particles we wished to put on shell, there are two possible
routes if $k''_{\perp}$ is large. As illustrated, it could flow through either 
one or two quark propagators. From (\ref{bxi6}) and (\ref{bxi7}), we see that 
the anomaly contribution is obtained from an expansion in powers of
$|k''_{\perp}/p'_{\perp}|$ (in which the first term does not contribute). 
Consequently, 
if $k''_{\perp}$ appears in only one propagator, rather than two,
the anomaly contribution 
will be reduced by a factor of $1/2$ (in the even signature amplitude where 
both chiralities are added for each propagator). The sign will also be opposite.
This is the normal ambiguity
of the ultra-violet anomaly, that occurs because of the choice of momentum
routing, which we expect to be determined by a Ward identity.

If $k'$ is small, the $k''$ mass-shell condition 
does not constrain the large transverse momentum ($p_{\perp}'$)
involved in the $k_-'$ mass-shell condition. As a result, the $k_-'$ mass-shell 
condition gives a constraint similar to (\ref{k''os}), i.e.
$$
k_-'~\sim ~ \frac{{p'}^2}{P_+} ~~\sim~~\frac{\epsilon~ M ~S^{\frac{1}{2}}}
{S^{\frac{1}{2}}}~~\sim ~~\epsilon~M
\auto\label{llmt2}
$$
Since the $k_+'$ integration
in $G_R$ has the lower limit (\ref{llmt}),
we will again obtain an infra-red divergence of the form of 
(\ref{ird2}). In this case, however, since there is a gluon attached
to both the left and right side quark lines, there is no
$CP$ conflict. Also, from Fig.~36, it is clear that 
neither gluon interferes with the kinematics
of the quark sub-amplitude, which remains such that the vector meson lines can 
consistently be placed on-shell.
\begin{center}
\epsfxsize=2.2in
\epsffile{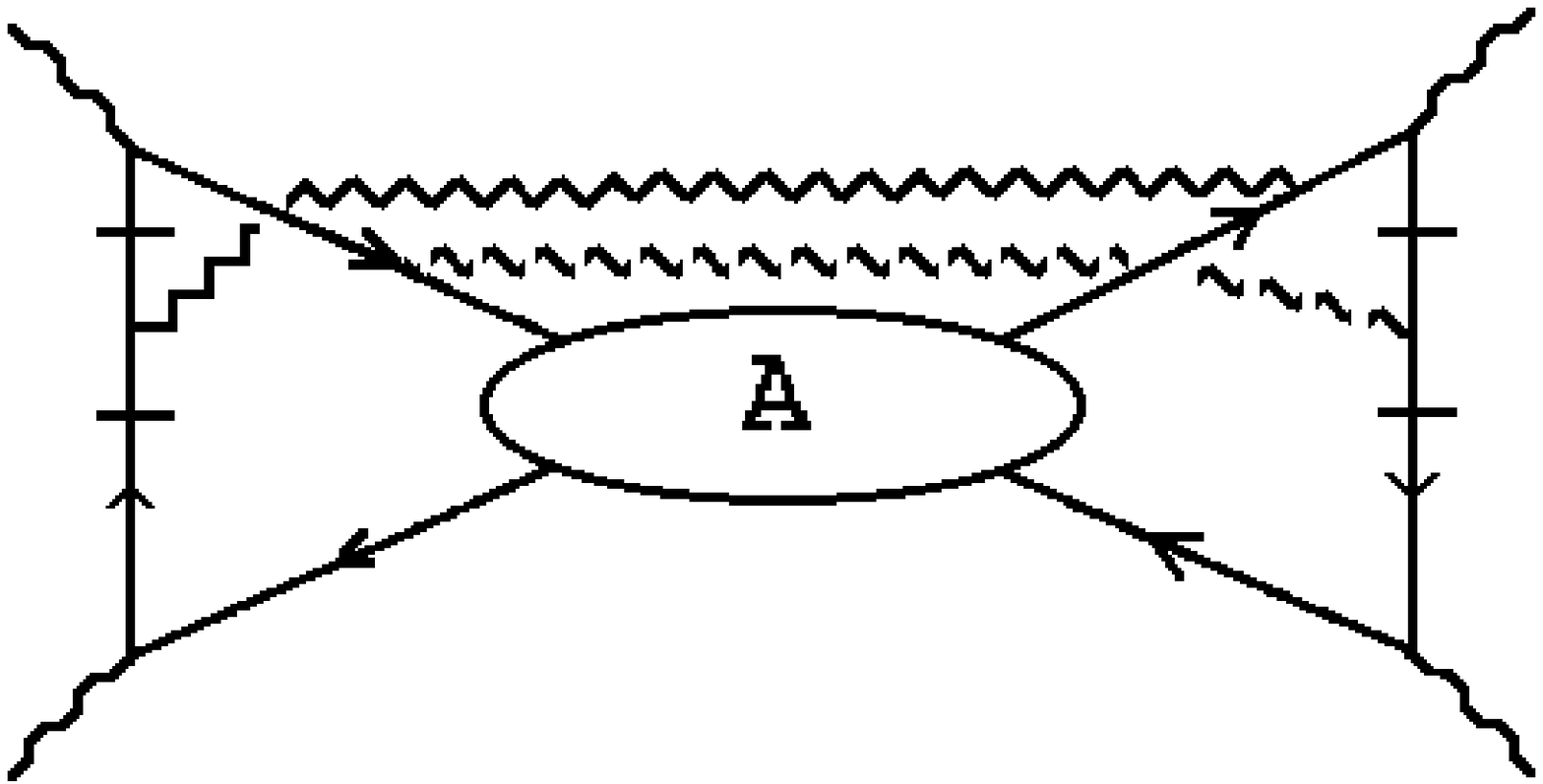}

Fig.~36 Two Gluons Accompanying the Quark Sub-Amplitude
\end{center}
In fact, if the anomaly has the sign and magnitude
obtained from the second routing of Fig.~35(b) (which is the ``normal''
routing), there will be a cancelation
with the divergence obtained from Fig.~27 when the diagram of Fig.~34 is
added and the two contributions, from $k''$ large and $k'$ small, and
$k'$ large and $k''$ small, are combined.

At this point we note that 
$k'$ is the total momentum flowing in at a ``vector'' vertex of
an effective triangle diagram with an anomaly. In this case, as we discuss in
the next Section (and illustrate for the usual triangle anomaly in Appendix C),
we expect that the appropriate Ward identity, which would determine which routing
in Fig.~35(b) is correct, will involve both the large internal
momentum region of the triangle diagram that generates the anomaly and a 
small internal momentum region that produces a very different kinematic
form, containing the ``anomaly pole''. 
The discussion of the anomaly pole and pion wee gluon couplings
that we have given in \cite{arw02} should, essentially,
carry over to an infra-red analysis of effective triangle vertices 
that would be the analogue of the ultra-violet analysis presented in this
paper. This analysis must be carried out before we can
establish, in detail, how the (reggeon) Ward identities are satisfied,
and that infra-red divergences are indeed eliminated when the diagram 
of Fig.~34(a) is added to that of Fig.~27.

\subhead{6.9 Anomalous Color Parity Gluons}

In the final form of two gluon anomaly contributions that we consider 
the second gluon transverse momentum $k''_{\perp}$ is
neither very small, nor grows with $S$. It is ``finite'', i.e. $O(M^2)$.
In the diagrams we consider, the above discussion implies that
the kinematics of the anomaly prevent the
second gluon from carrying very small transverse momentum.
In particular, we consider diagrams of the form 
shown in Fig.~37, in which the second gluon attachment is the same  
in both $G_L$ and $G_R$, and has the form of either Fig.~25(b) or Fig.~25(c). 
\begin{center}
\epsfxsize=1.5in
\epsffile{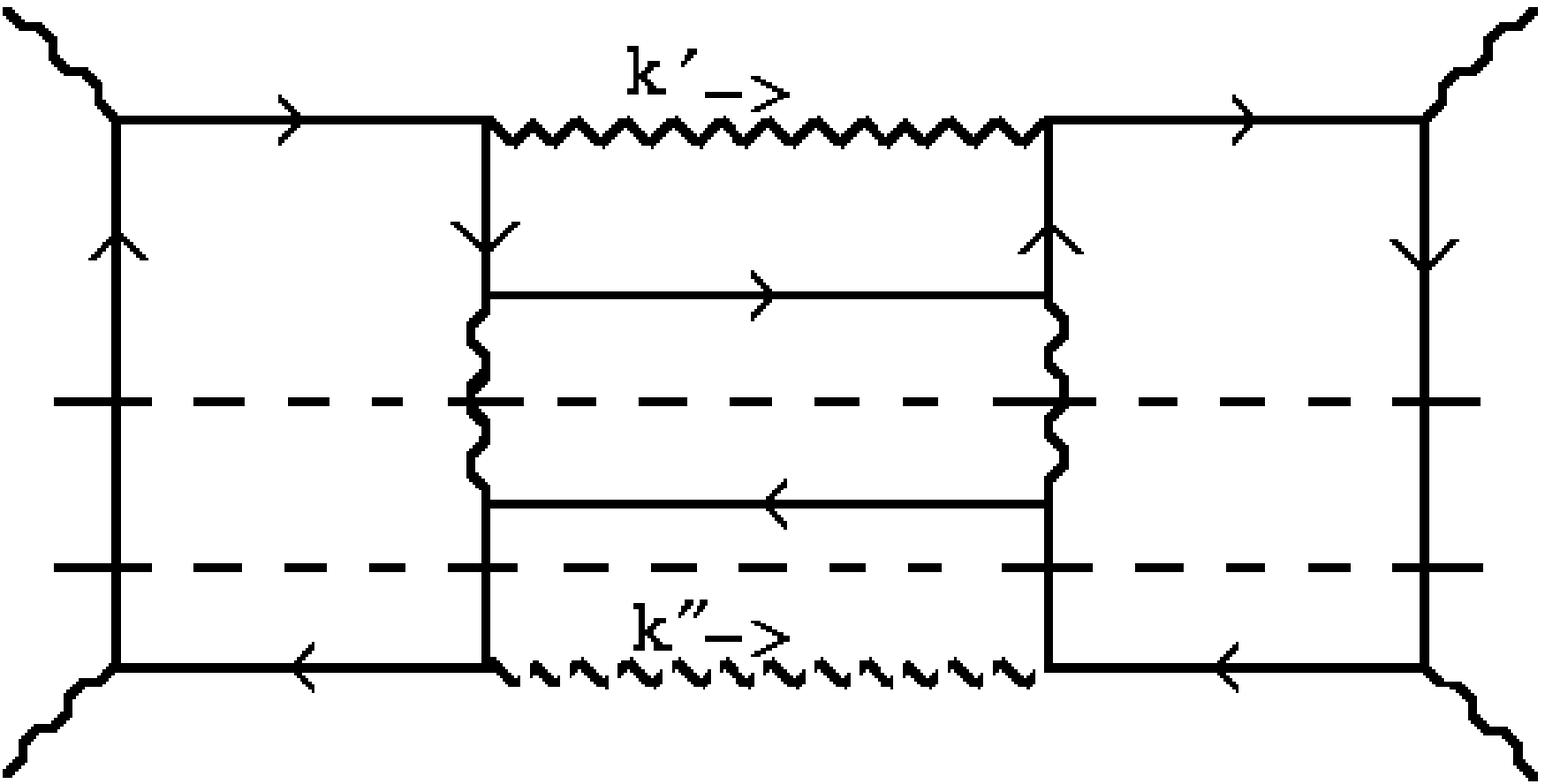}
\hspace{0.3in}
\epsfxsize=1.5in
\epsffile{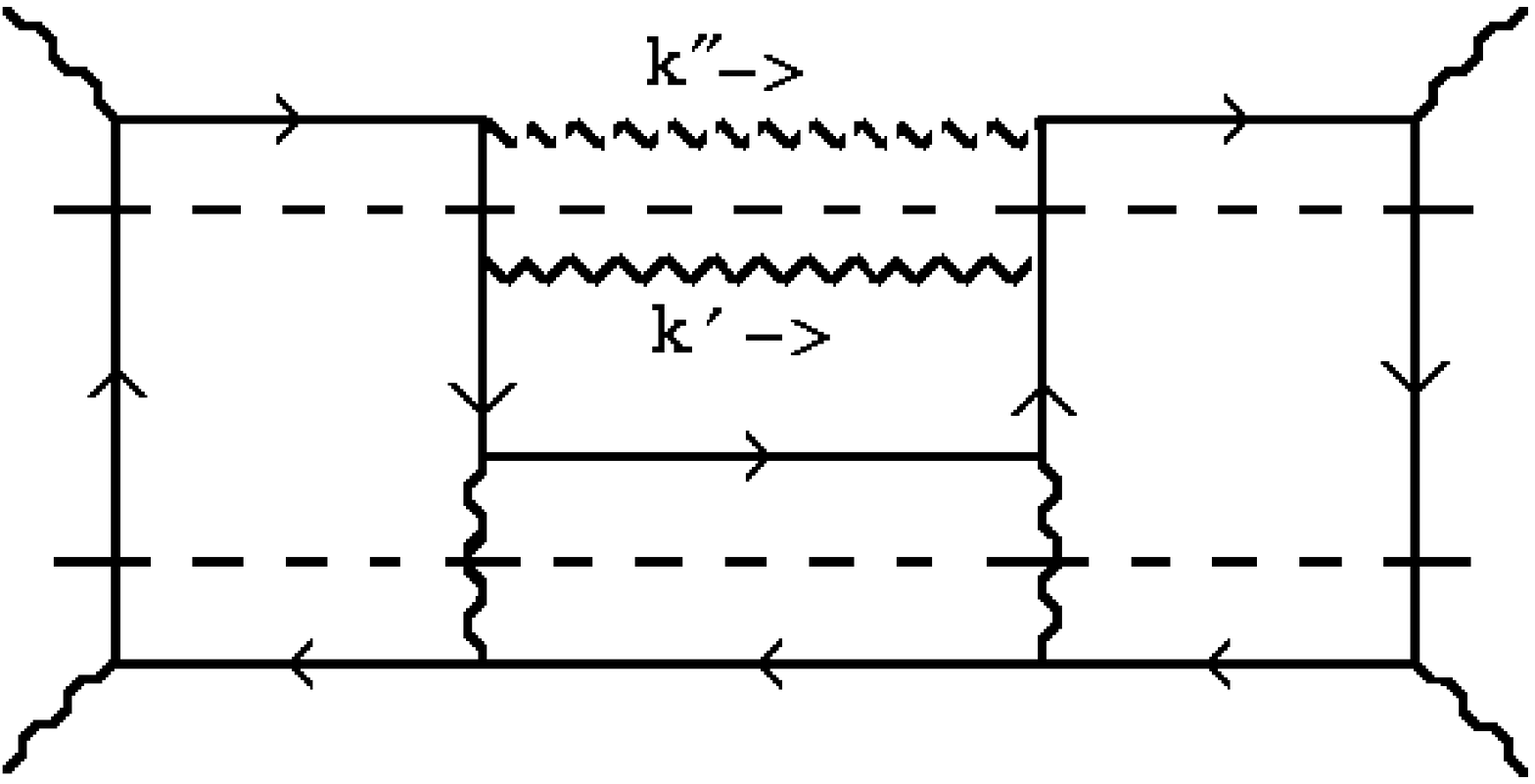}
\hspace{0.3in}
\epsfxsize=1.5in
\epsffile{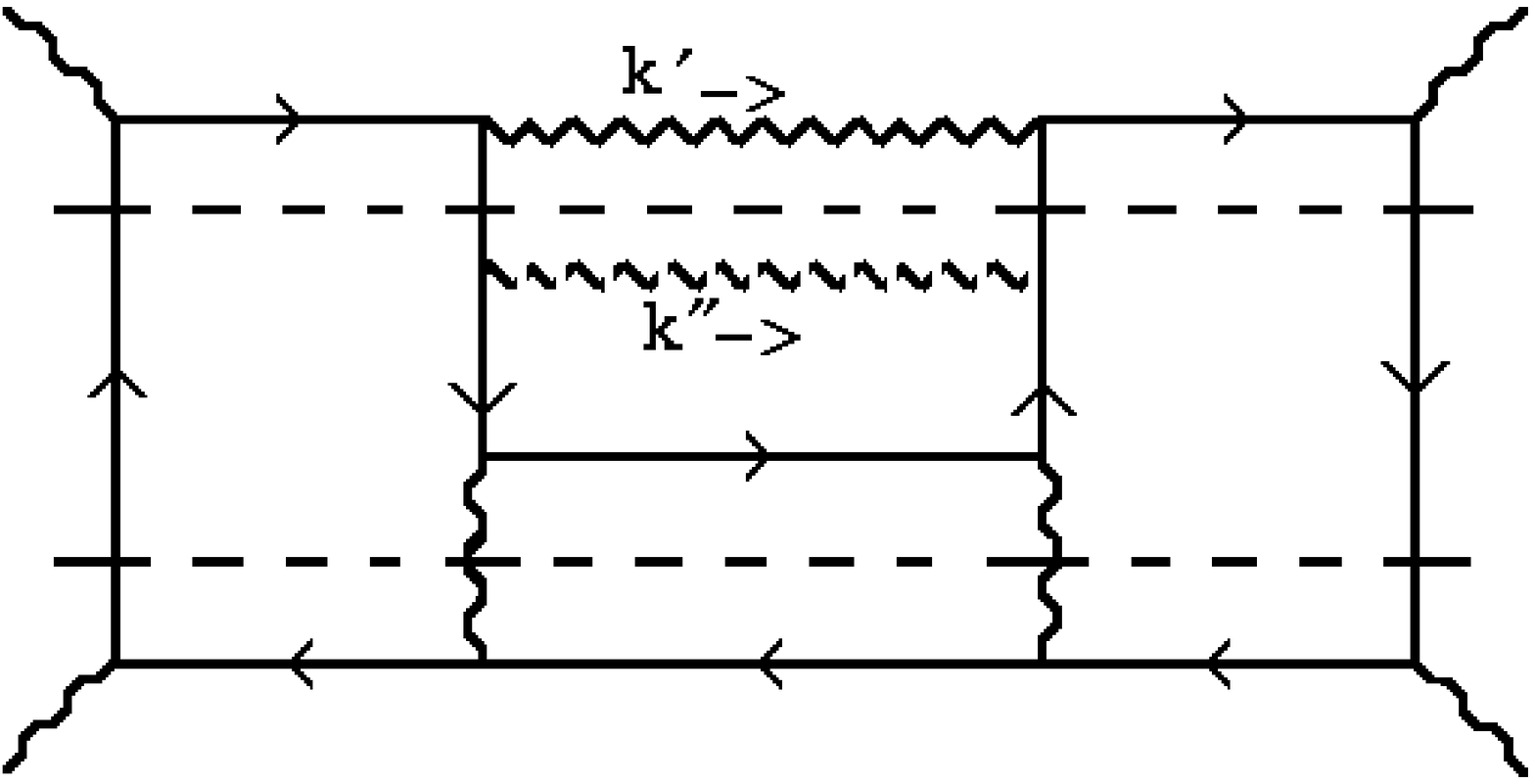}

(a)\hspace{1.8in}(b)\hspace{1.8in}(c)

Fig.~37 Gluon Attachments (a) - as in Fig.~25(b), or (b) and (c) - as in Fig.~25(c)

\end{center}
The broken gluon
line in these diagrams now indicates finite transverse momentum
and the lines put on-shell correspond, as illustrated, to two cuts through
each diagram. The combination of particle poles giving these two discontinuities,
together with the off-shell loop, that occurs either at the top, in the middle,
or at the bottom, in the three diagrams, is now responsible for the anomaly.
(As usual, a closely related 
set of diagrams is obtained by twisting simultaneously both the $G_L$ and 
the $G_R$ in each of these diagrams. For Fig.~37(a) the same diagram is
actually obtained, but the kinematic regions for the two gluons are 
interchanged.)

To form signatured amplitudes, as before, we consider the same helicity
scattering amplitudes shown in Fig.~38.
\begin{center}
\epsfxsize=1.5in
\epsffile{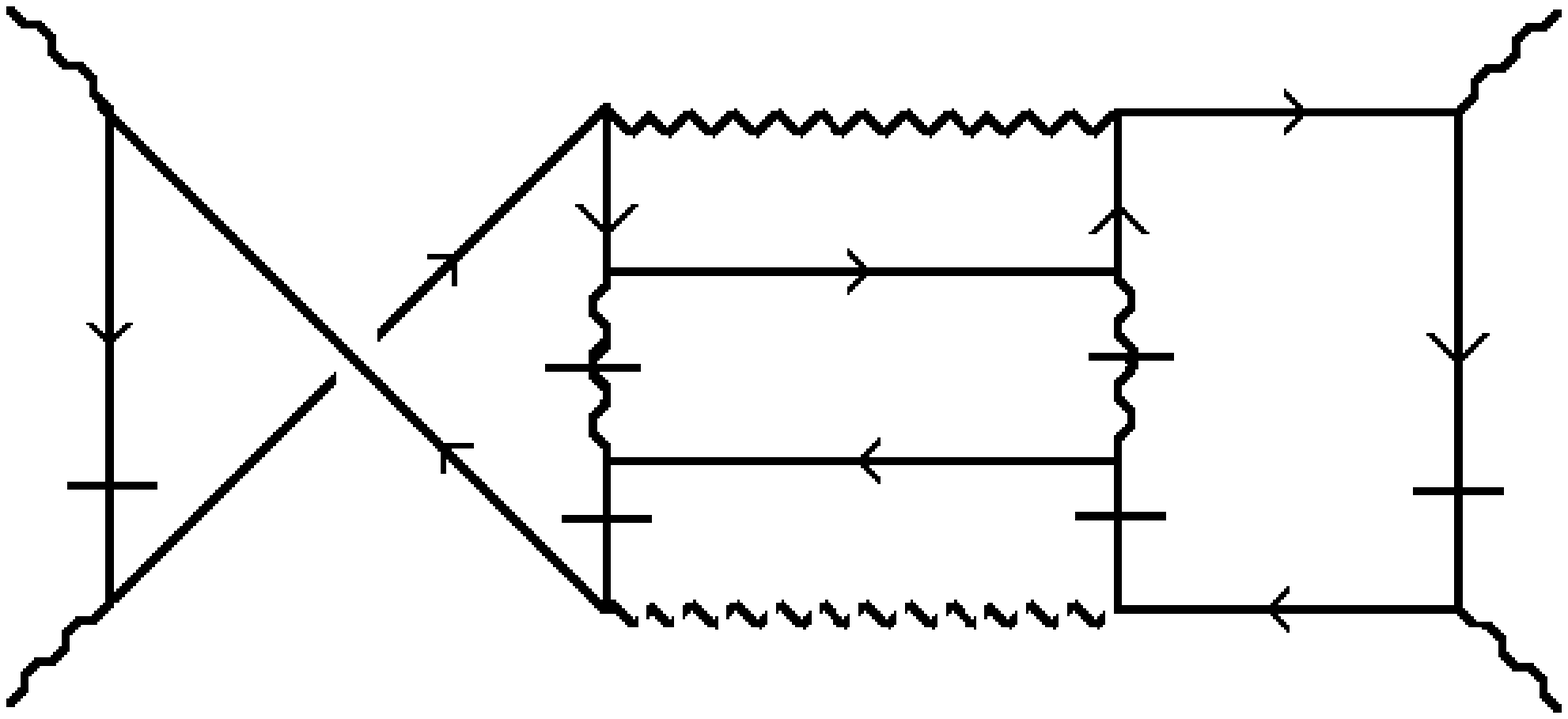}
\hspace{0.3in}
\epsfxsize=1.5in
\epsffile{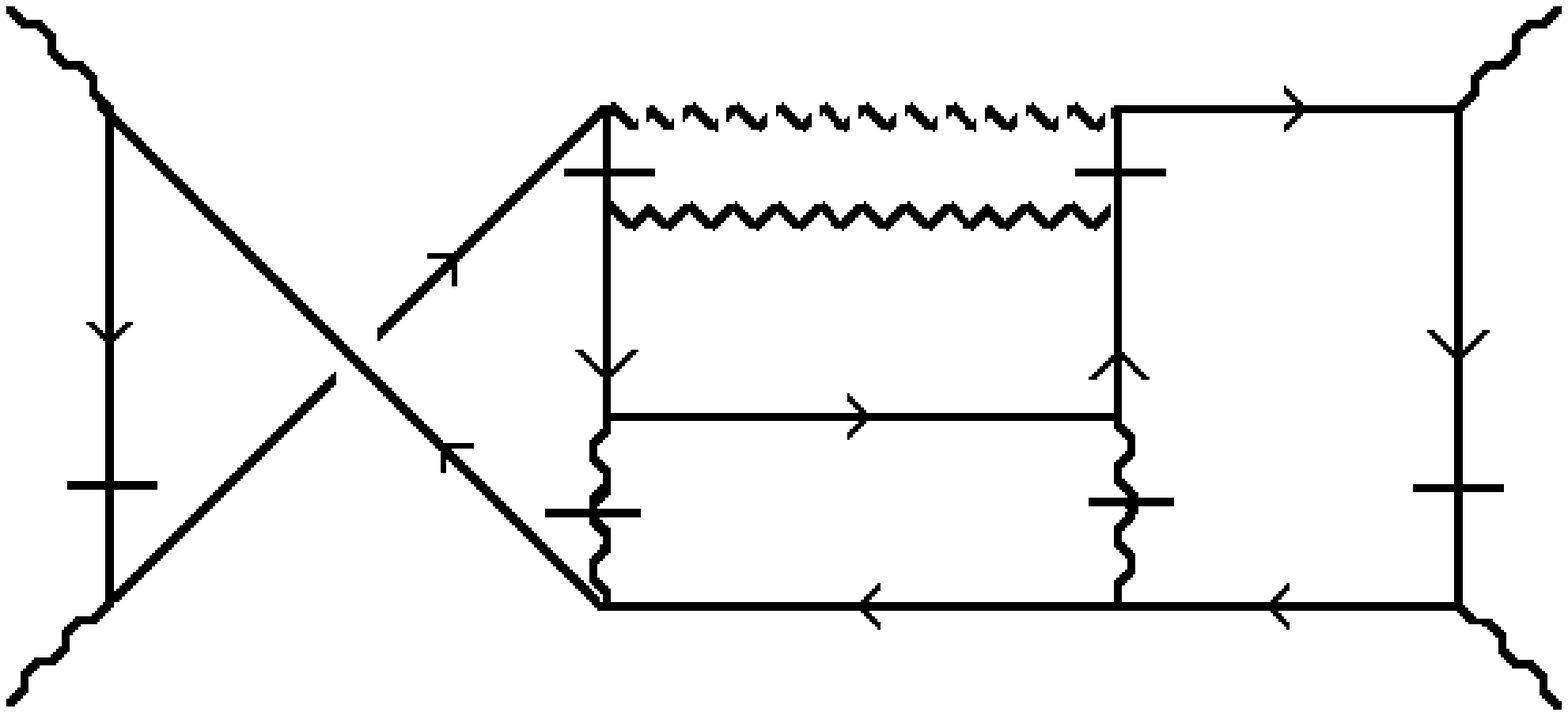}
\hspace{0.3in}
\epsfxsize=1.5in
\epsffile{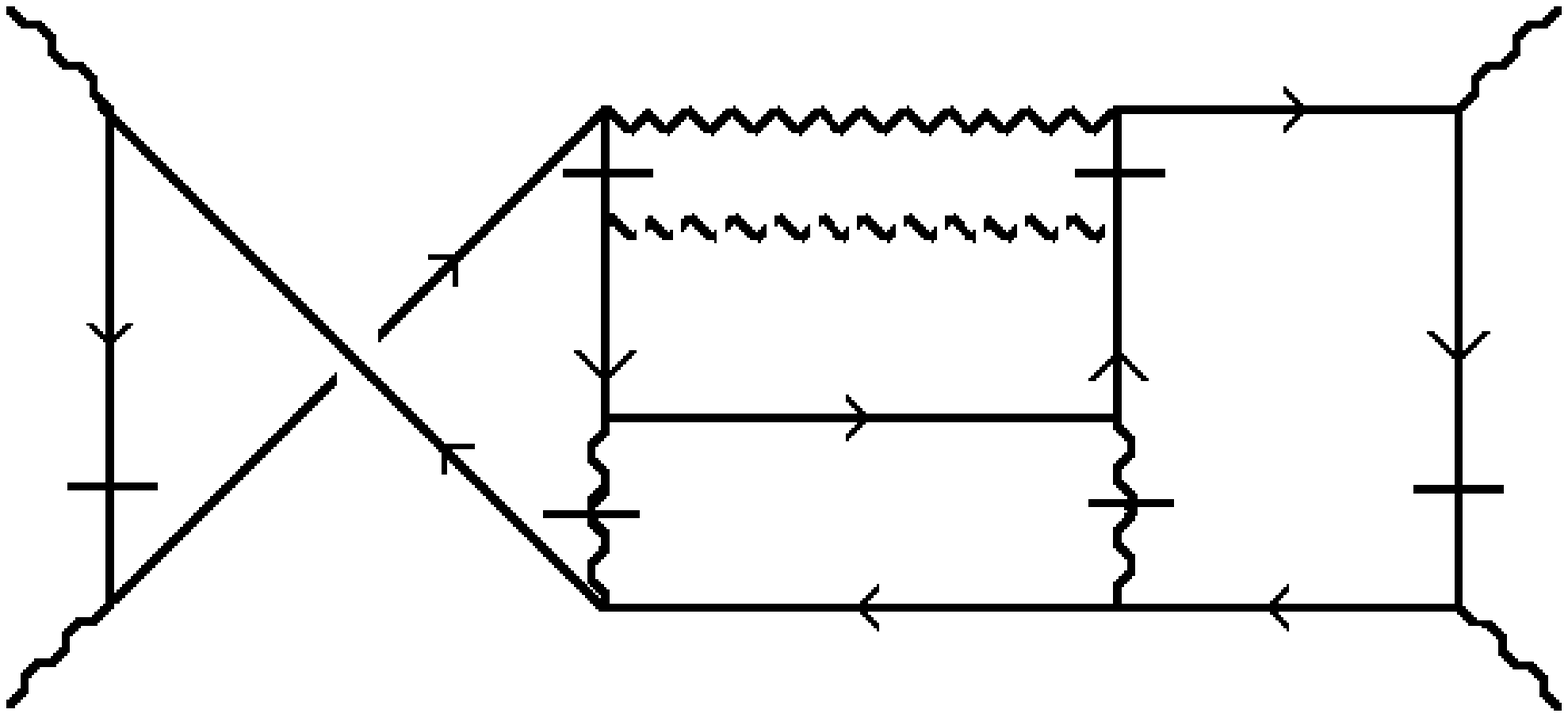}

(a)\hspace{1.8in}(b)\hspace{1.8in}(c)

Fig.~38 Same Helicity Diagrams Related to the Diagrams of Fig.~37 
\end{center}
Again we can argue that these diagrams are related to those of Fig.~37 by
$P_+ \to - P_+$ and so the same anomaly amplitude is obtained, but with
$S \to -S$. Therefore, these diagrams give anomaly contributions
that add to those of Fig.~37 in the even signature amplitude.
However, as in our
discussion of the diagrams of Fig.~23, because the finite gluon 
is not attached to the fast left side quark, the sign change $(iii)$ in
(\ref{sc2}) will not be present when $P_+ \to -P_+$ in the diagrams 
of Fig.~38. Consequently, the $G_L$ coupling obtained by adding the diagrams 
of Fig.~38 to those of Fig.~37, requires even $CP$ for the quark-antiquark
intermediate state. When the corresponding diagrams are added, 
this argument will similarly apply to $G_R$. As a result, the quark-antiquark
state is necessarily positive under $CP$. This can be
consistent with even signature for the complete intermediate state
only if the two gluon state is also
even under $CP$. This can, in turn, only be the case if the two gluon state 
carries antisymmetric octet color and so has,
``anomalous'', odd color charge parity (not equal to the number of gluons).

In the two gluon anomaly contributions that we have discussed in previous
sub-sections, the two gluon state has carried normal color charge parity
because the addition of twisted diagrams gave the symmetric color factor.
We must consider, therefore, whether there are also ``twisted diagrams''
related to those of Fig.~37 which cancel the antisymmetric part of the color
factor. 

Twisted diagrams related to the diagrams of Fig.~37 are shown in Fig.~39.
For Figs.~39(a) and (b), the color factor is indeed reversed compared,
relatively, to Figs.~37(a) and (b). Also, it is clear that,
in Figs.~39(a) and (b), the appropriate (hatched) lines can be consistently
placed on shell. (This is not true for other diagrams that could, potentially, 
be related, by twisting, to either of Fig.37(a) or (b).)
Therefore, as before, adding the diagrams of Figs.~39(a) and (b) to 
those of Figs.~37(a) and (b) gives the
symmetric color factor and so, because the quark-antiquark state has even $CP$, 
the anomaly contribution must cancel in the sums of these diagrams.
\begin{center}
\epsfxsize=1.5in
\epsffile{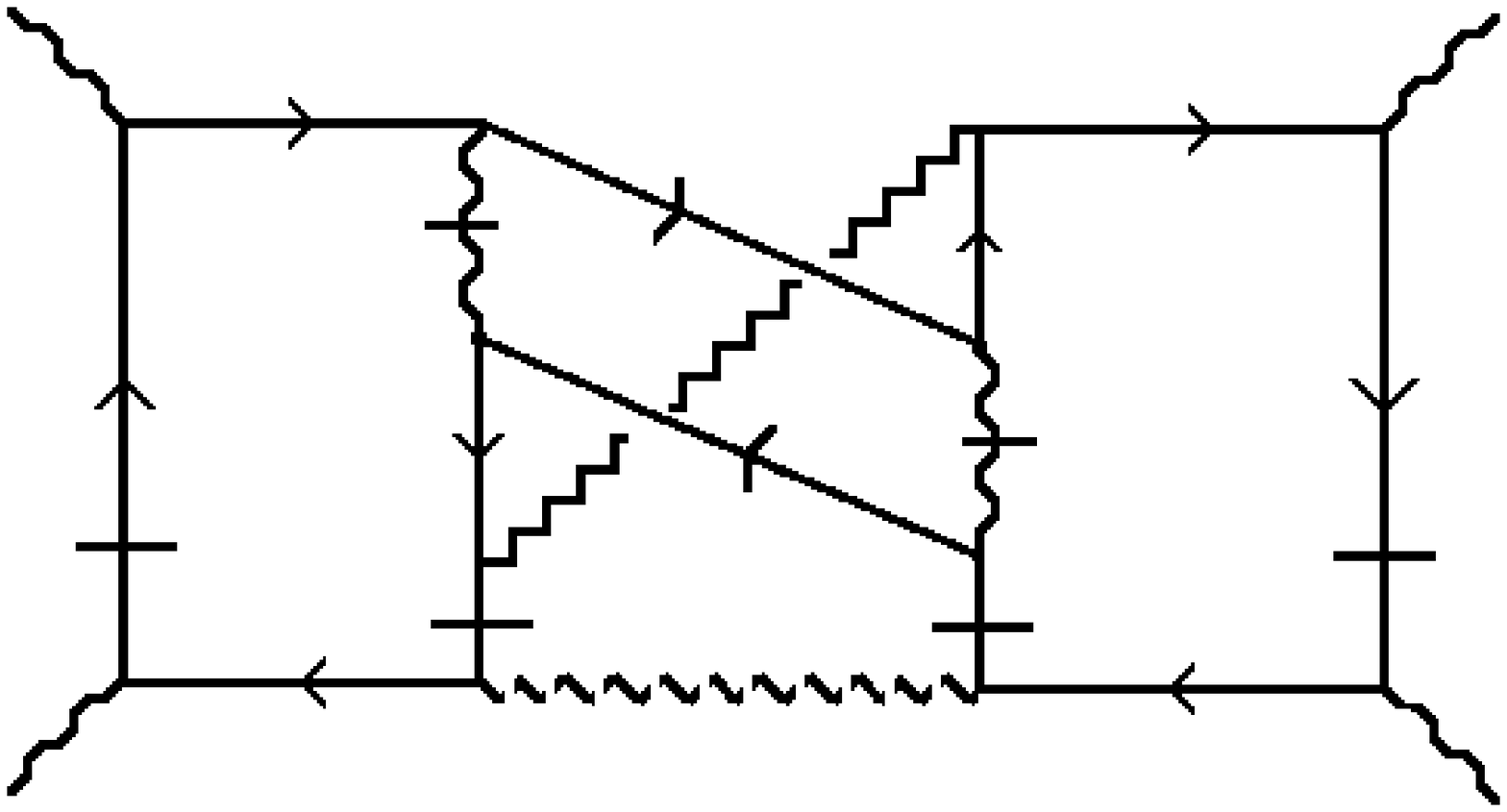}
\hspace{0.5in}
\epsfxsize=1.5in
\epsffile{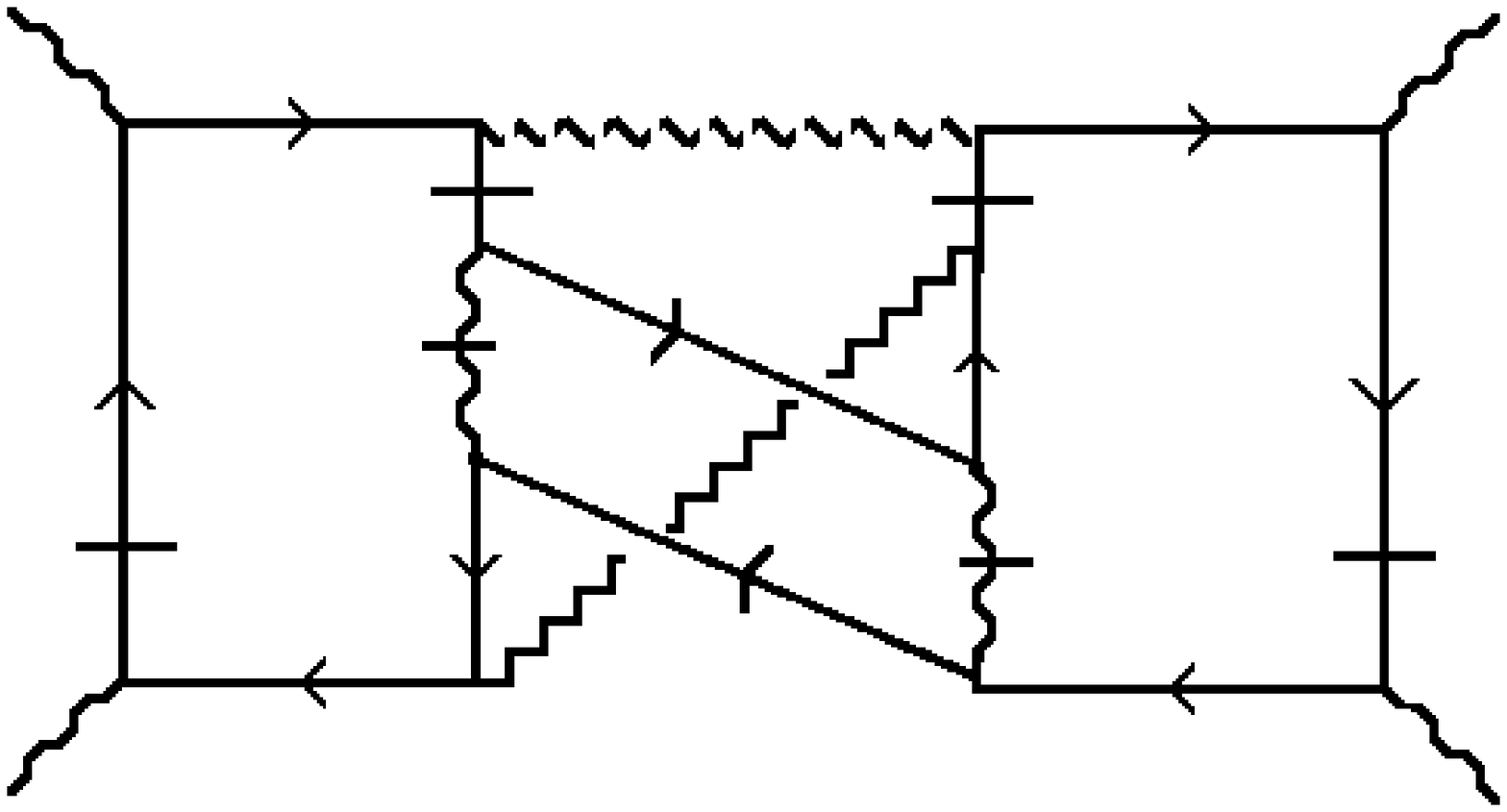}

(a)\hspace{2in}(b)

\epsfxsize=1.4in
\epsffile{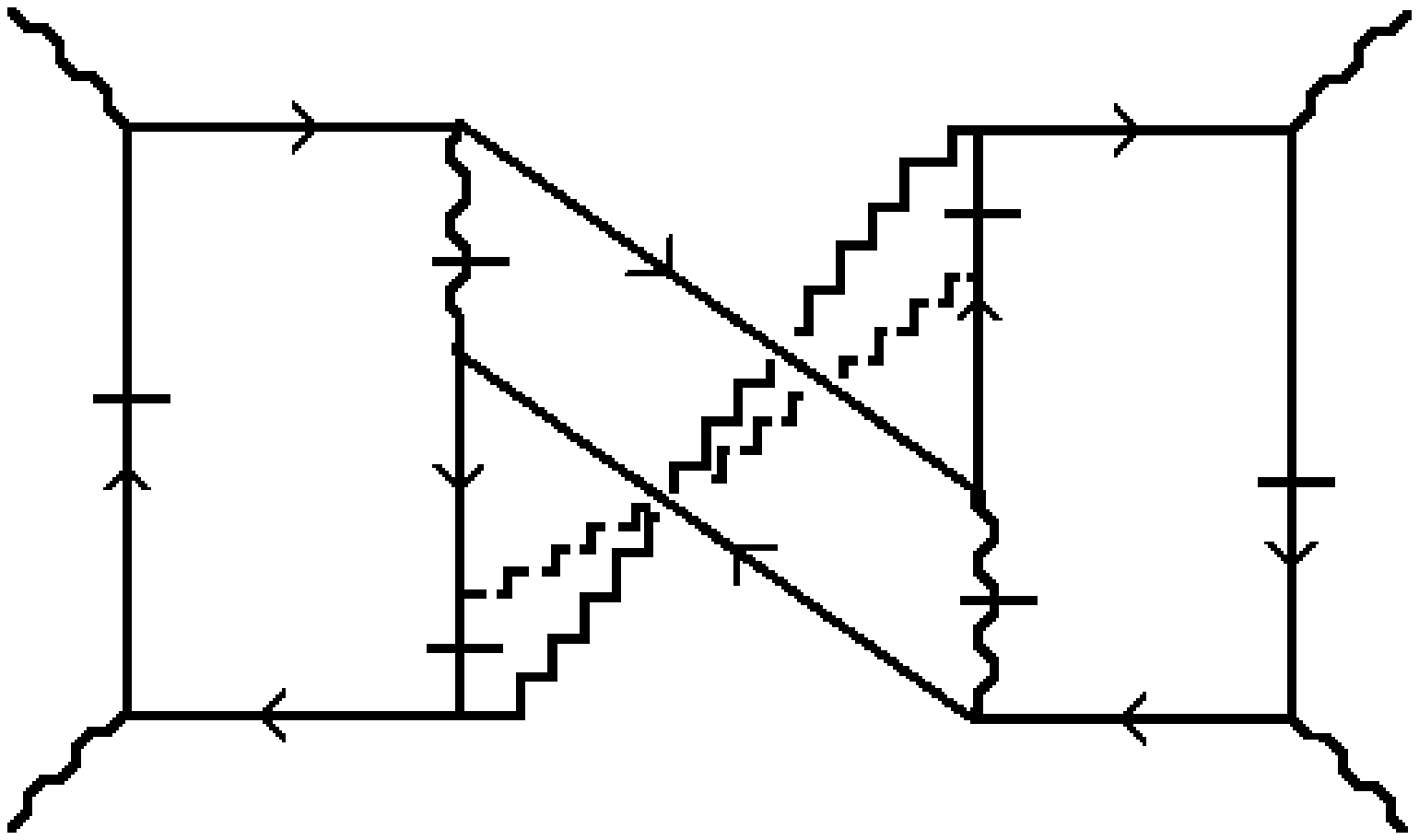}
\hspace{0.5in}
\epsfxsize=1.4in
\epsffile{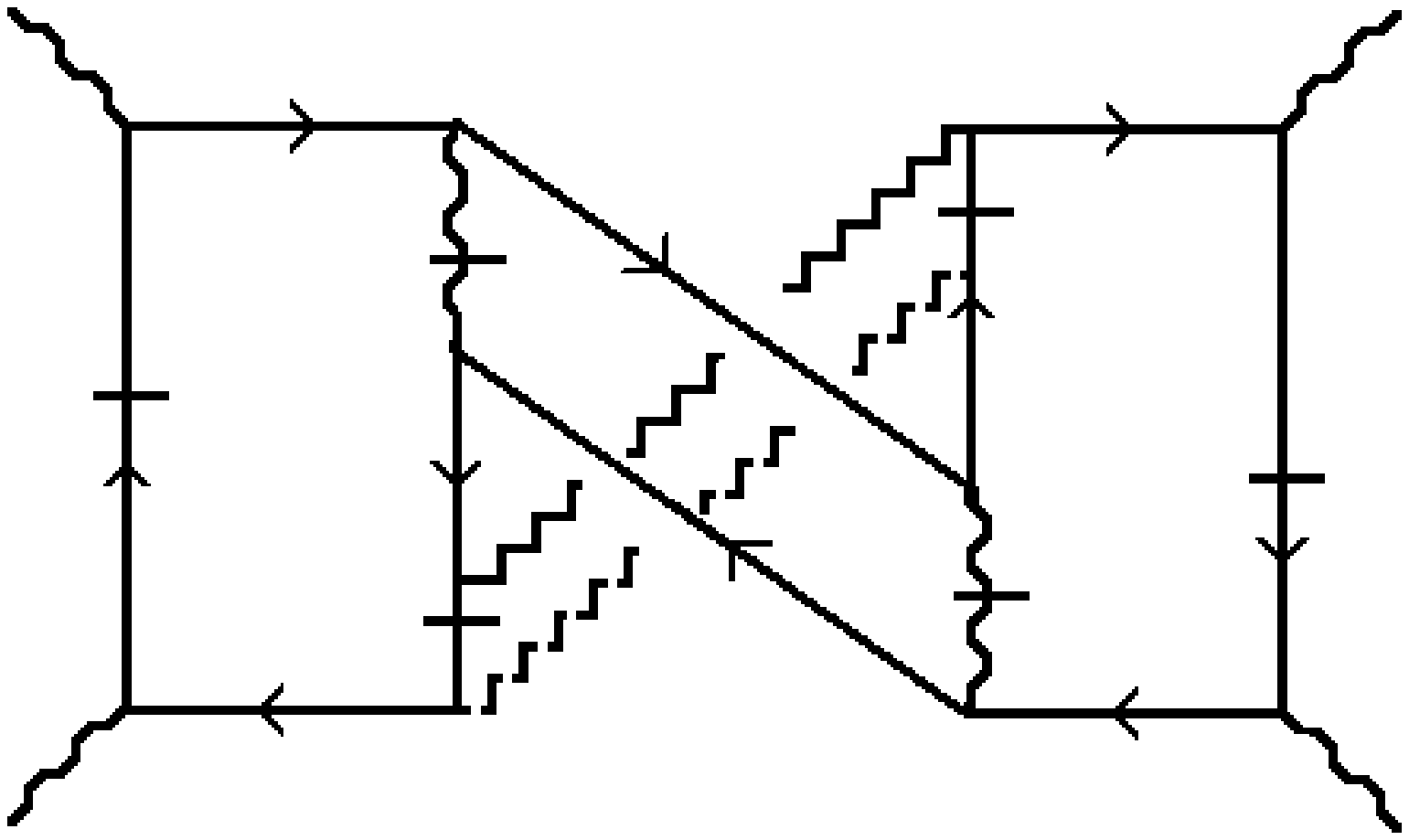}

(c1)\hspace{2in}(c2)

Fig.~39 Diagrams Related to the Diagrams of Fig.~33 by Twisting
\end{center}

Both Fig.~39(c1) and Fig.~39(c2) can be regarded as twisted relative to
Fig.~37(c). For Fig.~39(c1) the color factor is reversed
compared to Fig.~37(c), but the hatched lines clearly can 
not be consistently placed on-shell in the physical region.
For Fig.~39(c2) it seems probable 
that the hatched lines can be placed on-shell consistently, even though
the necessary cuts would cross. The issue is irrelevant, however, since
Fig.~39(c2) has the same color factor as Fig.~37(c).
In fact, there is no twisted diagram corresponding to Fig.~37(c) which has
a reversed color factor and in which all the
necessary hatched lines can be consistently placed on-shell. 

Since we can regard the color factor for Fig.~37(c) as
the sum of the symmetric and antisymmetric factors,
the antisymmetric, odd color parity,
component will be selected for the 
anomaly contribution from this diagram (and it's same helicity counterpart). 
Similarly for Fig.~39(c2), if it contributes.
Together with the contribution of corresponding diagrams obtained by 
twisting both $G_L$ and $G_R$, these will be the only
anomaly contribution from diagrams in which 
the second gluon carries ``finite''transverse momentum. In all cases, the
second gluon is 
attached as in Fig.~25(c), and not as in Fig.~25(b). The finite gluon 
contributions have the important property that the two gluon state carries 
``anomalous color parity''.  This is significant because,
as we emphasized in the Introduction, reggeized gluon exchanges that appear in
vector theory perturbative calculations\cite{fkl}-\cite{arw93} 
carry normal color parity. The appearance of anomalous color parity gluon 
states is, therefore, a direct consequence of the presence of the anomaly.

\subhead{6.10 General Multigluon Color Factors}

Before considering more complicated multigluon configurations
we give a general discussion of multigluon
color factors which generalizes the previously discussion
of two gluon color factors.

We note first that we can obtain (\ref{lji+}) and (\ref{lji-}) from 
(\ref{lij}) by a more general argument than just the symmetry
and antisymmetry of the $d$ and $f$ tensors. Consider
a product of $\lambda$ - matrices 
$$
P_{1n}~=~\prod_{j=1}^n ~\lambda_{i_j}
\auto\label{p1n}
$$
together with the product taken in the reverse order
$$
P_{n1}~=~\prod_{j=n}^1 ~\lambda_{i_j}
\auto\label{pn1}
$$
Using (\ref{lij}) extensively we can write
$$
P_{1n}~=~A_{1n}~+~
~\centerunder{$\hbox{\Large $\Sigma$}$}{\raisebox{-3mm}{$
{\scriptstyle k}$}} ~B_{1nk}~\lambda_k
\auto\label{p1n1}
$$
where $A_{1n}$ multiplies the unit matrix and 
both $A_{1n}= A_{i_1,i_2,..,i_n}$ and $B_{1n}= B_{i_1,i_2,..,i_n,k}$
contain combinations of $f$ and $d$ tensors. Similarly, we can write 
$$
P_{n1}~=~A_{n1}~+~
~\centerunder{$\hbox{\Large $\Sigma$}$}{\raisebox{-3mm}{$
{\scriptstyle k}$}}~B_{n1k}~\lambda_k
\auto\label{pn11}
$$
(\ref{p1n1}) and (\ref{pn11}) decompose $P_{1n}$ and $P_{n1}$, respectively,
into a sum of color singlet and color octet contributions.

It follows from the hermiticity of the $\lambda_i$
that 
$$
(P_{n1})^T~=~ \prod_{j=n}^1~(\lambda_{i_j})^T~=~ \prod_{j=n}^1~(\lambda_{i_j})^*
~=~(P_{1n})^*
\auto\label{tcc}
$$
where $(~)^T$ denotes transposition and $(~)^*$ denotes complex conjugation.
Equivalently,
$$
P_{n1}~=~\bigl(~(P_{1n})^T~\bigr)^*
\auto\label{tcc1}
$$
As a result 
$$
A_{n1}~=~A_{1n}^*~,~~~~~B_{n1k}~=~B_{1nk}^*
\auto\label{tcc2}
$$
or, equivalently,
$$
\eqalign{P_{1n}~+~P_{n1}~&= ~2~Re(A_{1n}) ~+~2~Re(B_{1nk})~\lambda_k \cr
P_{1n}~-~P_{n1}~&=~ 2i~Im(A_{1n})~+~2i~Im(B_{1nk})
~\lambda_k}
\auto\label{tcc3}
$$
which gives ((\ref{lji+}) and (\ref{lji-}), as a very simple case. 

Since the $f$ and $d$ tensors are both real, it
follows from (\ref{lij}) that a factor of $i$ is always
accompanied by an $f$ tensor. Therefore, the real and imaginary parts
of both $A_{1n}$ and $B_{1n}$ contain, repectively, 
even and odd numbers of $f$ tensors. If we then consider
$$
Tr\{\Sigma_{i_1,i_2,..,i_n}\bigl(P_{1n}~\pm~P_{n1}\bigr)\bigl(P_{1n}\bigr)\}
\auto\label{tra}
$$
the distinct color of $A_{1n}$ and $B_{1n}$ and the distinct
symmetry properties of the real and imaginary coefficients
implies that 
$$
Tr\{\Sigma_{i_1,i_2,..,i_n}\bigl(P_{1n}~+~P_{n1}\bigr)\bigl(P_{1n}\bigr)\}
~=~2~\Sigma_{i_1,i_2,..,i_n}\bigl([Re(A_{1n})]^2 ~+~[(Re(B_{1n})]^2~\bigr)
\auto\label{tra2}
$$
which is a sum of the squares of color factors for color zero and color octet
states which contain an even number of $f$ - tensors and so 
describe normal color parity gluon states. 
Similarly, 
$$
Tr\{\Sigma_{i_1,i_2,..,i_n}\bigl(P_{1n}~-~P_{n1}\bigr)\bigl(P_{1n}\bigr)\}
~=~-2~\Sigma_{i_1,i_2,..,i_n}\bigl(~[Im(A_{1n})]^2~+~[Im(B_{1n})]^2~\bigr)
\auto\label{tra3}
$$
which is a similar sum of squares of color factors which, because they 
contain an odd number of $f$ tensors,
describe gluon states with anomalous color charge parity.

In the two gluon states that we have so far considered, all color factors
apart from the first term
in (\ref{tra3}) have appeared. This term corresponds to a color zero
anomalous color parity multigluon configuration. It will appear in the 
three gluon diagrams that we discuss next.
We continue to
confine our discussion to a single large transverse momentum gluon and consider
only multiple soft or finite gluon contributions that do not
involve any factors of $ln S$. (In general, 
we anticipate that higher order logarithms 
lead to the separate reggeization of each of  
the transverse momentum gluons in the diagrams we study.)

\subhead{6.11 Two Soft Gluons}

We begin with two soft gluons and consider diagrams 
that contribute to the transverse momentum diagram of Fig.~40.
\begin{center}
\epsfxsize=1.8in
\epsffile{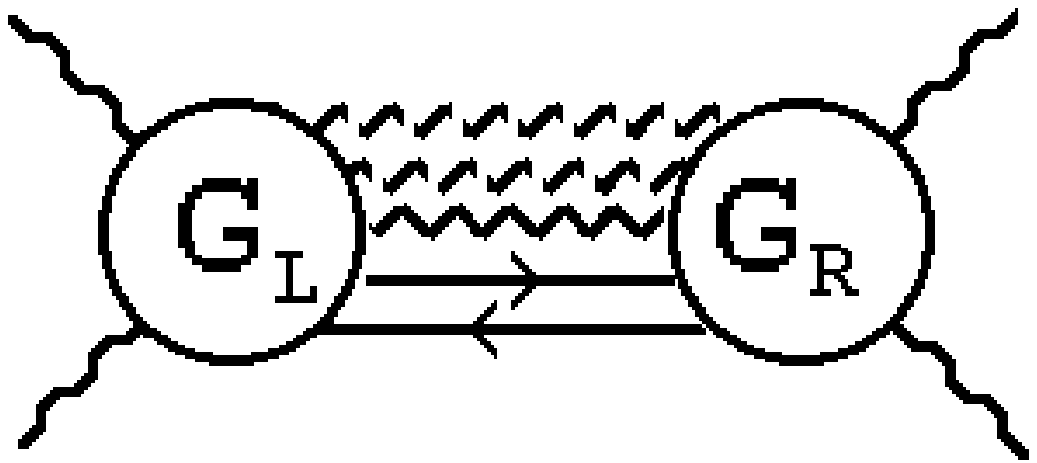}

Fig.~40 A Transverse Momentum Diagram With Two Soft Gluons 
\end{center}
We will straightforwardly
obtain an anomaly enhanced amplitude, as before, from diagrams in which
two soft gluons are attached to
the external effective point vertices, as illustrated in Fig.~41.
\begin{center}
\epsfxsize=2in
\epsffile{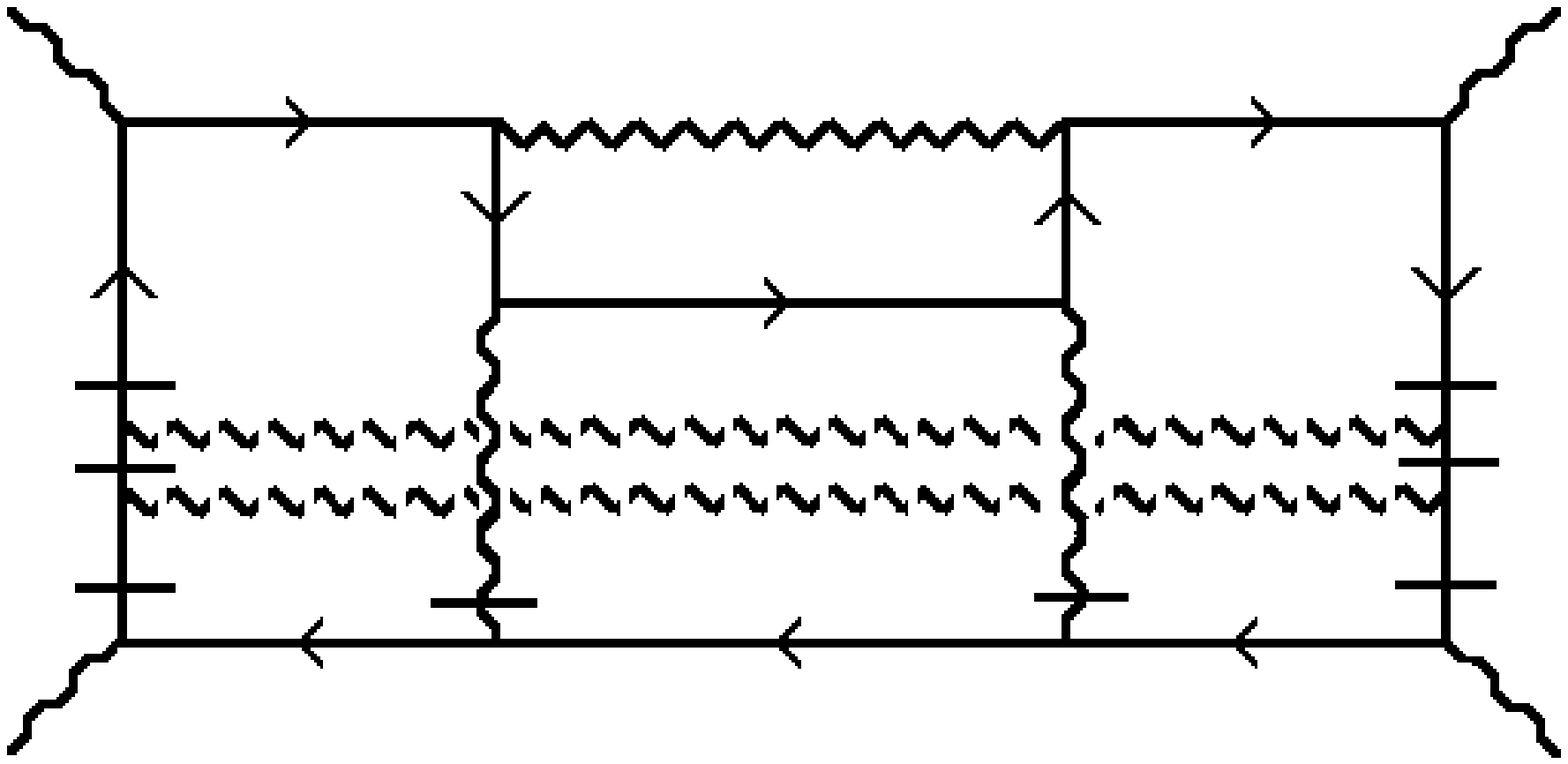}
\hspace{0.7in}
\epsfxsize=2in
\epsffile{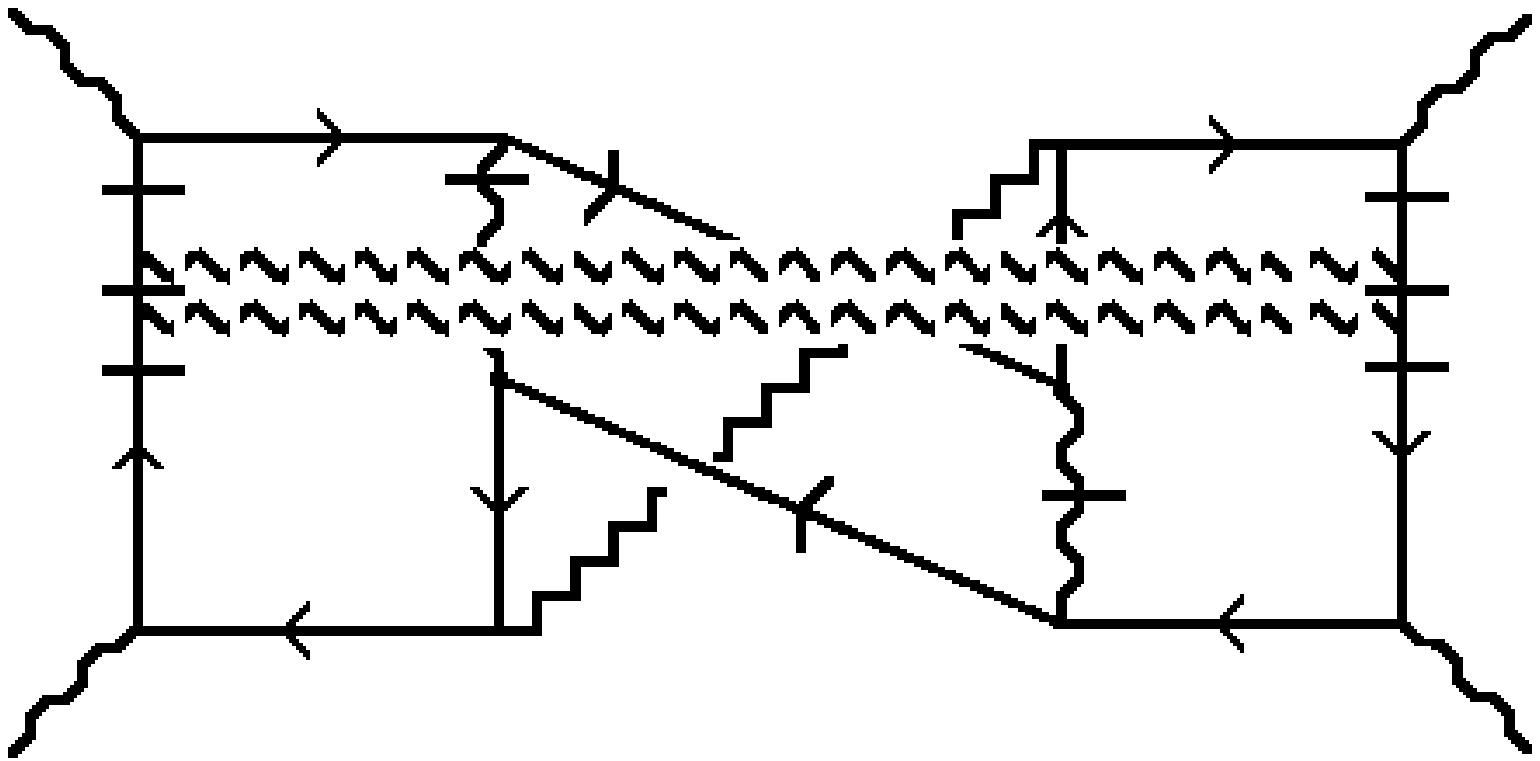}

(a)\hspace{2.8in}(b)

Fig.~41 Diagrams With Two Soft Gluons
\end{center}
Because all the large momenta pass through both soft gluon attachments and
are constrained by the $k_-'$ integration, the infra-red scale is the same for 
both and is $m^2/S^{\frac{1}{2}}$, as in (\ref{llmt}) and (\ref{llmt1}). 
Infra-red cancelations can, presumably, be discussed in the same manner as 
in our discussion of one soft gluon diagrams. 
The twisted diagram of Fig.~41(b) again reverses the color matrix multiplication
of Fig.~41(b) and so 
the sum of the two diagrams gives a color factor of the form of
(\ref{tra2}), corresponding to normal color parity for the complete
three gluon transverse state

To check that the anomaly amplitude obtained is even signature we note that
the anomaly
diagrams in the same sign helicity amplitude
will again be related to the opposite sign diagrams by $P_+ \to -P_+$.
For example, the diagram of Fig.~37(a) will be related to
a same sign helicity amplitude diagram as illustrated in Fig.~42
\begin{center}
\parbox{2in}{\epsfxsize=2in
\epsffile{ehes43.ps}}\parbox{1in}{
\centerline{$\longleftrightarrow$}}
\parbox{2.2in}{\epsfxsize=2.2in
\epsffile{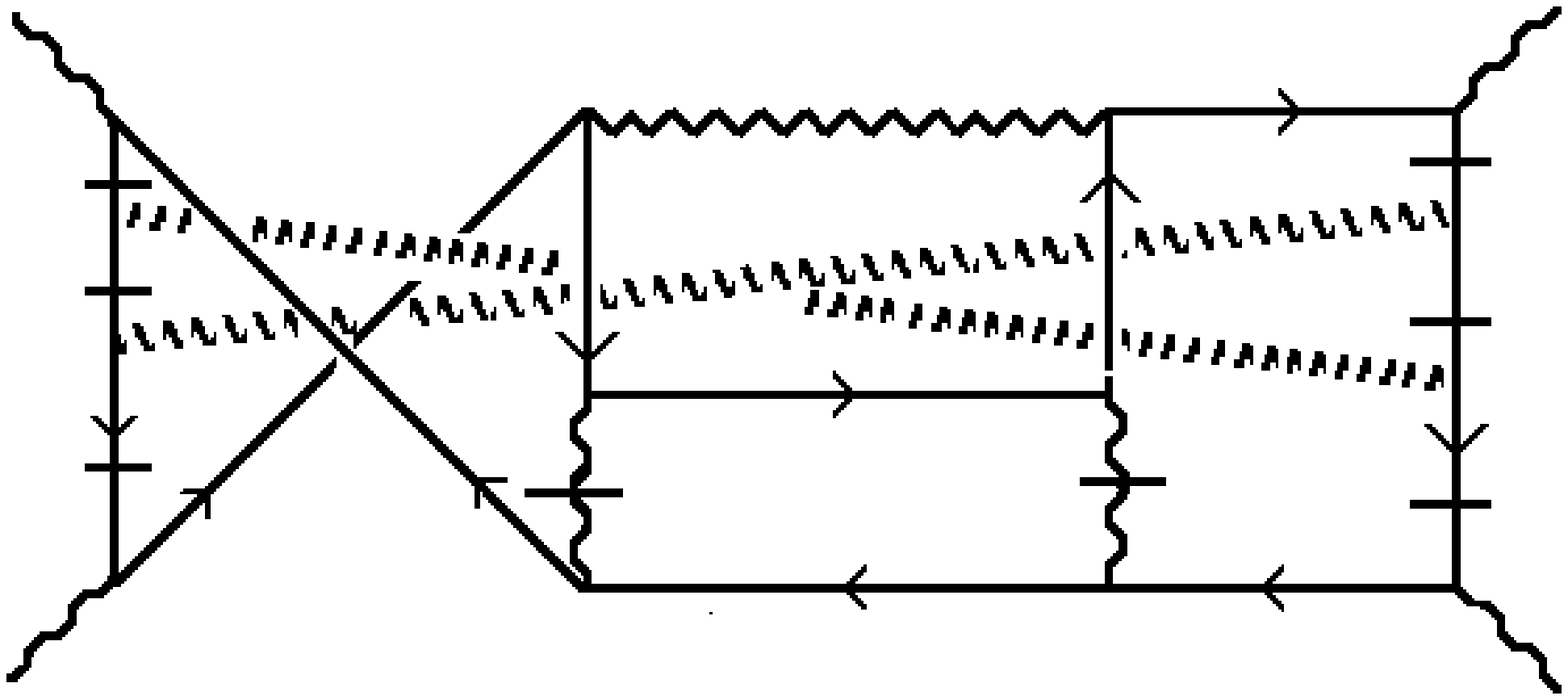}}

Fig.~42 Related Diagrams in Different Helicity Amplitudes 
\end{center}
Since the color factors are the same, these amplitudes are again related
by $S\to -S$ and will add in the even signature amplitude (and would 
cancel in a vector theory).

As before, the full
gluon state will have negative parity because of the single large
transverse momentum gluon. Combined with the negative (normal) color parity
this implies that the three gluon state is even under $CP$.
Repeating the discussion of the quark/antiquark state that we gave for 
the one soft gluon amplitudes we find that there is an additional change of
sign from $P_+ \to -P_+$ that results from the additional soft gluon. As a result,
quark/antiquark interchange gives no kinematic
change of sign and the color charge parity
transformation simply gives $\lambda_h \to \lambda_h^*$. Consequently,
the full quark/antiquark/multigluon intermediate
state is even under $CP$, as required for even signature.
(We should note, however, that although the color zero three gluon state 
has normal color charge parity, it is ``anomalous'' in that it has  
negative parity, giving an ``anomalous'' positive signature.)

\subhead{6.12 Color Zero Anomalous Color Parity }

Consider, next, adding a soft gluon
to the diagrams of Figs.~37(c) and 38(c), in which there 
is already one finite transverse momentum gluon present. 
If the soft gluon is attached to the left and right side
quark lines, the resulting
opposite sign and same sign helicity 
diagrams are shown, respectively, in Figs.~43(a) and (b).
\begin{center}
\epsfxsize=1.8in
\epsffile{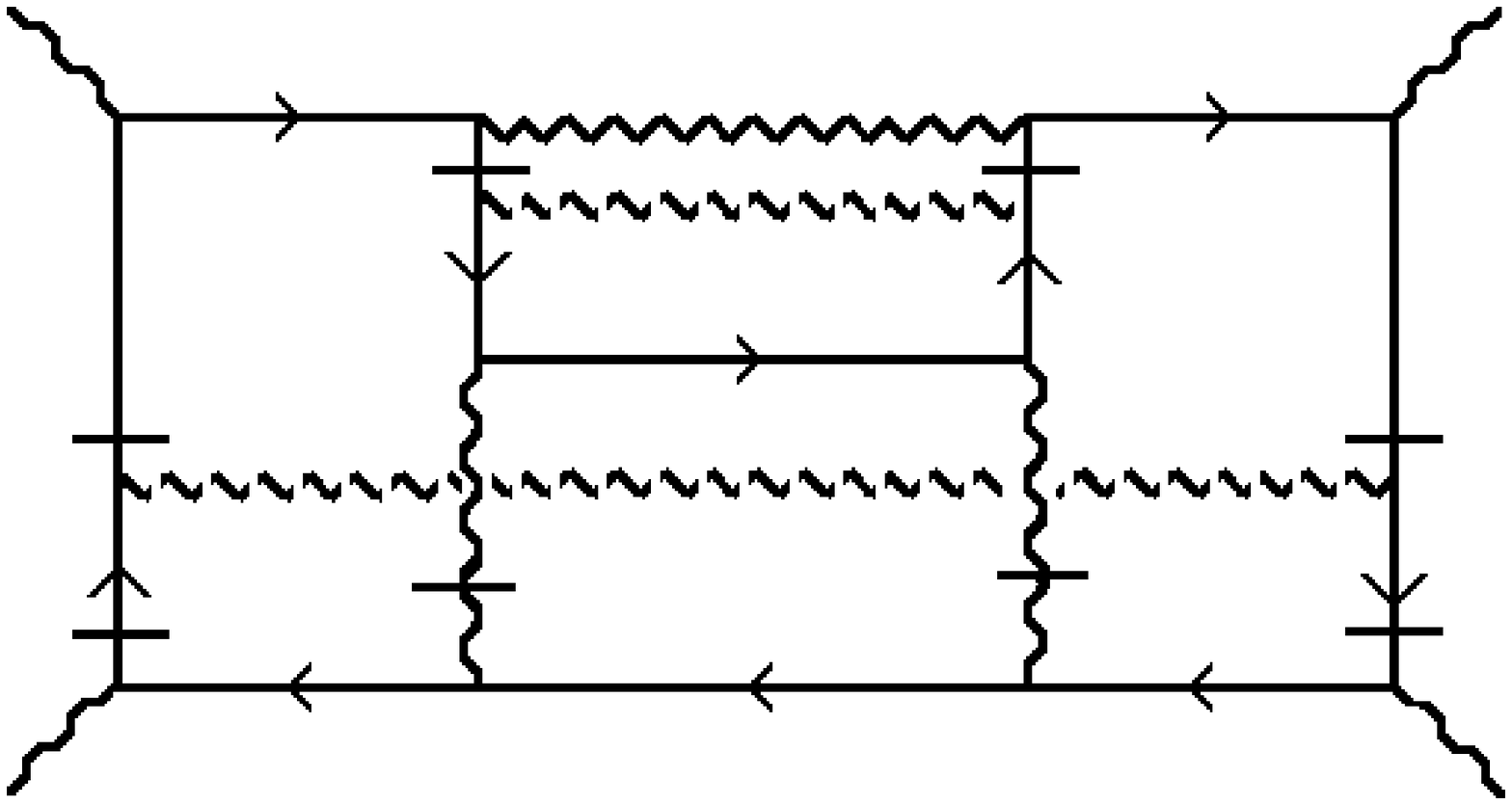}
\hspace{0.7in}
\epsfxsize=2.1in
\epsffile{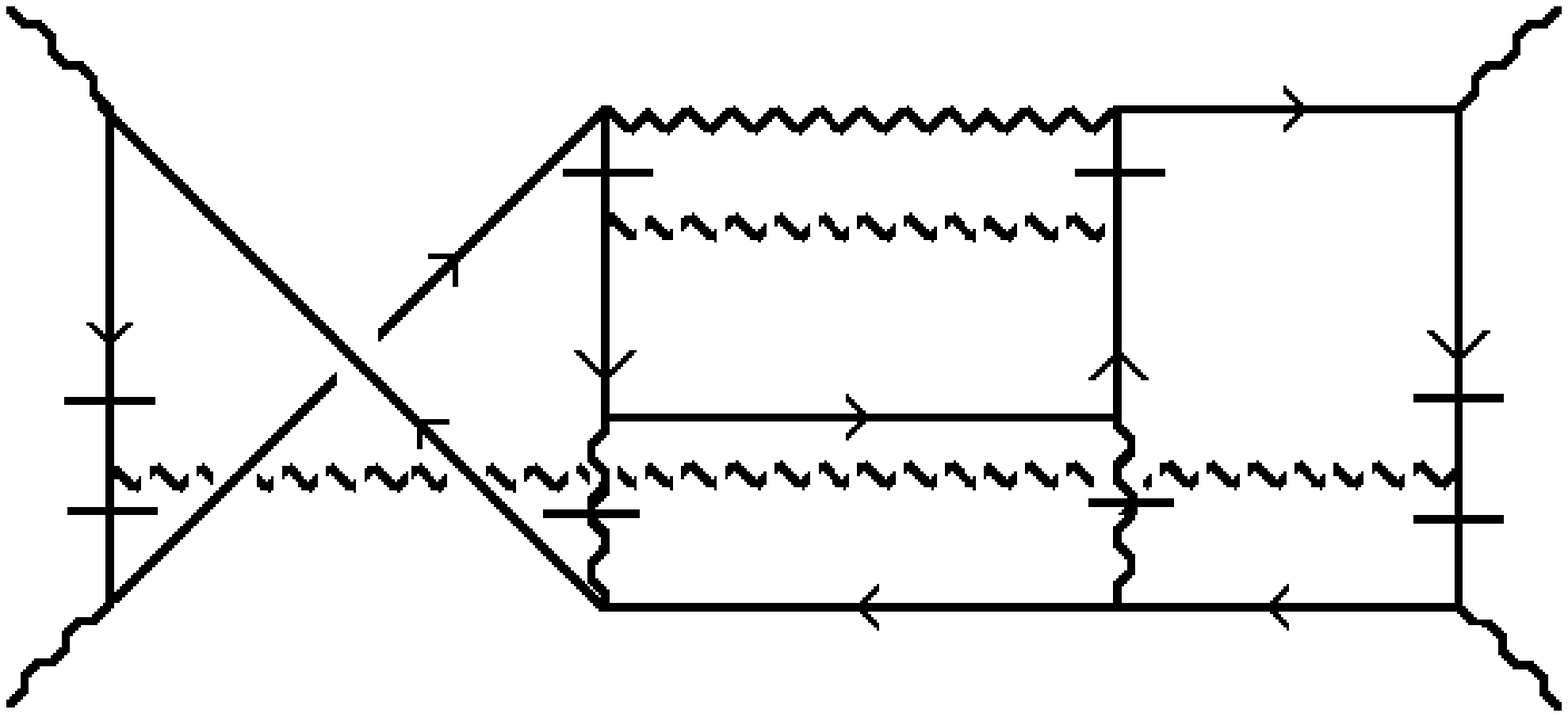}

(a)\hspace{2.8in}(b)

Fig.~43 Anomalous Color Parity Diagrams With One Soft and One Finite Gluon
\end{center}
The color factor for both diagrams is the same, i.e.
$$
 \centerunder{$\hbox{\Large $\Sigma$}$}{\raisebox{-3mm}{$
{\scriptstyle i,j,k}$}}
~~Tr\{\lambda_k\lambda_j\lambda_i\lambda_i\lambda_j\lambda_k\}
\auto\label{3gcf1}
$$

If we pick out the color zero intermediate state then the color factors on the 
left and right side of the diagrams must separately factor into traces. 
Therefore, we can write 
$$
\eqalign{ \centerunder{$\hbox{\Large $\Sigma$}$}{\raisebox{-3mm}{$
{\scriptstyle i,j,k}$}}
~~Tr\{\lambda_k\lambda_j\lambda_i\lambda_i\lambda_j\lambda_k\}
&\sim~ 
 \centerunder{$\hbox{\Large $\Sigma$}$}{\raisebox{-3mm}{$
{\scriptstyle i,j,k}$}}
Tr\{\lambda_j\lambda_i\lambda_k\}~
Tr\{\lambda_i\lambda_j\lambda_k\} ~~+~\cdots\cr
&\sim ~ \centerunder{$\hbox{\Large $\Sigma$}$}{\raisebox{-3mm}{$
{\scriptstyle i,j,k}$}}~f_{ijk}^2~+~\cdots } 
\auto\label{3gcf3}
$$
where the term we have shown explicitly is the color zero,
anomalous (even) color parity, term. 

Because, the diagrams of Fig.~43 contain one less gluon attached to the 
fast quark lines than the diagrams of Fig.~42, the quark-antiquark
component of the transverse state, in the even signature amplitude,
will be odd under $CP$. Therefore, to obtain even signature overall
the gluon component must also be odd under $CP$, implying that is
even under color charge conjugation. Consequently, the anomalous color parity term,
shown explicitly in (\ref{3gcf3}), 
is selected for the color zero component of the 
combination of the diagrams of Fig.~43 in the even signature amplitude.

\subhead{6.13  Multigluons}

Clearly we could generalise the foregoing 
discussion to a variety of multigluon configurations involving
combinations of soft and finite gluons, with effective vertices
of the form illustrated in Fig.~44.
\begin{center}
\epsfxsize=3.5in
\epsffile{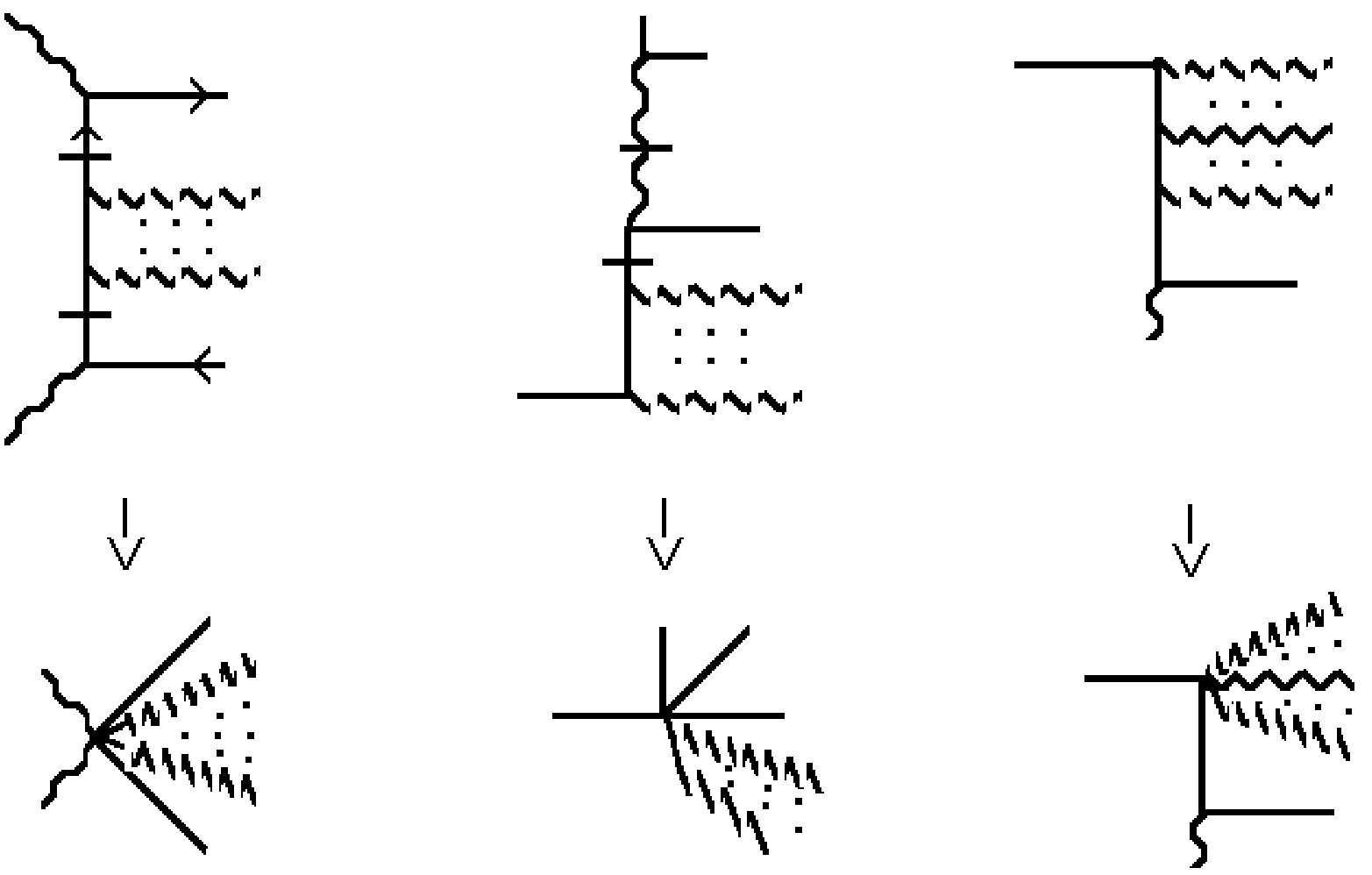}

(a)\hspace{1.2in}(b)\hspace{1.2in}(c)

Fig.~44 Multi-gluon Effective Vertices 
\end{center}
However, having established the coupling of color zero anomalous color parity
gluons, we have all the general properties that we require for the discussion
of the next Section.

Since a large transverse momentum gluon can give a scaling contribution of
the form
$$
\sim ~~~\int_{ (k_{\perp}'')^2\sim\sqrt{s}}~
\frac{d^2 k_{\perp}''}{(k_{\perp}'')^2} 
\auto\label{scc}
$$
it is also possible for additional large transverse momentum gluons
to participate in the enhancement effect. Potentially, this could be an 
elaborate phenomenon involving, presumably,
the reggeization of both quarks and gluons and, in higher orders, scaling
properties of reggeized gluon interactions, as well as 
the evolution of $\alpha_s$. However, since we will argue, in the next Section,
that large transverse momentum anomaly contributions are unphysical, there 
seems little point in exploring the issue any further. 
In part, we discuss the analagous infra-red phenomenon in the next Section.

\newpage 

\mainhead{7. REGGEON WARD IDENTITIES, CUT-OFFS, AND INFRA-RED DIVERGENCES}

Our calculations in the previous Sections have demonstrated that the 
anomaly enhanced diagrams, some of which contain
anomalous gluons, provide
the dominant contribution in the exchange channel we have considered.
However, as we remarked in the Introduction, we believe that the power enhancement 
involved should not be present in physical amplitudes. Assuming that there is no
perturbative cancelation, via some mechanism
that has yet to be elucidated, then obtaining ``physical'' amplitudes without the 
enhancement is, a priori, a challenging problem.

In this Section we will briefly outline how we anticipate the desired physical 
amplitudes can be obtained. The essential point will be 
that the contribution of the anomaly 
diagrams is very different if we take 
the regge limit before or after the removal of an 
ultra-violet transverse momentum cut-off. This cut-off introduces
infra-red divergences and if it is removed only at a 
very late stage, as we will propose, then  
the result obtained will also depend 
on whether all orders sums are performed before or after it is removed
and on how, and at what stage, 
infra-red cut-offs (in the form of gluon and quark masses) are removed.
This ambiguity 
is the essence of the anomaly and it would not be surprizing
if there is a unique procedure that is necessary to obtain the right 
``physical'' answer.

As we have already noted, both in the Introduction and in the previous
Section, a study of the 
infra-red anomaly contributions of the diagrams we have considered, that
matches the present study of ultra-violet contributions, will be necessary
before any detailed arguments can be carried through. To fully elucidate
infra-red anomaly contributions it will surely be necessary to abandon
the restriction to forward kinematics and transverse polarizations that has 
greatly simplified the foregoing discussion. Nevertheless, based on our 
experience with hadron scattering amplitudes\cite{arw02},
we believe that a complete procedure for obtaining physical 
amplitudes can be developed utilizing the following, briefly summarized, 
properties. Our hope is that since the present starting point is 
much simpler than the hadronic problem, the analysis will be correspondingly
more straightforward.

\subhead{7.1 Ward Identity Consequences and a Transverse Momentum Cut-Off}

Gauge invariance implies that a general amplitude $\VEV{A_{\mu}(q)~...~}$
with all (external) lines
on shell except for one gluon that carries four momentum $q$
satisfies the simple Ward identity
$$
q_{\mu}~\VEV{A_{\mu}(q)~...~}~=~0
\auto\label{wd2}
$$
This identity usually follows\cite{gth}, at a given order in perturbation theory,
only after the zero momentum gluon has been attached to the remainder
of the diagram at all possible points. It is well-known that this 
identity, in turn,
implies that the gluon amplitude vanishes at zero four momentum. 
Also, from analagous Ward identities\cite{gth}, a
similar result holds when more than one gluon carries vanishing four momentum.
(If the gauge symmetry is spontaneously-broken then, of course, the Ward identities
corresponding to (\ref{wd2}), such as the ``electroweak Ward identities'' 
referred to in sub-section {\bf 5.3}, have additional mass terms which prevent
the infra-red vanishing of amplitudes.)

The vanishing of a loop amplitude when external momenta are 
small compared to (``finite'') internal
momenta also implies, generally, a suppression of
internal momenta that are large compared to, finite, external momenta. 
If, however, there is an external axial current
producing an anomaly contribution in a loop, then the situation is different.
In this case, as we briefly review in Appendix C, in addition to the 
well-known anomalous Ward identity \cite{awb} for the axial current, vector
Ward identities require a cancelation between separate contributions,
with different kinematic structure, from large and 
small internal momentum regions. In particular, 
the large momentum anomaly contribution (\ref{uvco})
cancels with an infra-red term that, in special momentum
configurations (and only when the quarks are massless), 
reduces to a simple pole in the axial vector channel.
This is the ``anomaly pole'' that can, in the right circumstances, be
interpreted as a Goldstone boson pole, signaling chiral symmetry breaking.
In addition to the discussion in Appendix C,
a detailed analysis of the anomaly pole, and the 
internal momentum region generating it, can be found in\cite{arw02}.

In general, the above properties of gluon amplitudes, as functions
of four-dimensional momenta, transfer directly\cite{arw98} to corresponding
properties for the multi-gluon (multi-reggeon)
transverse momentum couplings that we discuss in this paper, as functions of
transverse momenta. The linear vanishing
when transverse momenta are scaled to
zero is sufficient to eliminate infra-red divergences in the 
transverse momentum diagrams that we consider. If there is an anomaly in a 
transverse momentum coupling then, as we already noted in our discussion
of infra-red cancelations in {\bf 6.8}, there will be
large/small internal momentum cancelations in the
associated transverse momentum (reggeon) Ward identities that parallel the 
cancelations that take 
place in the four-dimensional Ward identities. We expect that 
infra-red divergences will be avoided, in part, by cancelation of the ultraviolet
anomaly contributions we have found with infra-red ``anomaly pole'' contributions. 
In fact, the coefficient of the
anomaly in a transverse momentum coupling (which we did not determine)
should be fixed by this cancelation.

If we impose a transverse momentum cut-off in all internal loop
integrals of the diagrams we consider, this cut-off will be present in
all transverse momentum diagram integrals and also within the loop integrals giving
the external couplings. A cut-off in the transverse momentum diagram
integrals is gauge invariant. A priori, however, in the external couplings
such a cut-off is not gauge invariant. Therefore, if we take the regge limit with
a transverse momentum cut-off imposed, it will be a serious question whether
gauge invariance is restored by removing the cut-off after the limit. For
the present we note only that in \cite{arw02}
we argued that anomaly pole contributions to infra-red divergent amplitudes
are gauge invariant.

In the infra-red region, we anticipate that there will be 
effective triangle diagram contributions to $G_L$ and $G_R$ couplings
in which small transverse momentum gluons 
couple at all three vertices, as illustrated in Fig.~45.
\begin{center}
\leavevmode
\epsfxsize=2.5in
\epsffile{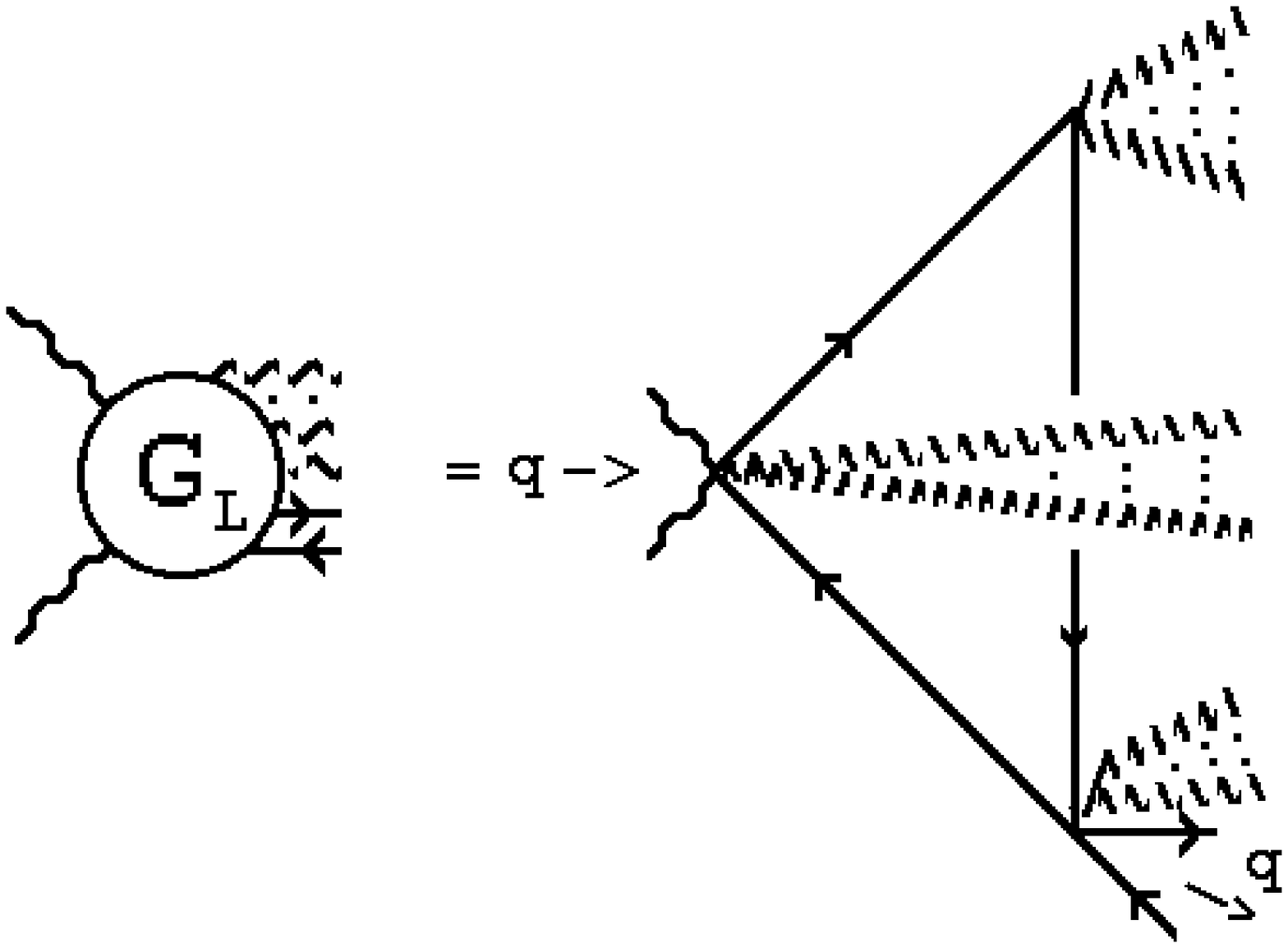}

Fig.~45 An Infra-Red Effective Triangle Diagram 
\end{center}
Based on the analysis of the previous Section, both 
normal and anomalous color parity multigluon
states should couple. With a transverse momentum cut-off imposed
we expect that, when the total gluon transverse momentum vanishes
the corresponding Ward identity will fail
and there will be a non-zero coupling involving (when the quarks are
massless) the anomaly pole.
As a consequence, in transverse momentum diagrams of the form of Fig.~40 (with
all gluons soft) there will be a logarithmic infra-red divergence of the form
$$
\int^{\lambda_{\perp}}~\frac{dQ^2}{Q^2}
\auto\label{ind}
$$
where $Q$ is the sum of all gluon transverse momenta. 
 
A priori, the anomaly pole can appear in both $G_L$ and $G_R$. However, at $t=0$,
where the pole should appear, a finite light-like momentum can be exchanged
which can be parallel to either $P_+$ or $P_-$. We suspect that this light-cone
momentum determines whether the pole appears via $G_L$ or $G_R$. Clearly,
a detailed study, of the kinematics/polarizations 
and kinematic forms associated with the appearance of the anomaly pole will be 
needed to be sure that, in the full amplitude, there 
is a simple pole with the appropriate residue to be associated with a pion.

\subhead{7.2 Transverse Momentum Infra-red Divergences}

Since the divergence (\ref{ind}) is not removed by external couplings 
(with a transverse cut-off), we must consider the effect of 
(all orders) interactions
amongst the gluons. In the lowest-order diagrams we expect the divergence to be 
present for both normal and anomalous color parity gluon states.
There may also be additional divergent transverse momentum configurations.
However, as we now describe,
when we sum all infra-red divergences to all orders we expect
that (\ref{ind}) is the only divergence that survives,
and then only for anomalous gluons.

We can summarize the general nature of gluon infra-red 
transverse momentum divergences and the role of a transverse 
momentum cut-off, very briefly, as follows. An expanded version of
this summary can be found in \cite{arw02}. For reasons that will become
apparent in the 
next sub-section we specifically discuss the case of SU(2) color, although
all the properties we describe remain the same for higher gauge groups.

The self-interactions of an $N$ gluon transverse momentum state 
$T_N$ are described by dimensionless ``kernels'' $
K^I_N(\hdots,k_i,\hdots,{k_j}', \hdots)$, where $I$ denotes
SU(2) color. (Each iteration of a kernel produces an additional factor of 
$ln~ S$, or $(J-1)^{-1}$ in the $J$ - plane, which we will not show explicitly.)
When the $t$-channel 
color is non-zero the  infra-red divergences related 
to reggeization do not cancel and
$$ 
\int~ \prod_{i=1}^N~ \frac{d^2k_i}{k_i^2 }
~K^I_N(\hdots,k_i,\hdots,{k_j}', \hdots)~\to ~\infty~,~~~Q^2, I \neq ~0
~~~~~\bigl(~Q=\Sigma_i k_i~\bigr) 
\auto\label{KIN}
$$
As a result, the sum of all diagrams in any colored 
channel exponentiates to zero as illustrated in Fig.~46. 
\begin{center}
\leavevmode
\epsfxsize=4.5in
\epsffile{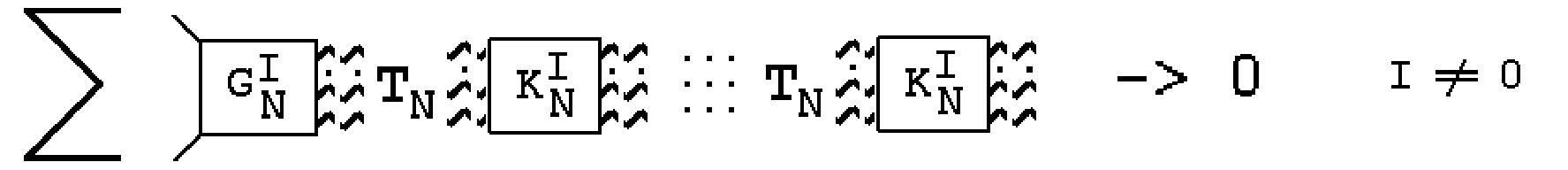}

Fig.~46 Iteration of a Gluon Kernel $K_N$.
\end{center}
$G^I_N$ is an external coupling
analagous to the $G_L$ and $G_R$ appearing in the previous Sections.

When $I=0$ and $Q^2 ~\neq 0$, there is a cancelation of
divergences in the $K^0_N$. (This is the infra-red
finiteness property which is extensively exploited in BFKL applications.)
At the leading-log level, infra-red finiteness leads directly to
conformal scale invariance. When renormalization effects are introduced,
scale invariance is lost in the ultra-violet region. 
Scale invariance properties may still be present in the infra-red region
(in particular, they will be present  
if there is an infra-red fixed point for the gauge coupling).
In this case, the kernels $K^0_N$ will scale canonically
as $Q^2 \to 0$ so that, with a transverse momentum cut-off $\lambda_{\perp}$, 
$$
\int_{|k_i|^2, |k^\prime_j|^2 ~<~\lambda_{\perp}} ~\prod_i \frac{d^2k_i}{ k^2_i}
~\prod_j \frac{d^2k^{\prime}_j}{ {k'}^2_j}~
K^0_N ( k_1, \cdots k_N,  k^{\prime}_1, \cdots k^{\prime}_N)
~ \sim ~ ~\int^{\lambda_{\perp}} \frac{dQ^2}{ Q^2}
\auto\label{irscl}
$$
The presence of the cut-off ensures that this divergence is unambiguously isolated
from ultra-violet divergences with which it might mix.

This is the same divergence as (\ref{ind}), which appears in the lowest-order 
diagrams. The kernels $K^0_N$ have Ward
identity zeroes which result in the special property that iteration of
any $K^0_N$ does not increase the degree of divergence. Instead, there is
a distinct contribution from each $T_N$
and the residue of the divergence 
can be written in a factorized form, as illustrated in Fig.~47. 
\begin{center}
\leavevmode
\epsfxsize=5.9in
\epsffile{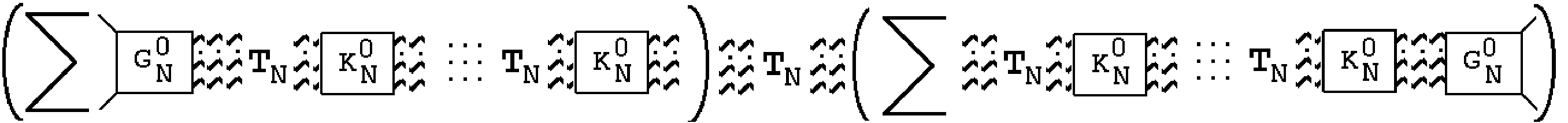}

Fig.~47  Isolation of the Divergence Associated with $T_N$.
\end{center}
If there is no 
Ward identity zero in the external couplings $G^0_N$, (\ref{irscl}) is 
a potential source of a simple infra-red divergence at $Q^2=0$.

Similar properties to the above hold for
the interactions of gluons with quarks. Crucially, however, there 
is no kernel describing a
transverse momentum interaction between a quark/antiquark pair and an
anomalous gluon state. This is because anomalous gluons couple
only through an anomaly and anomalies can not occur
within the two-dimensional kinematics that the kernels describe.

\subhead{7.3 SU(2) Color and Reggeon Field Theory}

If we consider all the diagrams discussed in previous Sections, generalized to 
include arbitrary numbers of gluons, and add both 
interactions amongst the gluons and between the quark-antiquark pair,
we arrive at the set of transverse momentum 
diagrams shown in Fig.~48. 
\begin{center}
\epsfxsize=4.5in
\epsffile{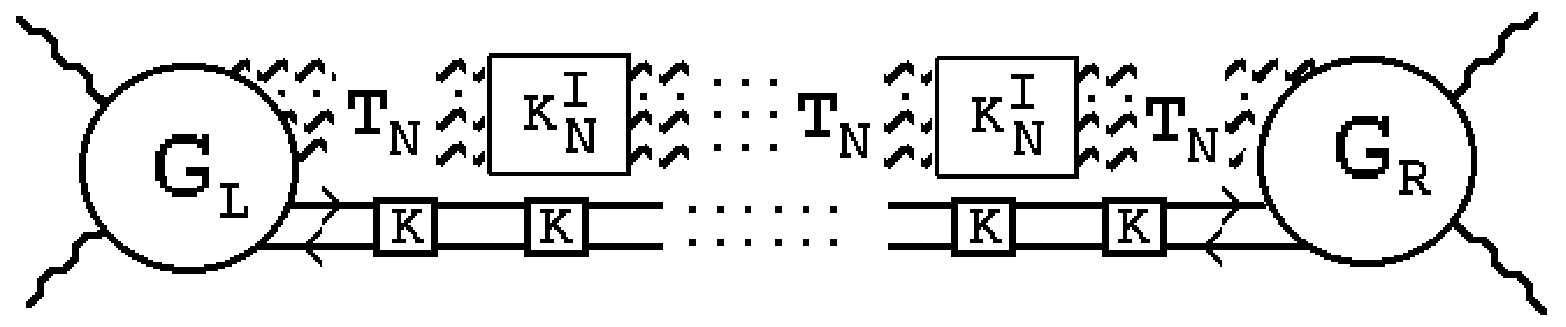}

Fig.~48 A Class Of Transverse Momentum Diagrams
\end{center}

If the gluons are anomalous and carry zero color, they will have no interaction
with the quark-antiquark pair and the 
divergence (\ref{irscl}) will occur when the transverse momentum
of all gluons is scaled to zero. As we discussed above, the anomaly pole
should appear in the coefficient of this divergence, presumably, with
the right kiematic structure to be interpreted as a pion pole.
All other similar diagrams, in which either the
color is non-zero or the gluons are not anomalous will be exponentiated to
zero by interactions that iterate
the divergence. However, since the cut-off has still to be removed
and it is unclear how to handle the infra-red divergence, the result is still far
from a sensible amplitude.

To obtain a more sensible result, we have to use a more sophisticated  
treatment of the infra-red divergences. In particular, we 
initially take the SU(3) gauge symmetry of QCD to be partially broken to 
SU(2). We could motivate this by noting, first, that the structure of 
the anomaly diagrams is much simpler. Only odd numbers of anomalous gluons
can carry color zero (because of the absence of
the $d$ - tensor). An overall logarithmic infra-red
divergence will still occur, as we have discussed in the previous sub-section,
because of the unbroken SU(2) gauge symmetry. However, some gluons (that are 
massive and outside the SU(2) subgroup) will interact with the quark-antiquark
pair. Also, we might hope to eliminate the divergence by
averaging over the direction of the SU(2) subgroup within SU(3),
as the transverse momentum cut-off is removed. 

With these last observations in mind, it is easy to appreciate why
Reggeon Field Theory (RFT) should be applied 
to the problem. In formulating the study of the QCD pomeron using 
RFT we have argued\cite{arw93,arw84,arw98}, 
that we should start from the reggeon diagrams (or, equivalently,
transverse momentum diagrams) in which
the gauge symmetry is completely broken. With a transverse momentum cut-off,
the gauge symmetry can first be restored to SU(2) and
the resulting reggeon diagrams can be described by Super-Critical Pomeron
RFT - provided all infra-red divergences 
can be absorbed into a ``pomeron condensate''.

For our present problem we anticipate applying RFT as follows. With 
SU(3) color broken to SU(2), we consider
all diagrams of the form illustrated in Fig.~49. In these diagrams, 
anomalous gluons (within an SU(2) subgroup) accompany 
a quark-antiquark pair that is interacting with massive, reggeized,
gluons. The massive gluons are outside the SU(2) subgroup and
carry non-zero transverse momentum.
\begin{center}
$~~~~$\epsfxsize=3in
\epsffile{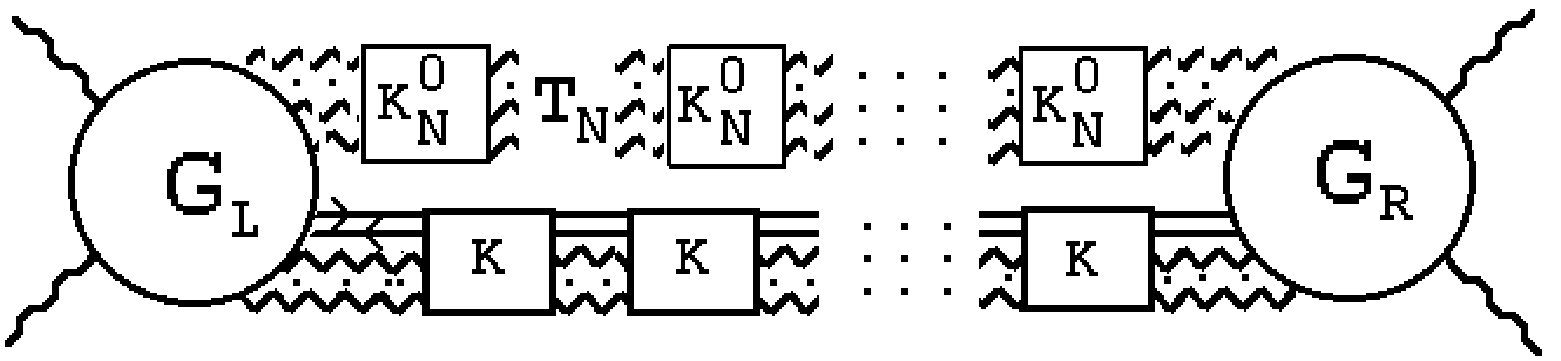}
\hspace{0.3in}
\epsfxsize=1.8in
\epsffile{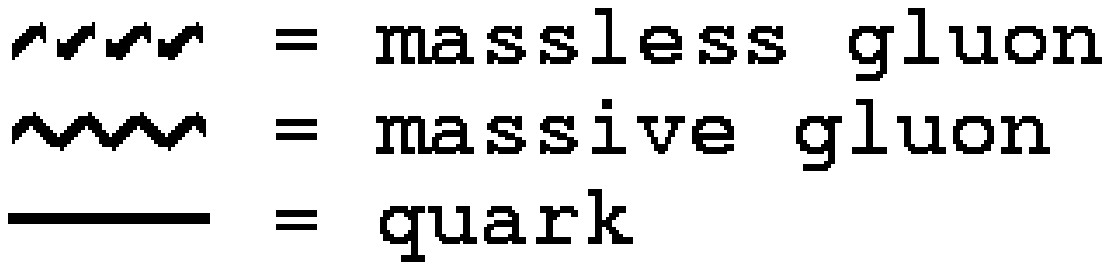}

Fig.~49 General Diagrams With a Divergence
\end{center}
This set should map completely on to Super-Critical RFT diagrams
containing both pomerons (with the pomeron being
a massive reggeized gluon plus
anomalous SU(2) gluons) and a reggeized Goldstone boson pion. In this mapping, 
the physical significance of the logarithmic infra-red 
divergence would be clear. It would be identified as responsible for
the appearance of a pomeron condensate.

The restoration of SU(3) gauge symmetry would be described
by the Critical Pomeron\cite{cri} interacting with a regge pole pion and (with 
the appropriate quark sector present\cite{arw93}-\cite{arw02})
the transverse momentum cut-off can be 
removed as part of the critical phenomenon. 
Also, as part of the critical phenomenon, 
the SU(2) direction of the pomeron condensate would be randomized within SU(3)
and disappear. In effect, the infra-red divergence, producing the condensate,
would be eliminated by averaging over the 
direction of the SU(2) subgroup within SU(3).

For hadron scattering it was important\cite{arw02} that the pomeron
condensate could be related to an anomalous 
gluon component of the scattering states. This was necessary, firstly,
because the $\gamma_5$ anomaly coupling of the pion to the pomeron 
is produced by a product of three orthogonal $\gamma$ matices. To obtain 
this product, it was essential to have anomalous gluon components
in both the scattering pion and the exhanged pomeron. As we have seen,
in electroweak
scattering this requirement is absent because the vector 
mesons have elementary $\gamma_5$ couplings, which allow the anomalous gluons
to appear in just the exchanged channel. However, the gluon components
of the scattering states also seemed to be important for 
the higher-order pomeron interactions needed to obtain the Critical Pomeron.
It may be, therefore, that 
RFT can only be consistently applied to the analysis
of infra-red divergences if the scattering vector mesons are also ``hadronic''.
That is, if they also have an anomalous gluon component, 
as they would have if they aquire their mass by absorbing Goldstone bosons
resulting from QCD chiral symmetry breaking - with the quarks being color sextet
quarks. The presence of the sextet quarks would produce\cite{arw93}-\cite{arw02}
an infra-red fixed-point (in the massless quark theory) that would guarantee
the infra-red scaling of gluon kernels producing (\ref{irscl}) and would also 
produce the ``quark saturation'' of QCD 
that we have argued is needed to obtain the Critical Pomeron with no transverse
momentum cut-off. Perhaps, 
all these features are needed to obtain a self-consistent description
of the regge limit for left-handed, massive, electroweak vector bosons. 

\newpage

\mainhead{8. CONCLUSIONS}

We have demonstrated that the triangle anomaly 
appears in the couplings of 
transverse momentum diagrams that describe the high-energy 
scattering of $W^{\pm,0}$ vector mesons. When the full amplitudes 
are directly evaluated, without any special cut-off procedure, the 
anomaly produces an enhancement, by a power of the energy, that threatens
the unitarity of the theory.
 
The most well-known consequence of a large momentum triangle anomaly is
the famous anomalous Ward identity for axial currents\cite{awb}. 
Less emphasized is the feature that, in the presence of the anomaly, vector
Ward identities are satisfied by a subtle cancelation between the contributions
of large and small internal momentum regions. In the vector meson scattering 
we have discussed, an effective current component 
with an anomaly appears and
it is the less emphasized feature that plays a crucial role. Even though
there are no anomaly-related cancelations between large and small internal
momenta in the finite momentum Ward identities, in a left-handed
gauge theory, it appears that the regge limit enhances large transverse
momentum regions such that there are cancelations of this kind in the  
transverse momentum (reggeon) Ward identities. There is 
then an ``anomaly problem'' in the sense that the regge limit result is 
very sensitive to the manipulation of ultra-violet and infra-red cut-offs,
as we have described. 

In Appendix D, we raise the possibility that
the occurrence of the anomaly
enhancement phenomenon in the diagrams that we have discussed is related, via
Ward identity contributions, to a more widespread phenomenon of large transverse 
momentum enhancement. 
If this is the case, then it is likely that the general transverse momentum
diagram formalism will fail. Since there would then be no reggeon diagram
formalism, $t$-channel unitarity
is also likely to fail. 
The conclusion, which is really the main conclusion of this paper,
is that to use the transverse momentum diagram formalism (and therefore
to ensure $t$-channel unitarity) it is essential to initially
employ a transverse momentum cut-off.

In previous papers we have 
found that, for bound-state amplitudes in QCD, the occurrence
of anomalies in multi-reggeon vertices (involving anomalous gluons)
leads to an analagous sensitivity to
infra-red and ultra-violet transverse momentum cut-offs. We have argued
that an ultra-violet cut-off should be kept
until after physical scattering amplitudes 
have been derived via an analysis of infra-red divergences. We anticipated
that, without an initial
cut-off, the ultra-violet anomaly effects would produce non-unitary power
enhancement of the energy behavior of bound-state amplitudes. 
However, as we noted in the Introduction,
accessing the anomalies in
hadron amplitudes is very complicated and, therefore, it is much more difficult 
to appreciate their significance. In the electroweak amplitudes we
have studied in this paper the anomaly appears immediately, because
of the presence of elementary left-handed couplings. As a result the choice 
between bad, large transverse momentum based, high-energy behavior and 
infra-red anomaly domination producing ``non-perturbative'' dynamics, is also 
immediately clear. 

Potential non-unitary properties
of electroweak high-energy scattering amplitudes may not be of great concern if,
as is currently believed by many physicists, the gravitational interaction
intervenes long before the relevant energies are reached. From this perspective,
our study of electroweak amplitudes can be viewed as simply a technical exercise 
in which left-handed vector mesons are used to study how, with the cut-off
manipulation we have described, the formation
of QCD bound states, including confinement and chiral symmetry breaking, 
can take place via regge limit infra-red anomaly effects.
Nevertheless, it seems hard to avoid the conclusion that  
if confinement and chiral symmetry breaking do 
not take place in this manner, then (assuming that it does not cancel)
the power enhancement
of quark-antiquark exchange by the  ultra-violet anomaly 
will dominate any electroweak symmetry
breaking mechanism that is perturbatively based.

Our point of view is that the unitarity of the electroweak
part of the Standard Model is a deep constraint. Indeed,
it could be that obtaining
consistent high-energy scattering amplitudes for massive vector mesons, with
left-handed couplings to quarks, may actually
require QCD confinement and chiral symmetry breaking to take place via the anomaly,
and may even, perhaps, require
that the chiral symmetry breaking (of higher color quarks) is responsible for 
electroweak symmetry breaking.

We were led to the present investigation as an outcome 
of our study of the QCD pomeron. For a long period of time 
we understood the crucial
role of the anomaly in producing unitary high-energy amplitudes within QCD,
but were unable to find a simple starting point from which to begin construction
of such amplitudes. Then, in our most recent paper\cite{arw02}, we showed that
wee gluon properties of the pion, obtained from the anomaly, provide
such a starting-point, at least in part. At the same time we realized that
such properties should appear if the pion is extracted from
the wee parton structure of an electroweak vector meson. This led us to 
the, a priori much simpler, problem of how an exchanged 
pion appears within
the scattering of vector mesons. We now believe that the results of this paper 
will lead to an understanding of the pion which will,
eventually, provide a simple starting point 
for the construction of QCD high-energy amplitudes. 
 
\subhead{Acknowledgement}

I am grateful to Jochen Bartels for useful criticism of a preliminary version
of this paper. I am also grateful to Bill Bardeen for valuable
discussion of electroweak Ward identity cancelations.

\newpage

\renewcommand{\theequation}{A.\arabic{equation}}
\setcounter{equation}{0}
\vskip 1cm \noindent
\noindent {\large\bf Appendix A. ~ Complex Notation for
Transverse $\gamma$ - Matrices}
\vskip 3mm \noindent
 
To discuss high-energy vector meson
scattering amplitudes involving (massless) fermion 
exchange, it is convenient to use a complex number notation\cite{kms}
for both transverse momenta and $\gamma$ matrices. In this formalism,
the consequences of 
chirality conservation and a left-handed gauge interaction are particularly 
apparent.

In addition to using conventional 
light-cone momenta $k_{\pm} = (k_0 \pm k_1)/\sqrt{2}$ ,
we write 
$$
\kt~ =~ k_2~ +~i~k_3~, ~~~~~ \kt^*~ =~ k_2~ -~ik_3
\auto\label{cta}
$$ 
to describe
transverse momenta. We then have
$$
k_{\perp}^2 ~=~\kt\kt^*
\auto\label{cta0}
$$
and
$$
2 k_{\perp}\cdot q_{\perp}~=~\kt\qt^* ~+ ~\kt^*\qt
\auto\label{cta01}
$$ 
We can also write 
$$
\eqalign{\kt\qt^*~=~ (\kt^*\qt)^*~&= ~k_2q_2 + k_2q_3 ~+~i( k_2q_3 -k_3q_2)\cr
&=~k_{\perp}\cdot q_{\perp}~+~i~k_{\perp}\times q_{\perp}}
\auto\label{cta1}
$$
where 
$$
k_{\perp}\times q_{\perp}~=~|k_{\perp}|~| q_{\perp}|~\sin \theta
\auto\label{theta}
$$
with $\theta$ the angle between the two vectors.

To describe transverse $\gamma$ - matrices, we similarly write
$$
\gam~=~(~\gamma_2~+~i~\gamma_3~)/\sqrt{2}, 
~~~~\gam^*~=~(~\gamma_2~-~i~\gamma_3~)/\sqrt{2}
\auto\label{cga}
$$
We then have 
$$
(\gam)^2~=~ (\gam^*)^2 ~=~0~, ~~~~ 
\gam~\gam^*~+~ \gam^*~\gam ~= ~2
\auto\label{cga1}
$$
and we can write
$$
2 \st{k}_{\perp}~=~\kt\gam^* ~+~ \kt^*\gam
\auto\label{cga2} 
$$

In the regge limit the transverse part of
an exchanged fermion propagator dominates, i.e. for a massless fermion
$$
\frac{\st{k}}{ k^2} ~~\to ~~ \frac{1}{ 2} \biggl(\frac{\gam^*}{ \kt^*}
+ \frac{\gam} {\kt} \biggr)
\auto\label{frp}
$$  
where the two terms represent the two different chiralities.
For example, the transverse momentum integration of 
a quark-antiquark state with transverse momentum $q_{\perp}$ and
equal chiralities (opposite sign helicities) takes the form
$$
\int~ d^2\kt_1 d^2\kt_2 ~\delta^2(\qt-\kt_1-\kt_2)
\biggl(  \frac{\gam}{ \kt_1 }\otimes \frac{\gam^* }{ \kt_2^* }
~~+~~ \frac{\gam^*}{ \kt_1^* }\otimes \frac {\gam}{ \kt_2 }\biggr)
\auto\label{f2tr}
$$
where the $\otimes$ sign indicates that the two $\gamma$ - matrices are 
separately associated with the two fermion lines.
The contribution of a two fermion state with opposite chiralities is clearly
analagous. However, the distinct 
combinations of same sign and opposite sign chiralities 
are exchanged and interact separately\cite{kms,rk}.
As we elaborate on briefly in Appendix B,
the very different properties of the interaction of 
same sign and opposite sign chirality
exchanges is of fundamental importance.

If we also define
$$
{\Pit}_+ ~=~ -\frac{1}{2}\gam\gam^*~,~
~{\Pit}_-~ =~ -\frac{1}{2}\gam^*\gam~,~
~\Pi_+ ~= ~\frac{1}{2}\gamma_-\gamma_+~,~
~\Pi_- ~=~ \frac{1}{2}\gamma_+\gamma_-
\auto\label{prj1}
$$
then we can write 
$$
\gamma_5 = (\Pi_+ - \Pi_-) ({\Pit}_+ - {\Pit}_-)
\auto\label{prj2}
$$
Spinors in the subspaces $\Pi_-{\Pit}_+$ and 
$\Pi_- {\Pit}_-$ (or $\Pi_+{\Pit}_+$ and 
$\Pi_+ {\Pit}_-$) carry opposite chirality, as
is evident from the following relations 
$$
\eqalign{\gamma_-\gam[1-\gamma_5]\gam^*  ~&=~\gamma_-[1+\gamma_5]\gam \gam^*
~=~ \gamma_-[1- 
 \Pi_- {\Pit}_+] \gam \gam^* \cr
&=~  \gamma_- (1 + \frac{1}{2}\gam \gam^* )\gam \gam^*
~=~  2 \gamma_- \gam \gam^* \cr
~& \cr
\gamma_-\gam^*[1-\gamma_5]\gam ~&=~ \gamma_-[1+\gamma_5]\gam^*\gam 
~=~ \gamma_-[1+ 
 \Pi_- {\Pit}_-]\gam^* \gam \cr
&=~  \gamma_- (1 - \frac{1}{2}\gam^* \gam )\gam^* \gam
~=~  0 
}
\auto\label{prj3}
$$
Similarly, we can show that
$$
\gamma_+\gam^*(1-\gamma_5)\gam ~= ~0 ~,~~~~~
\gamma_+\gam(1- \gamma_5)\gam^*  ~= 
~2 \gamma_+ \gam\gam^*
\auto\label{prj4}
$$
$$
\gamma_-\gam(1+\gamma_5)\gam^* ~= ~0 ~,~~~~~
\gamma_-\gam^*(1+\gamma_5)\gam  ~= ~2 \gamma_-\gam^* \gam 
\auto\label{prj5}
$$
$$
\gamma_+\gam^*(1+\gamma_5)\gam ~= ~0 ~,~~~~~
\gamma_+\gam(1+\gamma_5)\gam^*  ~= ~2 \gamma_+  \gam\gam^*
\auto\label{prj6}
$$

For a vector particle, with momentum along the 1-axis,
the polarization vectors for states with helicity $\lambda = \pm 1$ are 
$$
\eqalign{\epsilon^{\mu}(\lambda = +1)~&=~ - \frac{1}{\sqrt{2}}(0,0,1,i) \cr
\epsilon^{\mu}(\lambda = -1)~&=~  \frac{1}{\sqrt{2}}(0,0,1,-i) }
\auto\label{pov}
$$
A vector boson with helicity $\lambda=-1$
can make a transition to a left-handed intermediate state quark via the 
emission of an antiquark. To calculate the scattering of a vector boson 
with helicity $\lambda=-1$  we introduce an initial coupling of
$\bar{\psi}\gam^*(1-\gamma_5)\psi$ and a final state coupling of
$\bar{\psi}\gam^(1-\gamma_5) \psi$. Utilising the above relations 
we find that, as illustrated in Fig.~A1, 
there is only one non-zero coupling to potential quark-antiquark 
transverse momentum states that could be exchanged.
\begin{center}
\epsfxsize=4.7in
\epsffile{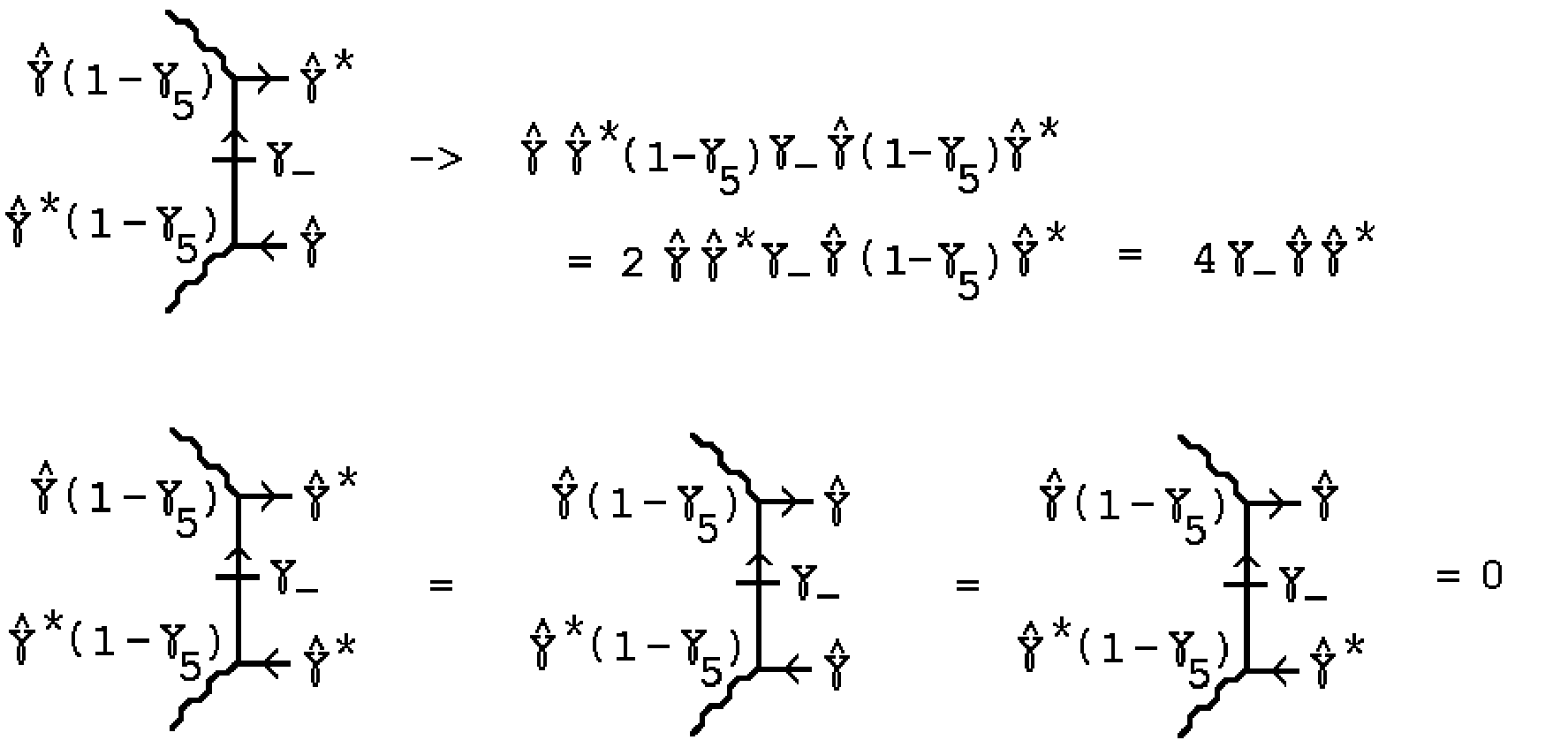}

Fig.~A1 Couplings to Quark-Antiquark Transverse Momentum States
\end{center}
As a consequence,
if we consider the scattering of opposite helicity states there is only one 
possible lowest-order diagram, which is that shown in Fig.~A2(a).
\begin{center}
\epsfxsize=5.5in
\epsffile{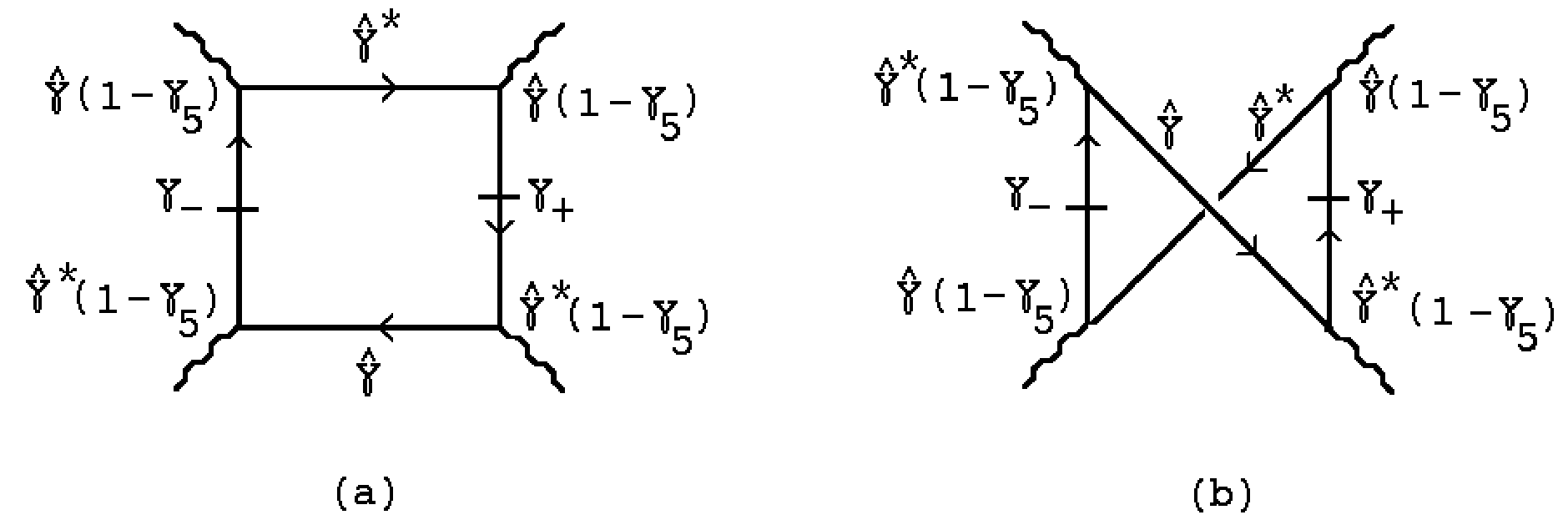}

Fig.~A2 Lowest-Order Diagrams
\end{center}
The initial $\gam^*(1-\gamma_5)$ vertex on the right-hand side of Fig.~A1(a)
represents the coupling of a vector boson with the same polarization, but  
opposite helicity (since it has opposite momentum along the $1$-axis) to that
of the left-hand side vector boson.

A simple way to see that the diagram of Fig.~1(a) contributes to 
opposite helicity scattering is to note that, because of the direction of the 
quark arrow, the intermediate state consists of a left-handed quark, which must
be produced by a negative helicity vector boson, and a right-handed antiquark,
which must be produced by a positive helicity vector boson. The direction of the 
arrow is fixed by choosing the left-hand 
vector meson to be the one with negative helicity.
The diagram of Fig.~1(a) contributes to the $A_{-+}$ helicity amplitude while the 
diagram with the arrow reversed contributes to the $A_{+-}$ helicity amplitude.
By similar reasoning, the diagram of Fig.~2 contributes to the 
$A_{-+}$ helicity amplitude.

The diagram shown in Fig.~A2(b) is the only possibility for
the scattering of states with equal, negative, helicities. In the cross-channel,
in which the incoming and outgoing, right hand, vector mesons are interchanged, 
the diagram untwists to become the diagram of Fig.~A1(a).
Fig.~A2(b) contributes to the $A_{--}$ helicity
amplitude, while the corresponding diagram with the quark arrow reversed 
contributes to the $A_{++}$ helicity amplitude. For the amplitudes
with $\pi^0$ quantum numbers in the $t$ - channel, that we discuss
in this paper, it follows from $CPT$ invariance that
$$
A_{++}~=~A_{--}~~~~~~\hbox{and} ~~~~~~A_{-+}~=~A_{+-}
\auto\label{eha}
$$
Note that, in all diagrams, only same sign chirality states are exchanged.

\newpage

\renewcommand{\theequation}{B.\arabic{equation}}
\setcounter{equation}{0}
\vskip 1cm \noindent
\noindent {\large\bf Appendix B. ~ Review of Leading and Non-Leading Logs}
\vskip 3mm \noindent

As far as we know, the diagrams we discuss in this paper
have not been discussed in
detail in the literature. However, if we were to make the (wrong) 
assumption that the
left-handed coupling does not affect the extraction of high-energy 
logarithms, or (more simply) if we impose a transverse momentum cut-off, 
there are a number of well-known results that would carry over, 
almost directly, into our problem. Just to put the discussion of this paper in 
context, we give here a very brief, non-technical, overview\cite{rk}
of these results.

All the results concern the extraction of leading and non-leading logarithms. 
If we organize the 
quark-antiquark exchange diagrams into distinct series, as 
illustrated in Fig.~B1, depending on the power of $\alpha_s$ (the QCD coupling)
involved, then typical diagrams giving such logarithms are 
\newline \parbox{0.5in}{$~$}
\parbox{4.6in}{\begin{center}
\epsfxsize=4.5in
\epsffile{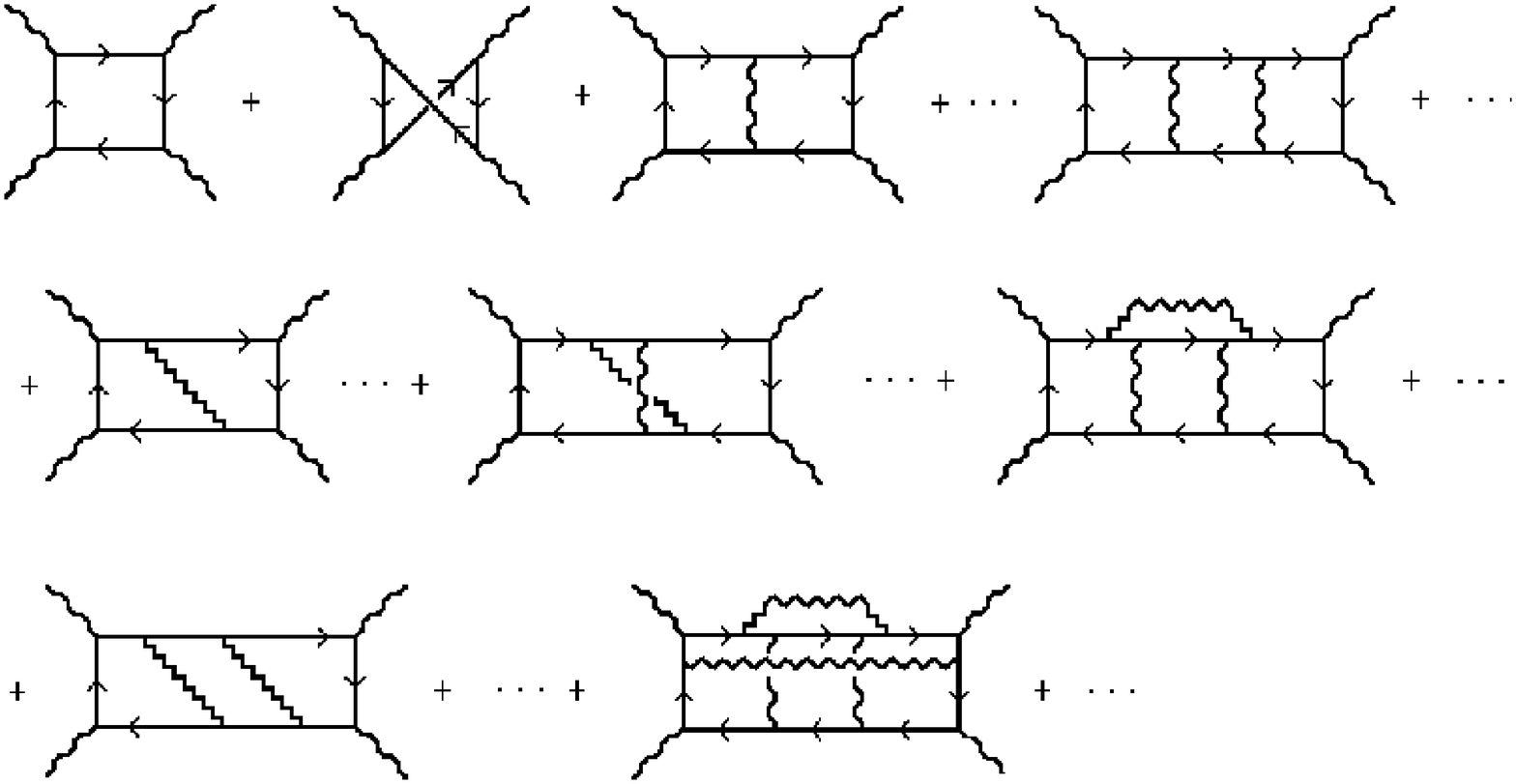}
\end{center}}
\parbox{0.6in}{\begin{center}
$$
O(\alpha_s^0)
$$

$$
O(\alpha_s)
$$

$$
O(\alpha_s^2)
$$

\end{center}}
\begin{center}
Fig.~B1 Quark-Antiquark Exchange Diagrams Organized According to Powers of
$\alpha_s$.

\end{center}
The first series contains purely electroweak diagrams that have 
a logarithmic expansion in $\alpha_w$ (the ``electroweak'' coupling).
The second series contains $O(\alpha_s)$
corrections to the first series, the third series contains $O(\alpha_s^2)$
corrections to the first series, etc.

All diagrams of the form shown in 
Fig.~B1 would be expected to give high-energy amplitudes of the form
$$
A(S,0)~\centerunder{$\sim$}{\raisebox{-5mm}{$S\to \infty $}}
~ \centerunder{\hbox{\large $\Sigma$}}{\raisebox{-3mm}
{${\scriptstyle n,m,r} $}}
~~ a_{nmr}~\alpha_{w}^n~\alpha_s^m~ [lnS]^r
\auto\label{ser}
$$
To make our discussion straightforward we can suppose that 
we, initially, introduce a transverse momentum cut-off so that we can ignore
ultra-violet transverse momentum divergences - including 
both the anomaly power divergences that we discuss in this paper,
and the logarithmic divergences that we discuss below. As a result, 
all the coefficients $a_{nmr}$ can be represented as (sums of) transverse momentum 
diagrams of the form illustrated in Fig.~B2. 

\noindent \parbox{0.6in}{$~$}\parbox{4.4in}{
\epsfxsize=4.3in
\epsffile{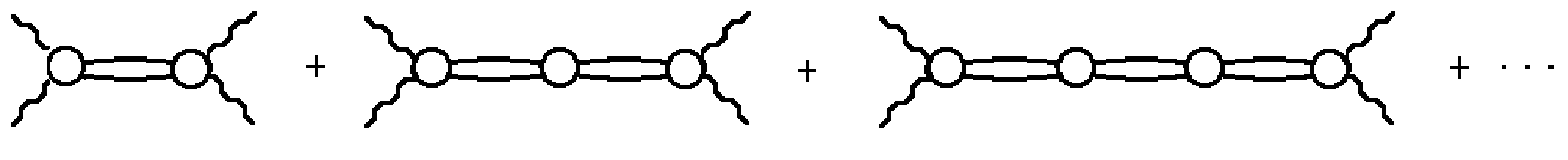}
}
\parbox{0.9in}{\begin{center}
$$
O(\alpha_s^0)
$$
\end{center}}
\newline  \parbox{0.6in}{$~$}\parbox{4.4in}{
\epsfxsize=4.3in
\epsffile{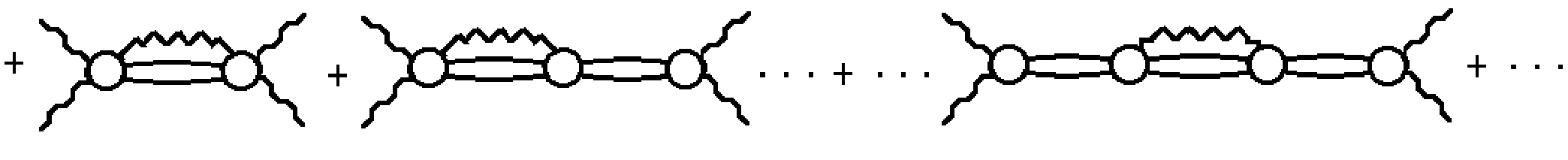}}
\parbox{0.9in}{\begin{center}
$$
O(\alpha_s)
$$
\end{center}
}
\newline  \parbox{0.6in}{$~$}\parbox{4.4in}{
\epsfxsize=4.3in
\epsffile{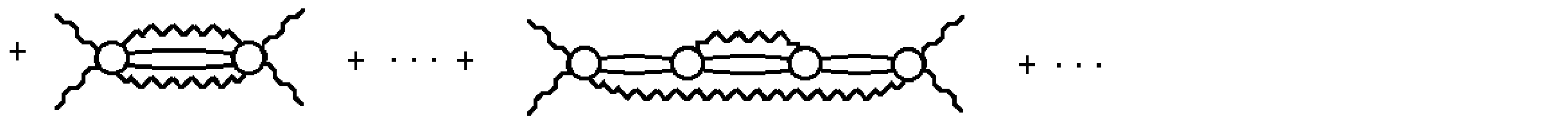}}
\parbox{0.9in}{\begin{center}
$$
O(\alpha_s^2)
$$
\end{center}}
\begin{center}
Fig.~B2 Transverse Momentum Diagrams Originating from the Diagrams of Fig.~B2
\end{center}

With the transverse momentum cut-off in place, the
first two diagrams in Fig.~B1 give a leading log amplitude which contains
the first diagram in Fig.~B2 multiplied by $ln S$ and a 
next-to-leading log amplitude which contains
the same transverse momentum diagram but with no factor of $ln S$.
The third diagram in Fig.~B1 gives a leading log amplitude which contains
the second diagram in Fig.~B2 multiplied by $ln^2 S$ 
and a next-to-leading log amplitude which contains
the first diagram in Fig.~B2 multiplied by $ln S$ and so on. In general,
the external couplings and the internal vertices in the transverse
momentum diagrams aquire more and more structure (involving loop
integrals) as first leading logs,
then next-to-leading logs, next-to-next-to-leading logs,  etc., 
are included in the sum (\ref{ser}). 

The diagram of Fig.~2, appearing in Section 3, 
is the last diagram shown explicitly in the second row of Fig.~B1.
It is first-order in $\alpha_s$ and, conventionally, as noted in Section 3,
we would expect that it's
leading-log contribution would contain the last $O(\alpha_s)$
transverse momentum diagram
shown explicitly in Fig.~B2 - with simple vertices. This diagram  
being obtained by placing all vertical lines on-shell, as in Fig.~5 using 
longitudinal momentum integrations. At the next-to-leading log
level the second  $O(\alpha_s)$
transverse momentum diagram should be generated and the first  $O(\alpha_s)$
transverse momentum diagram, which is the 
diagram that appears in Fig.~1, should be generated by Fig.~2
at the next-to-next-to-leading log level. 

The transverse momentum diagram of Fig.~17 is the first $O(\alpha_s^2)$ 
diagram appearing in Fig.~B2 and would be generated, at leading log, by
the first $O(\alpha_s^2)$ diagram in Fig.~B2. The diagram of Fig.~20 is the
second $O(\alpha_s^2)$ diagram appearing explicitly in Fig.~B1. The
anticipated leading log result for this diagram would be $ln^5 S$
multiplied by the second $O(\alpha_s^2)$ transverse momentum 
diagram appearing explicitly in Fig.~B2 and 
the first $O(\alpha_s^2)$ diagram in Fig.~B2 would be generated at 
the next-to-next-to-leading log level, i.e. multiplied by a factor of $ln^3 S$.

In the leading and non-leading-log studies of
pure vector gauge theories\cite{fl} there is no problem with ultra-violet
divergences, either in transverse momentum, or more generally.  
Only the normal 
(ultra-violet) divergences associated with renormalization have appeared
in the non-leading-log vertices. The Ward 
identities of the gauge theory produce cancelations that lead always to
convergent transverse momentum integrals, with the accompanying
logarithms just those predicted by regge theory. (Even though, as is very
well known by now, individual feynman
diagrams produce transverse momentum divergences that, at first sight,
produce additional logarithms beyond those anticipated by regge theory.)
Equivalently, the complete sum of logarithms and transverse momentum 
diagrams can be rearranged\cite{bs} into sub-series represented by 
reggeon diagrams.

When fermions are involved there is, as we already noted in Section 3, 
the extra subtlety of the logarithmic divergence of fermion transverse momentum
integrals\cite{rk,kms}. Therefore, if the transverse momentum cut-off is removed,
extra powers of $ln S$ will be generated and 
the series (\ref{ser}) must be rearranged appropriately.
However, fermion reggeization is not affected
by the divergences (since the relevant transverse momentum integrals involve
combinations of fermions and gluons). Consequently, as we noted in Section 3, 
when the transverse momentum diagrams are organized into reggeon diagrams,
the presence of the reggeon propagator reduces the divergence
from log to log[log] form.
Also, the reggeon kernel for opposite sign helicities gives convergent integrals.
(As we noted in the previous Appendix, the distinct 
combinations of opposite and same sign helicities 
are exchanged and interact separately.) Only
the kernel for opposite sign helicities (same sign chiralities)
produces logarithmic divergences at large transverse momentum.

In the diagrams discussed in this paper, the quark-antiquark states we consider
are same sign chirality states. However, the anomaly enhancement overwhelms
the logarithmic divergence that would otherwise result. 
We believe this is important, physically. If we start with a transverse momentum 
cut-off both the anomaly power divergence and the logarithmic divergence 
will be absent. 
When the confinement and chiral symmetry breaking described in Section 5 is
implemented via the extraction of infra-red divergences, 
it may be (and the results of \cite{arw02} directly suggest this) that
only (transverse momentum) convergent same sign helicity exchanges are involved
in forming bound states. Since ``double logs''
are, a priori, in conflict with regge theory,  
this is probably necessary for the bound states to be described by regge theory.

\newpage

\renewcommand{\theequation}{C.\arabic{equation}}
\setcounter{equation}{0}
\vskip 1cm \noindent
\noindent {\large\bf Appendix C. ~ The Anomaly and Vector Ward Identities}
\vskip 3mm \noindent

To understand the special nature of Ward identities in the presence of
the anomaly, it is helpful to recall some well-known 
properties of the one loop contribution, shown in Fig.~C1, of massless fermions to
an axial-vector/two-vector three current vertex $T_{\mu \alpha \beta}(k_1,k_2)$. 
\begin{center}
\leavevmode
\epsfxsize=2in
\epsffile{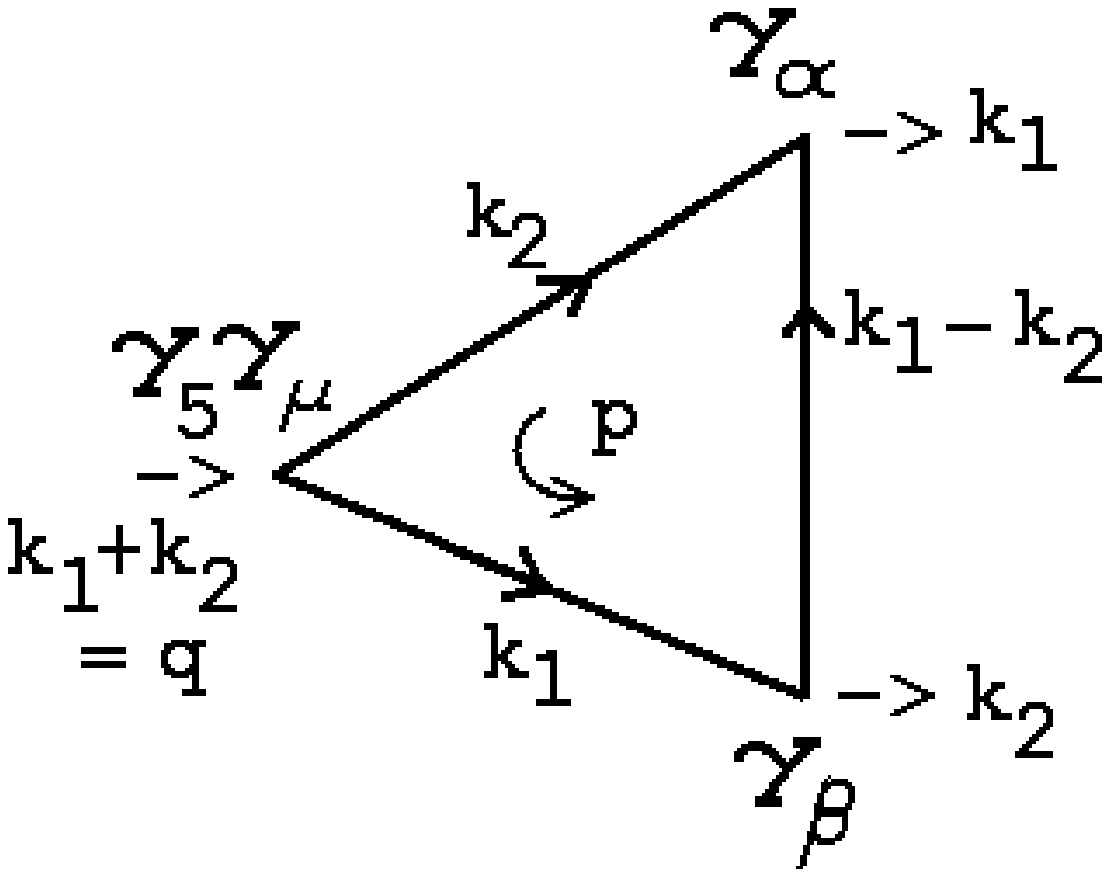}

Fig.~C1 The Fermion Loop Contribution to $T_{\mu \alpha \beta}(k_1,k_2)$

\end{center}
After decomposition into invariant amplitudes  
$$
\eqalign{T_{\mu \alpha \beta}(k_1,k_2) ~&= ~ A_1~
{\hbox{\large $\epsilon$}}_{\sigma\alpha\beta\mu}~ k_1^{\sigma}  ~+~ A_2~ 
{\hbox{\large $\epsilon$}}_{\sigma\alpha\beta\mu} ~k_2^{\sigma} 
~+~A_3~
{\hbox{\large $\epsilon$}}_{\delta \sigma\alpha\mu}~ 
k_{1\beta}k_1^{\delta} k_2^{\sigma}  \cr
~~~& +~A_4~  {\hbox{\large $\epsilon$}}_{\delta \sigma\alpha\mu}
~ k_{2\beta}k_1^{\delta}
k_2^{\sigma}~+~A_5~  {\hbox{\large $\epsilon$}}_{\delta \sigma\beta\mu}
~k_{1\alpha}k_1^{\delta}
k_2^{\sigma}~+~A_6~ {\hbox{\large $\epsilon$}}_{\delta \sigma\beta\mu} 
~ k_{2\alpha}k_1^{\delta}
k_2^{\sigma} }
\auto\label{inde}
$$
the vector Ward identities
$$
k_1^{\alpha}~\Gamma_{\mu \alpha \beta}~=0 ~,~~~
k_2^{\beta}~\Gamma_{\mu \alpha \beta}~=0
\auto \label{vwi}
$$
require 
$$
A_2~=~k_1^2~A_5 ~+~k_1\cdot k_2 ~A_6
\auto\label{vwi1}
$$
$$
A_1~=~k_2^2~A_4 ~+~k_1\cdot k_2 ~A_3
\auto\label{vwi2}
$$

The large momentum region (with appropriate regularization) 
gives an ``anomaly'' contribution to $A_1$ and $A_2$ of the form
$$
T_{\mu \alpha \beta}(k_1,k_2) ~= ~ \frac{1}{ 4 \pi^2}~
{\hbox{\large $\epsilon$}}_{\sigma\alpha\beta\mu}~ k_1^{\sigma}  ~+~ 
\frac{1}{ 4 \pi^2}~ 
{\hbox{\large $\epsilon$}}_{\sigma\alpha\beta\mu} ~k_2^{\sigma} ~~+~~\cdots
\auto\label{uvco}
$$ 
leading to the well-known ``anomalous'' divergence equation 
$$
(k_1 + k_2)^{\mu}~T_{\mu \alpha \beta}~=~
\frac{1}{ 2 {\pi}^2 }~{\hbox{\Large $\epsilon$}}_{\delta\sigma\alpha\beta} 
~k_1^{\delta} k_2^{\sigma}
\auto\label{awi}
$$
For the vector Ward 
identities to hold in the presence of (\ref{uvco}), there must be
related, infra-red singular, contributions to the other $A_i$.
For example, when $k_1^2=0$,  (\ref{vwi1}) becomes
$$
A_2~=~k_1 \cdot k_2~A_6 ~=~\frac{q^2 - k_2^2}{2} ~A_6
\auto\label{vwi11}
$$
implying that there must be a pole in $A_6$, arising from the region
of small internal momentum. In appropriate circumstances, this pole can be 
interpreted as a Goldstone boson pole, signaling chiral symmetry breaking.

If we consider $k_1 \to 0$, and assume that all the
$A_i$ are sufficiently non-singular, then (\ref{inde}) gives
$$
T_{\mu \alpha \beta}(k_1,k_2) ~\centerunder{$\to$}{\raisebox{-5mm}{$k_1 \to 0$}}
~~ ~ A_2~
{\hbox{\large $\epsilon$}}_{\sigma\alpha\beta\mu} ~k_2^{\sigma} 
\auto\label{k10}
$$
which, if we keep only the ultra-violet anomaly term (\ref{uvco}), gives 
$$
T_{\mu \alpha \beta}(k_1,k_2) ~
~\centerunder{$\to$}{\raisebox{-5mm}{$k_1 \to 0$}}
~ ~~\frac{1}{4\pi^2}~
{\hbox{\large $\epsilon$}}_{\sigma\alpha\beta\mu} ~k_2^{\sigma}~~\neq~0 
\auto\label{k101}
$$
Alternatively, if we use (\ref{vwi1}), together with (\ref{k10}), we obtain 
$$
T_{\mu \alpha \beta}(k_1,k_2) ~
~\centerunder{$\to$}{\raisebox{-5mm}{$k_1 \to 0$}}
~~~ (k_1^2A_5 + k_1\cdot k_2 A_6)~
{\hbox{\large $\epsilon$}}_{\sigma\alpha\beta\mu} ~k_2^{\sigma} ~~\to 0
\auto\label{k102}
$$
For consistency, again, there must be infra-red singular contributions
to $T_{\mu \alpha \beta}(k_1,k_2)$ that cancel the ultra-violet
anomaly contribution (\ref{uvco}) and produce the ``Ward identity zero''
(\ref{k102}).

From our point of view, therefore, the
presence of the ultra-violet anomaly (\ref{uvco})
has two consequences. The first is the
anomalous Ward identity (\ref{awi}). The second is that the vector
Ward identities require a cancelation between separate contributions
(with different kinematic structure) from large and 
small internal momentum regions. As a consequence,
if an explicit ultra-violet cut-off is introduced, (\ref{uvco}) will be 
modified and the vector Ward identities will no longer hold. The contribution,
to the vector current divergence, of the pole term in
$A_6$ will survive, however, since it is generated in the infra-red 
region\cite{arw02}. 

\newpage

\renewcommand{\theequation}{D.\arabic{equation}}
\setcounter{equation}{0}
\vskip 1cm \noindent
\noindent {\large\bf Appendix D. ~ Electroweak Ward Identity Cancelations}
\vskip 3mm \noindent

In this Appendix we consider whether Ward identity cancelations can 
remove the longitudinal polarization contributions of vector mesons 
that produce the anomaly enhanced high-energy behavior in the diagrams we
have discussed. 
As described in sub-section {\bf 3.13}, 
we are interested in Ward identity implications when we add all the diagrams 
that effectively replace 
the tree diagram that forms the lower part of Fig.~1 by another tree diagram.
As in Section 3, it will be sufficient for our purposes to consider only 
the diagrams of an abelian theory.

We focus on the same region of phase space as in Section 3, which will be
the basis for all approximations we make. If we ignore (1-$~\gamma_5~$) factors
(which are irrelevant for the present discussion), the lower part of Fig.~1
gives the amplitude shown in Fig.~D1
\begin{center}
\epsfxsize=5in
\epsffile{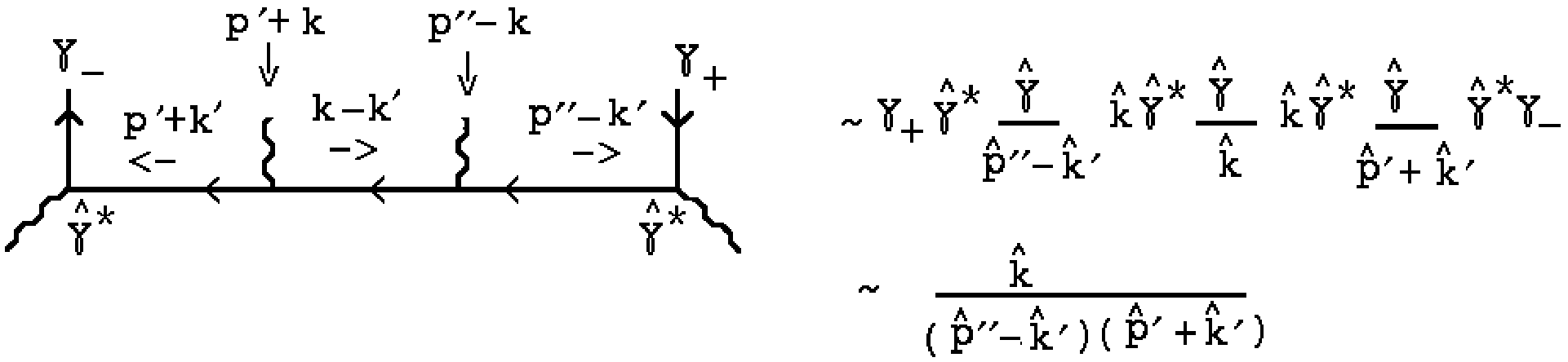}

Fig.~D1 The Tree Diagram Obtained From Fig.~1.
\end{center}
We consider, first, the addition of tree
diagrams in which the internal left-side vector boson line
is attached at all possible points. 
We begin with the diagrams obtained by moving this line to the right.

The subdiagram forming the left part of Fig.~D1
can be split into two pieces as illustrated in Fig.~D2. 
\begin{center}
\epsfxsize=4.5in
\epsffile{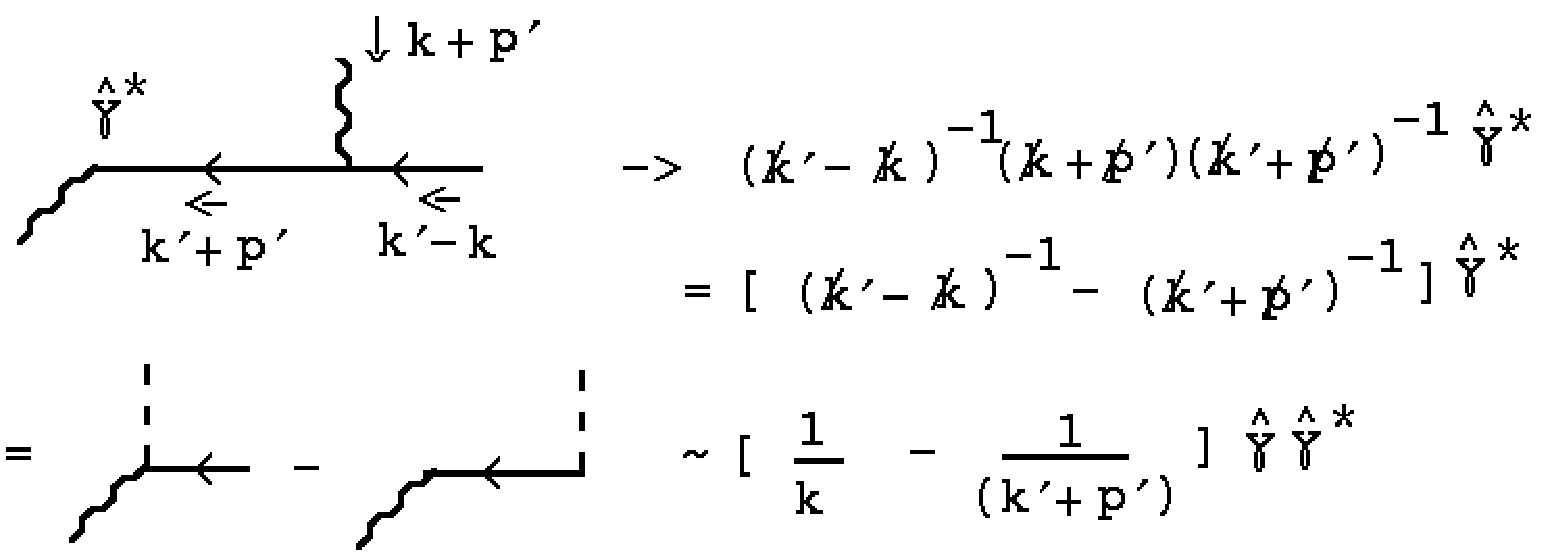}

Fig.~D2 Splitting Fig.~D1
\end{center}
where the dashed line indicates that additional momentum flows in to a vertex 
without changing the algebraic structure.
It is the Ward identity cancelation for the second piece of Fig.~D2 
that involves moving the vector boson line to the right. 
Note, first, that if we combine this term 
with the right-side of the full diagram we obtain the amplitude shown in Fig.~D3
\begin{center}
\epsfxsize=4in
\epsffile{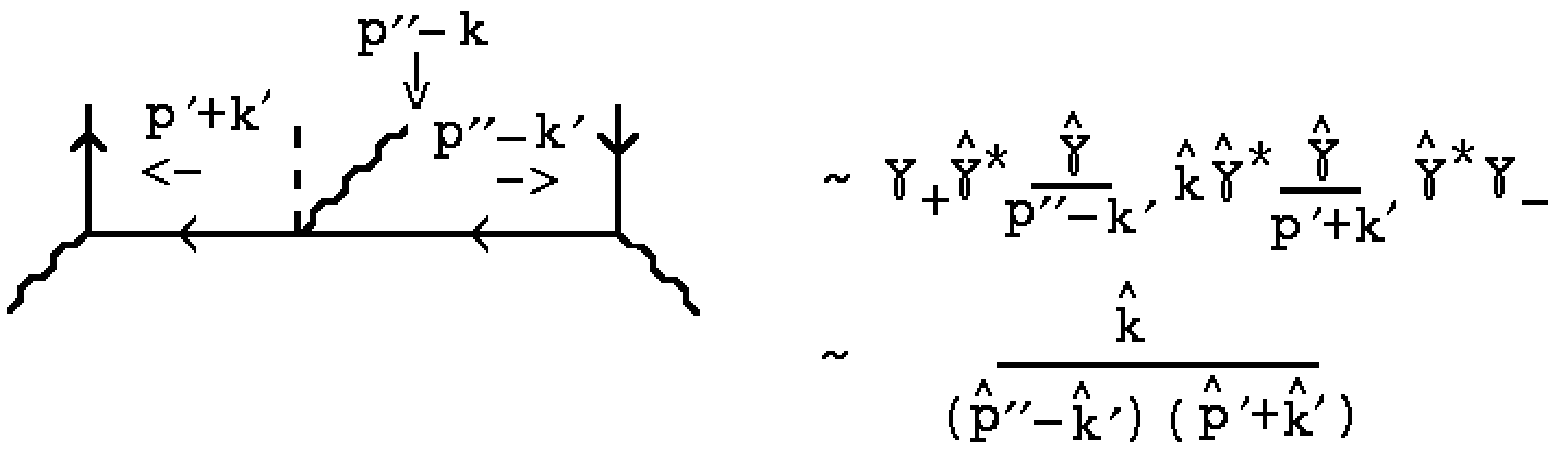}

Fig.~D3 The Amplitude Obtained From the Second Term of Fig.~D2 
\end{center}
and so the relevant piece of Fig.~D1 is retained.

The first tree
diagram obtained by moving the left side internal vector meson to the
right is the second diagram appearing in Fig.~16,
which is the lower part of the feynman diagram appearing in Fig.~17.
We consider the right part of this tree diagram and divide it into 
two pieces as in Fig.~D4.
\begin{center}
\epsfxsize=5in
\epsffile{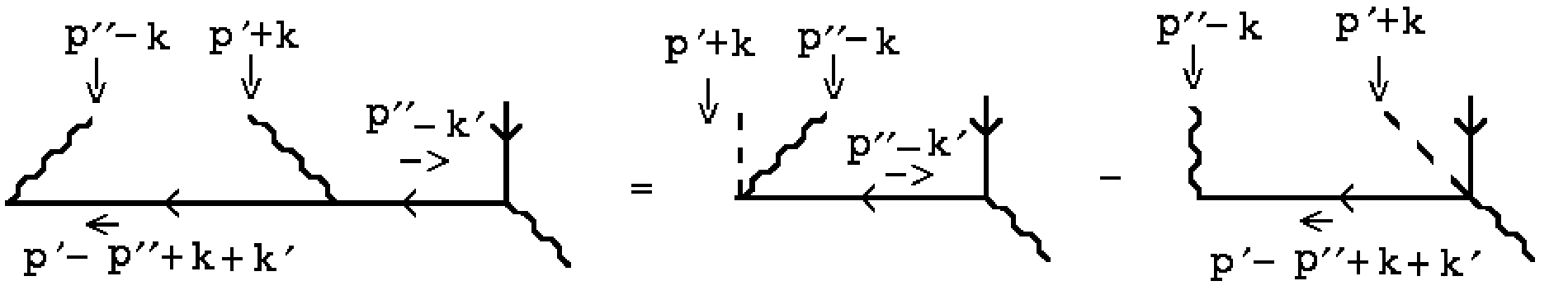}

Fig.~D4 Splitting of Another Tree Diagram 
\end{center}
The first piece gives an amplitude that directly
cancels the amplitude of Fig.~D3. Therefore, it would appear that 
the amplitude involved in the anomaly enhancement is immediately
eliminated. However, there are further cancelations that remain to be 
discussed.

The second piece of Fig.~D4 has to be combined with the 
contribution of the tree diagram shown in Fig.~D5.
\begin{center}
\epsfxsize=2.4in
\epsffile{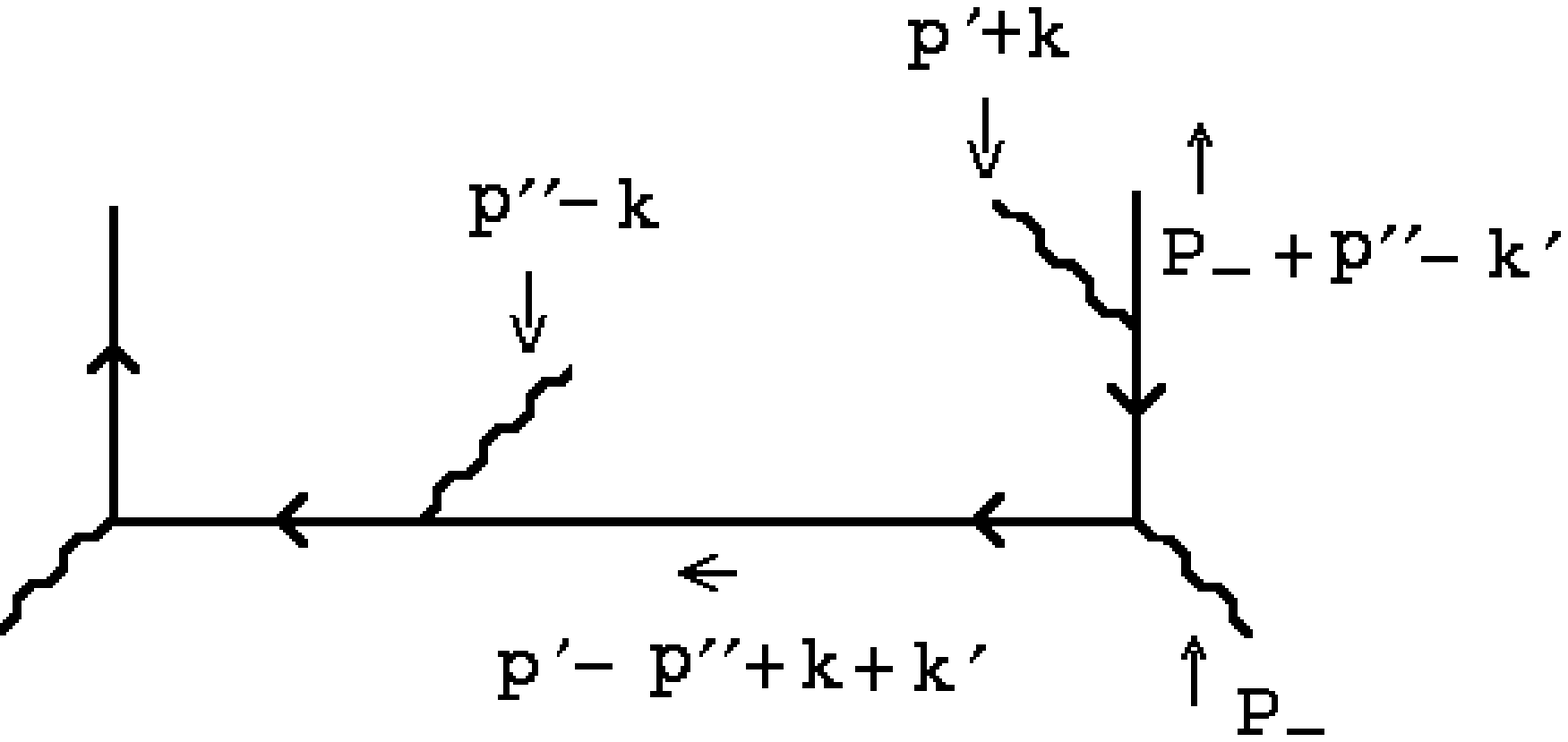}

Fig.~D5 Another Tree Diagram 
\end{center}
Making the usual separation (into two pieces) of the right side of the tree diagram
appearing in Fig.~D5 and removing the piece that 
cancels with the second piece of Fig.~D4 leaves the piece shown in Fig.~D6.
\begin{center}
\epsfxsize=5.8in
\epsffile{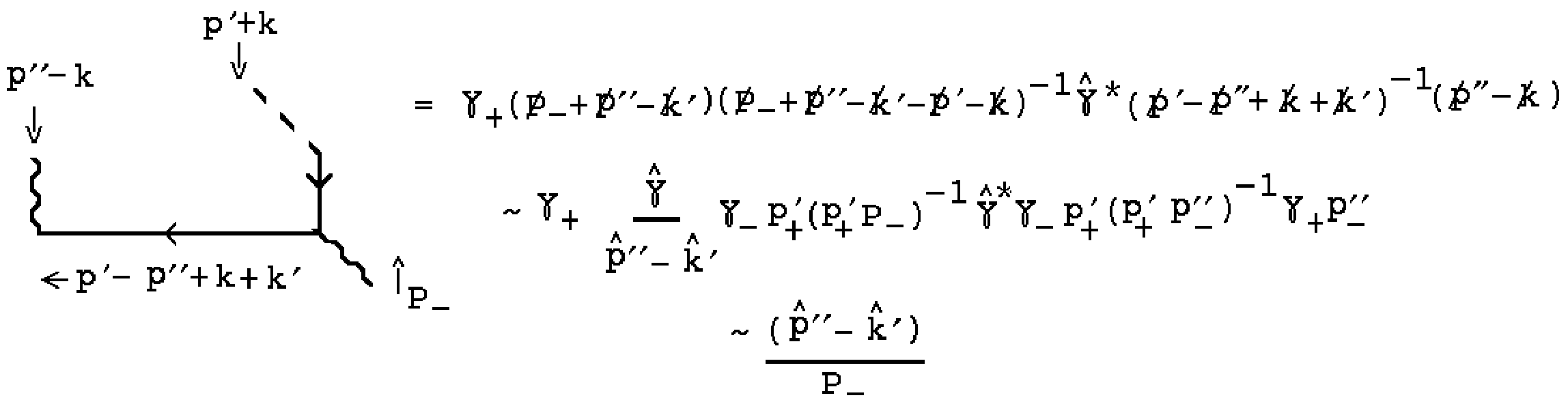}

Fig.~D6 Part of the Tree Diagram Appearing in Fig.~D5
\end{center}
This piece would be zero if the
vertical antiquark were exactly, and not just approximately, on-shell.
 
Since the amplitude of Fig.~D6 goes to zero as $P_- \to \infty$,
with all internal momenta fixed, it is, superficially, a non-leading
asymptotic contribution. However, it has worse large transverse momentum
behavior than the original amplitude of Fig.~D2.
In effect, we have replaced a leading asymptotic
contribution that has manageable internal momentum behavior with a
superficially non-leading contribution with bad internal momentum behavior.
At this point, this substitution does not actually lead to any important 
effects, although this will not be the case for an analagous substitution
that we make later. We obtain the maximal
contribution from the amplitude of Fig.~D6 if we
use the mass-shell condition $P_-k_-' \sim (p_{\perp}'' -k_{\perp}')^2$
and combine the resulting amplitude with the second term in 
Fig.~D2. This gives
$$
\bigl(~\frac{{k'_-}^2}{k_-}~\bigr)~~\frac{1}{(\pt'+\kt')
(p_{\perp}'' - k'_{\perp})}
\auto\label{d1}
$$
This amplitude 
does not have the growth at large $k_{\perp}$ that the amplitude of
Fig.~D1 has and so can be neglected. 

We now consider the contribution of the first term in Fig.~D2. This has
to be combined with tree diagrams obtained by moving the left side internal vector
meson line to the left.
There is
only one such diagram, which is shown in Fig.~D7. 
\begin{center}
\epsfxsize=1.8in
\epsffile{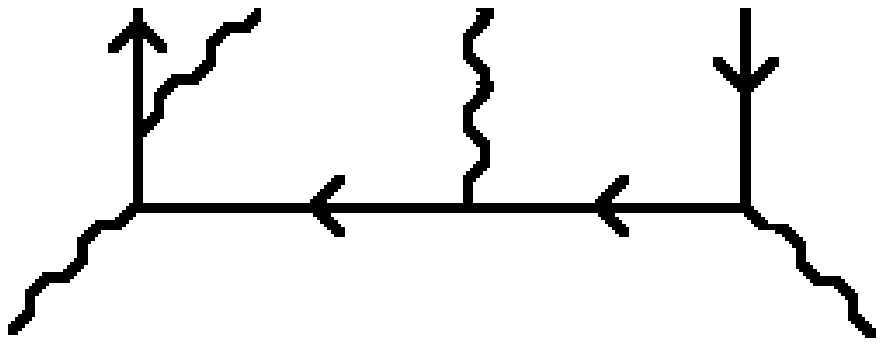}

Fig.~D7 Another Tree Diagram
\end{center}
Normally this contribution to the high-energy
limit would be ignored because an off-shell propagator
carries the large $P_+$ momentum. However,
if we split this diagram into two pieces as in Fig.~D8,
\begin{center}
\epsfxsize=4.5in
\epsffile{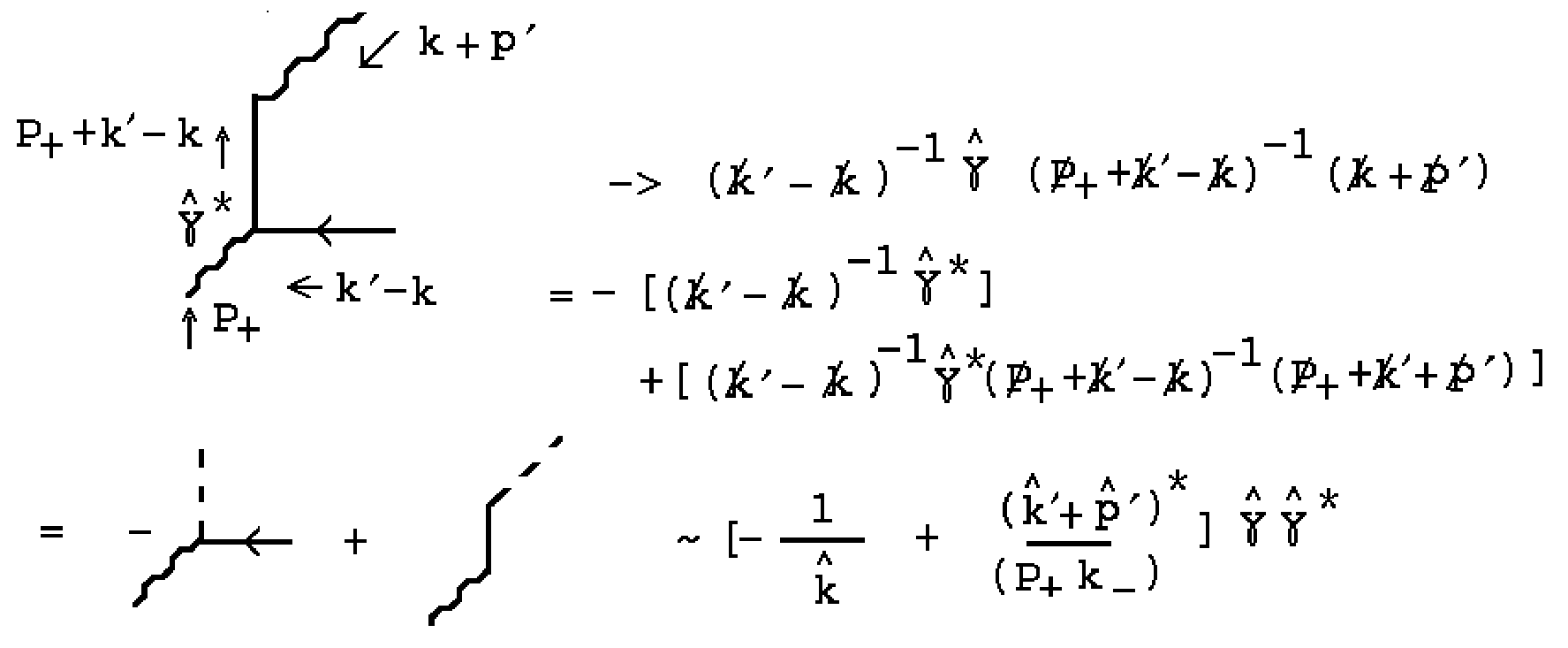}

Fig.~D8 Splitting The Diagram of Fig.~D7
\end{center}
the first piece cancels with the first piece of Fig.~D2. 

If the vertical
quark line were on shell so that the full numerator, and not just the asymptotic 
$\gamma_-$ piece, were present, the second piece of Fig.~D8 would be zero.
In fact, if we use the mass-shell condition
$P_+k_-' \sim (p'_{\perp} + k'_{\perp})^2$ we obtain
$$
\frac{(\pt'+\kt')^*}{P_+k_-}~\sim ~\frac{k'_-}{k_-}~
\frac{(\pt'+\kt')^*}{(p'_{\perp} + k'_{\perp})^2}~= ~
\frac{k'_-}{k_-}~
\frac{1}{(\pt' + \kt')}
\auto\label{d2}
$$
Since both $k_-'$ and $k_-$ are finite in the momentum region 
we are considering, (\ref{d2})
is of the same form as the second term in Fig.~D2. In this case, therefore,
a superficially non-leading
asymptotic contribution, with bad large transverse momentum
behavior, gives a contribution that can not be nelected.

We now consider the additional tree 
diagrams that would be involved 
in a Ward identity for the right side internal
vector meson line. From the above discussion it follows that, after we have added
all such diagrams and carried out the analogous cancelations to those above, 
there will be one surviving 
contribution that will give an amplitude of the form of Fig.~D1. This 
will come from the tree diagram shown in Fig.~D9.
\begin{center}
\epsfxsize=3.5in
\epsffile{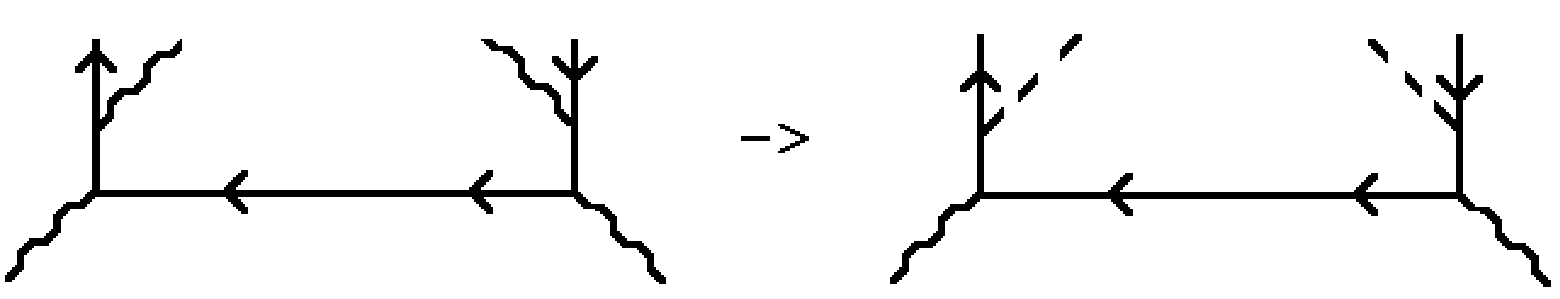}

Fig.~D9 The Tree Diagram Giving the Surviving Amplitude 
\end{center}
The piece of this diagram that we have picked out 
would vanish if both the quark and antiquark
vertical lines were on-shell. From Fig.~D8 and (\ref{d2}) it is clear that 
this piece gives a ``superficially non-leading'' amplitude of the form 
$$
\frac{(\pt'+\kt')^*}{P_+k_-}~\kt~\frac{(\pt''-\kt')^*}{P_-k_+}~
\auto\label{d3}
$$
which, after we use the mass-shell conditions for $P_+$ and $P_-$,
gives the amplitude
$$
\bigl(~\frac{k'_+k'_-}{k_+ k_-}~\bigr)~~~
\frac{\kt}{(\pt' + \kt')(\pt'' - \kt')}
\auto\label{d4}
$$
Since both $k'_+k'_-$ and $k_+ k_-$ are finite, this is, indeed, an amplitude
of the form of Fig.~D1.

We conclude that the large transverse momentum behavior of the 
amplitude in Fig.~3, which combines with the loop
amplitude in the top half of Fig.~1 to give the anomaly, does not cancel
after the imposition of Ward identities. In this respect, therefore,
nothing is gained by implementing the Ward identity cancelations.
However, the lack of cancelation is
entirely due to the asymptotic on-shell nature of the quark and antiquark lines. 
This raises a general issue of principle. Including the remaining amplitude 
that would put these lines exactly on-shell would apparently cancel the behavior 
(\ref{d4}). Yet this amplitude would normally be neglected as contributing only to
non-leading high-energy behavior. It can contribute to the leading
behavior only if there are large transverse momentum divergences. 
In fact, as we have emphasized, to carry out
the Ward identity cancelations we have actually included several diagrams
that would normally be considered non-leading. 

The normal procedure is to effectively assume in advance (and then 
justify a posteori) that transverse momenta will be 
sufficiently cut-off after the summation over all diagrams. The 
leading parts of diagrams can then be safely extracted without worrying 
about transverse momentum divergences. The occurrence of the anomaly
enhancement phenomenon in the diagrams that we have discussed could imply that 
in many other diagrams large transverse 
momenta are also sufficiently important that the normal methods are inadequate.
If this is the case, then it is likely that the general transverse momentum
diagram formalism will fail. Since there would then be no reggeon diagram
formalism, $t$-channel unitarity
is also likely to fail. 
The conclusion, which is really the main conclusion of this paper
(as we already stated in Section 8),
is that to use the transverse momentum diagram formalism (and therefore
to ensure $t$-channel unitarity) it is essential to initially
employ a transverse momentum cut-off.


\newpage

\begin{thebibliography}{99}

\bibitem{fkl}  V.~S.~Fadin, E.~A.~Kuraev and L.~N.~Lipatov, {\it Sov. Phys.
JETP} {\bf 45}, 199 (1977).

\bibitem{bs} J.~B.~Bronzan and R.~L.~Sugar, {\it Phys. Rev.} {\bf D17}, 
585 (1978), this paper organizes into reggeon diagrams the results from 
H.~Cheng and C.~Y.~Lo, Phys. Rev. {\bf D13}, 1131 (1976), 
{\bf D15}, 2959 (1977). 

\bibitem{fs} V.~S.~Fadin and V.~E.~Sherman, Sov. Phys. JETP {\bf 45}, 
861 (1978).

\bibitem{fl1} V.~S.~Fadin and L.~N.~Lipatov, {\it Nucl. Phys.} {\bf B406},
259 (1993).

\bibitem{fl} V.~S.~Fadin and L.~N.~Lipatov, {\it Nucl. Phys.} {\bf B477},
767 (1996) and further references therein.

\bibitem{jb} J.~Bartels, {\it Z. Phys.} {\bf C60}, 471 (1993) and further
references therein.

\bibitem{arw93} A.~R.~White, {\it Int. J. Mod. Phys.} {\bf A8}, 4755 (1993).

\bibitem{arw02} A.~R.~White, {\it Phys. Rev.} {\bf D66}, 056007 (2002).

\bibitem{arw021} A.~R.~White, {\it Phys. Rev.} {\bf D66}, 045009 (2002).

\bibitem{arw84} A.~R.~White, {\it Phys. Rev.} {\bf D29}, 1435 (1984).

\bibitem{rk} For a detailed review and further references, see R.~Kirschner,
{\it Nucl. Phys. Proc. Suppl.} {\bf 51C}, 118 (1996).

\bibitem{gth} G. 't Hooft, {\it Nucl. Phys.} {\bf B33}, 173 (1971).

\bibitem{awb} S.~l.~Adler and W.~B.~Bardeen, {\it Phys. Rev.}
{\bf 182} 1517-1536 (1969).

\bibitem{arw98} A.~R.~White, {\it Phys. Rev.} {\bf D58}, 074008 (1998). 

\bibitem{cri}  A.~A.~Migdal, A.~M.~Polyakov and K.~A.~Ter-Martirosyan, 
{\it Zh. Eksp. Teor.  Fiz.} {\bf 67}, 84 (1974); 
H.~D.~I.~Abarbanel and J.~B.~Bronzan, {\it Phys. Rev.} {\bf D9}, 2397 (1974).

\bibitem{kms} R.~Kirschner, L.~Mankiewicz, A.~Schafer and L.~Szymanowski, 
 {\it Z. Phys.} {\bf C74}, 501 (1997). 
 

\end{thebibliography}
\end{document}